\RequirePackage{fix-cm}
\documentclass[twocolumn,epjc3]{svjour3}  
\RequirePackage{graphicx}
\usepackage{mathtools}
\usepackage[utf8]{inputenc}
\usepackage[percent]{overpic} % in preamble

\RequirePackage{mathptmx}      % use Times fonts if available on your TeX system
\RequirePackage{textcomp,gensymb}
\RequirePackage{amsmath}
\RequirePackage{microtype}
\RequirePackage[numbers,sort&compress]{natbib}
\RequirePackage[colorlinks,citecolor=blue,urlcolor=blue,linkcolor=blue]{hyperref}
\RequirePackage{siunitx}
\newcommand{\be}{\begin{equation}}
	\newcommand{\ee}{\end{equation}}
\newcommand{\bea}{\begin{eqnarray}}
	\newcommand{\eea}{\end{eqnarray}}
\newcommand{\bem}{\begin{multline}}
	\newcommand{\eem}{\end{multline}}

\journalname{Eur. Phys. J. A}
\begin{document}

\title{Partial conservation of seniority in semi-magic nuclei%\thanksref{t1}
}


\author{Chong~Qi\thanksref{addr1}
        }


\thankstext{e1}{e-mail: chongq@kth.se}
\authorrunning{Chong Qi} % if too long for running head

\institute{Department of Physics, KTH Royal Institute of Technology, Roslagstullsbacken 21, Stockholm, SE-106 91, Sweden \label{addr1}
}

\date{Received: date / Accepted: date}

\maketitle
\begin{abstract}
	The concept of seniority plays a central role in nuclear structure physics by classifying many-body states according to the number of unpaired nucleons. While exact seniority conservation holds in single-$j$ systems with $j \leq 7/2$, deviations arise for higher-$j$ orbitals where residual interactions can mix states of different seniority. Surprisingly, certain states in systems with $j \geq 9/2$ exhibit partial conservation of seniority, remaining solvable even when the symmetry is expected to break. This paper reviews the theoretical foundation of the seniority scheme, its connection to pairing interactions and coefficients of fractional parentage, and the conditions under which solvability persists. Particular emphasis is placed on the $j=9/2$ case, where two $v=4$ states with $I=4$ and $I=6$ remain unmixed under arbitrary interactions. We discuss analytical proofs of their existence, numerical studies, and supporting experimental evidence from semi-magic nuclei across five regions of the nuclear chart. Extensions to symbolic shell-model approaches are also presented, highlighting their utility in exploring wave functions and symmetries in many-body systems.

\end{abstract}

\section{Introduction}
\label{intro}
Seniority, often denoted as $v$, is a quantum number introduced in Racah's pioneering work in atomic spectroscopy~\cite{racah1942a,racah1943b,racah1949c,racah1951e,racah1952d}, which quantifies the number of particles in a many-body system that are not coupled into pairs with zero total angular momentum. 
From a broader perspective, the concept of seniority provides a powerful framework for classifying and understanding many-body wave functions and the structure of complex quantum systems. It therefore represents a fundamental quantity in quantum many-body theory. 

While the seniority model originated in atomic physics, its most significant applications in nuclear physics, particularly for describing the structure and decay properties of systems with identical nucleons (neutrons or protons) within a single-$j$ shell~\cite{talmi1952,talmi1993,dean2003,Grawe:2004}. These are systems that are expected to be dominated by the pairing correlation among like particles. The seniority model can provide
analytical solutions and characteristic patterns for energy spectra as well as electromagnetic transitions, offering significant advantages in explaining semi-magic nuclei and open-shell nuclei close to major shell closures. Due to its simplicity, the model is still extensively applied even today in both theoretical~\cite{RevModPhys.84.711,PhysRevC.106.024308,PhysRevC.98.061303,PhysRevC.100.014318,PhysRevC.90.067305,PhysRevC.106.024303,sym16121685,qbwh-c3gp} and experimental works~\cite{PhysRevLett.134.182501,PhysRevResearch.6.L022038,VALIENTEDOBON2021136183,PhysRevLett.129.112501,PhysRevC.108.064313,Ertoprak2020,PhysRevC.105.L031304,MORALES2018706,physics4010024,physics4030048,PhysRevC.110.034303,PhysRevLett.121.022502,Ertoprak2018,wq9m-trj8,PhysRevC.110.034320,Yang_2023}. Historically, the seniority model is closely connected to the nuclear shell model, or more specifically, the independent particle model, which was proposed over 75 years ago~\cite{Mayer1949,HaxelJensenSuess1949,Mayer1950,HaxelJensenSuess1950,MayerJensen1955,Talmi1962,Mayer1948}. It was implicitly applied in the first shell-model description of the ground-state spin-parity and low-lying spectra of nuclei close to shell closures. One usually assigns the seniority quantum number zero, $v=0$, to the ground states of even-even nuclei (meaning all nucleons are paired to spin zero), $v=1$ for the ground state of odd-$A$ nuclei, and $v=2$ for the ground state of odd-odd nuclei. Most applications of the seniority model are restricted to single-$j$ systems~\cite{talmi1993,physics4030048}. Although the seniority scheme is traditionally employed with the nuclear shell model to classify low-lying configurations, the seniority quantum number arises naturally from the pairing Hamiltonian, whose eigenstates are correlated many-body wave functions extending beyond single-particle couplings.
The development of various pairing models~\cite{ZHAO20141,Pan_2020,LIU2021107897} and generalized seniority schemes~\cite{PhysRevC.85.034324,MAHESHWARI2021122277,MAHESHWARI2019121619,jia2016generalized} has further extended its applicability beyond single-$j$ shells, enabling the study of more complex nuclear configurations. 

Beyond nuclear physics, the seniority model has shown valuable applications in other scientific domains, including quantum chemistry~\cite{Men1975,bytautas2011,alcoba2014,chen2015,limacher2015,limacher2016,Perez2018,BYTAUTAS201874,PhysRevA.106.032203,Kossoski2022,VanHootegem2016}, quantum computation~\cite{Gunst2021,halder2025efficientquantumstatepreparation,Elfving2021}, and quantum entanglement~\cite{PhysRevC.106.024303}. Similar to those of nucleons, the seniority-zero configuration can effectively capture static electron correlation for many atomic and molecular systems as well.
The quantum mechanical characterization of all these systems is highly nontrivial due to their
many-body nature. Seniority helps classify ground state phases and provides analytical solutions for specific spin values, offering computational efficiencies for complex systems. The model's widespread applicability underscores its universal power in revealing the underlying symmetries and correlations that govern diverse quantum many-body problems.

However, one can state that the seniority model is an approximate symmetry in many realistic scenarios. In nuclear physics, this leads to phenomena like seniority mixing, especially in deformed nuclei or for nucleons in orbitals with angular momenta $j\geq7/2$. There is extensive ongoing research exploring these limitations, which often utilize seniority isomers and anomalous transition strengths to probe the interplay between single-particle and collective nuclear dynamics. The seniority symmetry would be expected to break even in simple systems with $j\geq 9/2$. However, an interesting phenomenon, which will be referred to as partial conservation of seniority, is that the seniority symmetry is still preserved for states with certain seniority numbers~\cite{Van_Isacker2014-xc}. So far, this has been found only in systems with four particles and $j=9/2$. The property has been proven theoretically~\cite{Escu,Zamick, Isacker1,Isacker2,qi2011partial,qi2017partial,PhysRevC.98.061303}.There are some preliminary experimental evidences as well indicating its presence in nuclear systems~\cite{das2022nature,PhysRevResearch.6.L022038}.

In this article we provide an overview of recent theoretical and experimental efforts in studying this phenomenon.
For completeness, we start in Sec. \ref{sec2} with a brief introduction to the seniority model and some of its applications in nuclear many-body systems, including the origin of seniority, its algebraic foundations, and applications to pairing and solvability in single-$j$ shells.
Sec. \ref{partial} explains the conditions under which certain states, notably in $j=9/2$, remain solvable and unmixed under arbitrary interactions, supported by proofs and examples.
Sec. \ref{sec4} explains the structure of the wave functions for the partially seniority-conserved states as we know them so far and the development of symbolic shell-model methods to construct and analyze wave functions that reveal hidden symmetries and conserved quantum numbers.
This is continued in Sec. \ref{transition}, which analyzes selection rules and $E2$ transition strengths in the seniority scheme, emphasizing how partial conservation manifests in transition hindrance and isomerism.
In Sec. \ref{expt-p} we summarize the experimental progress on measuring the spectra and $E2$ transition properties of semi-magic nuclei involving $j=9/2$ orbitals and present experimental evidence for partial seniority conservation in semi-magic nuclei, outlining prospects for future measurements.
A brief comparison of seniority-coupled and spin-aligned neutron-proton coupling schemes is given in Sec. \ref{np}. Some open directions for further theoretical and experimental study are discussed in the Summary.

\section{The Seniority Symmetry in Nuclear Many-Body Systems}
\label{sec2}

The study of quantum many-body systems, ranging from atomic nuclei to complex molecules and condensed matter, involves intricate interactions among numerous constituent particles. Understanding the emergent properties of these systems necessitates theoretical frameworks that can simplify their description while retaining essential physical insights. The seniority model stands as one such powerful framework, in particular for systems like atomic nuclei, which are governed by strong attractive pairing interactions. Consequently, seniority offers a classification scheme for both the energy spectrum and the eigen wave function. Comprehensive descriptions of the theoretical background and applications of the seniority model in nuclear systems may be found in Ref.~\cite{talmi1993} and in more recent reviews~\cite{Isacker2,sym14122680,physics4030048,VanIsacker2024}.

\begin{figure}[!ht]
    \raggedright
    \includegraphics[width=0.49\textwidth]{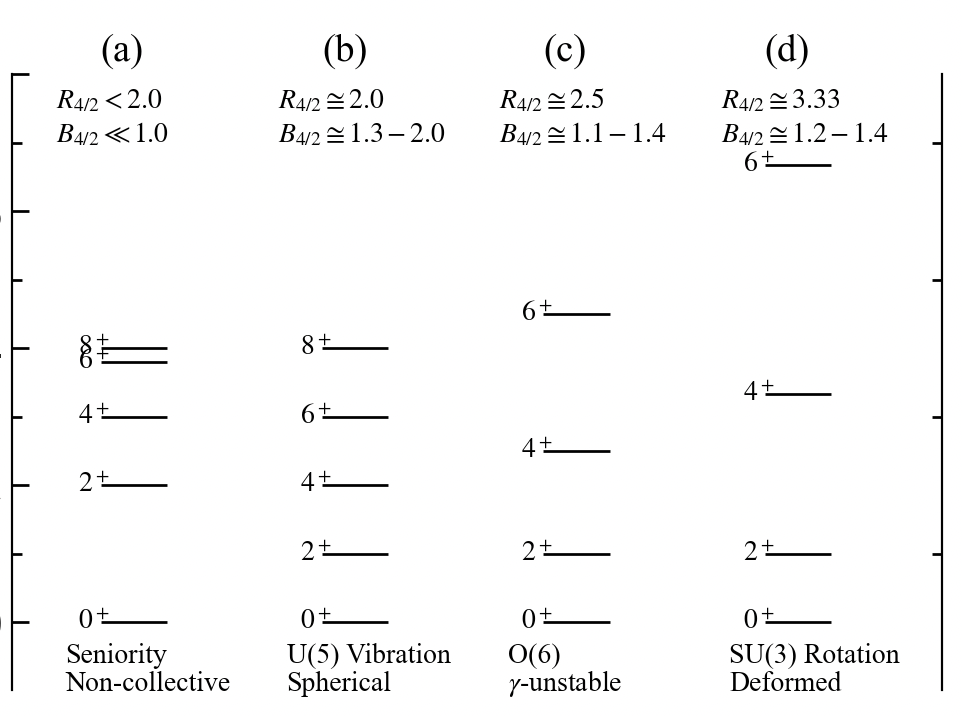}
    \caption{Typical spectral patterns for even-even atomic nuclei: \textbf{(a)} Seniority spectrum characterized by relatively high $2^+$ energy and suppressed excitation energies for other yrast states. The $J=2j-1$ state with maximum spin for a pair often appears as an isomeric state due to the small transition energy and nearly-vanishing transition strength~\cite{sym14122680}. These states are also described as non-collective states; \textbf{(b)} Surface vibrations with equally spaced spectrum; \textbf{(c)} Triaxially soft rotor; and \textbf{(d)} Well-defined collective rotation from deformed nuclei. Those three groups are usually associated with collective motions.\\
    The ratio of excitation energies of the $2_1^+$ and $4_1^+$ states, $R_{4/2}$, is typically $< 2$ for a seniority spectrum, about $2$ for vibrational nuclei, and close to $3.33$ for rotational nuclei. 
    The ratio of E2 transition strengths, $B_{4/2}$, is mostly $>1$ for collective motions but is near zero for seniority symmetry, as transitions between states of the same seniority are almost forbidden.}
    \label{fig:spectra}
\end{figure}

In Fig.~\ref{fig:spectra}, several characteristic spectral patterns of even-even nuclei are shown. The majority of these nuclei are considered as ``deformed" in the intrinsic framework and exhibit regular collective behavior, which can be interpreted as vibrational or rotational motion associated with various nuclear shapes. The associated collective states are typically connected by strong E2 transitions, leading to large values of
\begin{equation}
B_{4/2} \equiv \frac{B(E2; 4_1^+ \rightarrow 2_1^+)}{B(E2; 2_1^+ \rightarrow 0_1^+)}.
\end{equation}
The different collective patterns can be classified algebraically by the symmetry groups U(5), O(6), and SU(3): U(5) corresponds to unitary transformations describing nearly spherical, vibrational nuclei; O(6) is the orthogonal group associated with $\gamma$-soft, triaxial-unstable shapes; and SU(3) is the special unitary group describing axially deformed nuclei with rotational collectivity, with each symmetry reflecting characteristic patterns of energies and transition strengths.

In contrast, semi-magic nuclei and some open-shell nuclei near shell closures exhibit a more irregular, non-collective spectrum. $E2$ transitions among these states are weak, leading to very small $B_{4/2}$ values, particularly among excited states. Many such nuclei show remarkably similar patterns, which can be well understood within the framework of the seniority model.

Another common diagnostic is the ratio of excitation energies of the $2_1^+$ and $4_1^+$ states, 
\begin{equation}
R_{4/2} \equiv \frac{E(4_1^+)}{E(2_1^+)}.
\end{equation}
It is typically $< 2$ for a seniority spectrum. Here, a large  energy gap separates the $I=0$ ground state from a crowding of excited states with different spins. 

\begin{figure}[ht]
\centering
\includegraphics[width=0.45\textwidth]{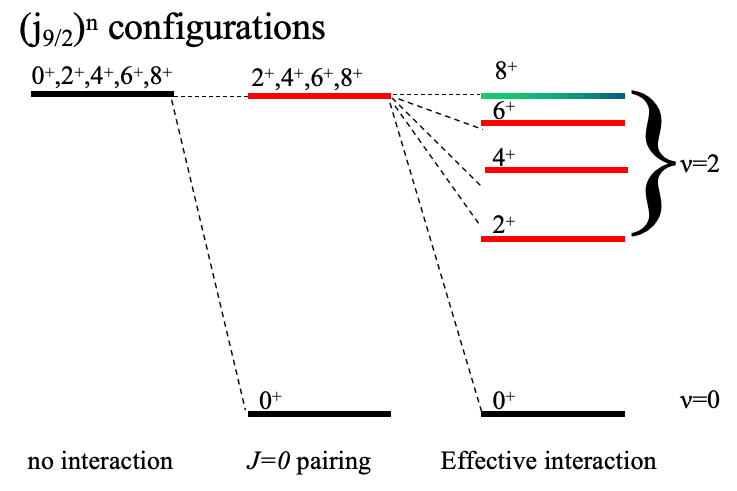}
\caption{A schematic comparison between the spectrum of a single-$j$ even-$n$ system with no interaction (left), a $J=0$ pairing interaction (middle), and a realistic shell-model effective interaction (right). In the non-interacting case, all states with different spins are degenerate, while with strong pairing, the $I=0$ state with $v=0$ emerges as the ground state. Meanwhile, $v=2$ states with even angular momenta $I=2,4,\dots,2j-1$ can be formed by coupling two unpaired particles. Levels are colored to indicate $v=2$ broken-pair states, with the isomeric state highlighted in a different color. Higher-seniority states appear (mostly at higher energies) when more pairs are broken.}
\label{scheme}
\end{figure}

It is straightforward to imagine
that all states with different angular momenta will be degenerate for a system of free particles in a single-$j$ shell, as illustrated in Fig.~\ref{scheme}.  
Due to two particular features of atomic nuclei, distinct shell structure as a result of strong spin-orbit interactions~\cite{Mayer1949,HaxelJensenSuess1949,Mayer1950,HaxelJensenSuess1950,MayerJensen1955,Talmi1962,Mayer1948} and the strongly attractive monopole pairing interaction, many features of atomic nuclei can be easily described by the coupling of few valence nucleons within an isolated single-$j$ shell combined with the seniority model.

The ground state for an even-even system always carries angular momentum zero and $v=0$, while one can expect low-lying states with $v=2$ and relatively higher-lying $v=4,6,\dots$ states. The experimental low-lying spectrum can be accurately reproduced by introducing realistic effective two-body interactions within the shell-model framework, which can be derived either from realistic nucleon-nucleon interactions using perturbation theory or by fitting to experimental data. Under certain conditions, these realistic interactions can still preserve seniority symmetry.
Meanwhile, most mean-field models and theoretical studies of nuclear ground states consider only the monopole pairing interaction, where seniority symmetry is always conserved. 

\begin{table}[]
    \caption{Distribution of total angular momenta $I$ for states in the $j=9/2$ shell with different seniority $v$ and particle/hole number $n$ (maximum $n=5$). Superscripts denote the number of states for a given $I$ (no superscript if there is only one state). States with the same seniority have the same excitation energy under a pairing interaction, as shown in Fig.~2 of Ref.~\cite{RevModPhys.84.711}.
}
    \label{tab:number}
    \centering
\begin{tabular}{llc}
\hline
$v$ & $n$ & $I$ \\
\hline\vspace{0.1cm}
0 & 2,4 & 0 \\\vspace{0.1cm}
1 & 1,3,5 & 9/2 \\\vspace{0.1cm}
2 & 2,4 & 2,4,6,8 \\\vspace{0.1cm}
3 & 3,5 & 3/2, 5/2, 7/2, 9/2, 11/2, 13/2, 15/2, 17/2, 21/2 \\\vspace{0.1cm}
4 & 4 & 0,2,3,4$^2$,5,6$^2$,7,8,9,10,12 \\\vspace{0.1cm}
5 & 5 & 1/2, 5/2, 7/2, 9/2, 11/2, 13/2, 15/2, 17/2, 19/2, 25/2 \\
\hline
\end{tabular}

\end{table}

\begin{figure}[ht]
\centering
\includegraphics[width=0.45\textwidth]{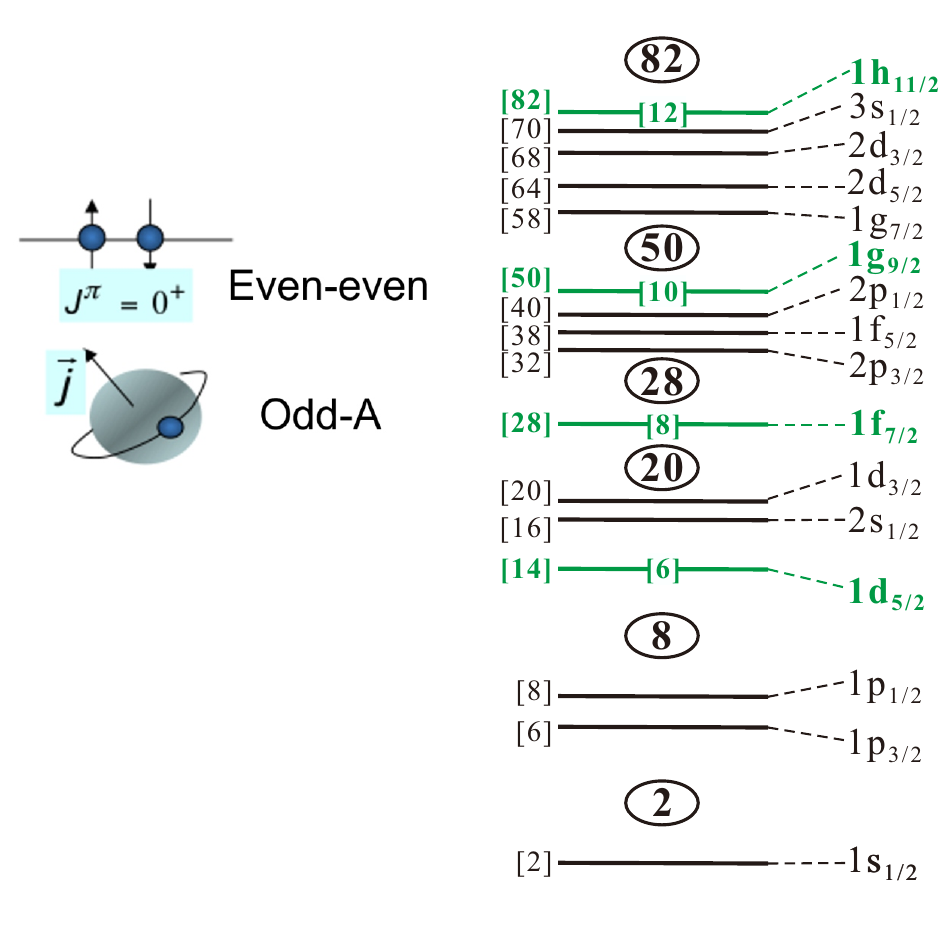}
\caption{Illustration of the nuclear shell structure up to $N/Z=82$, characterized by strong spin-orbit splitting and $jj$ coupling. This leads to a spin-zero ($v=0$) ground state for even-even nuclei and a spin-$j$ ($v=1$) ground state for odd-mass nuclei. Orbitals with higher $j$ values, highlighted in green, are of special interest and involve nuclei just above $N/Z=8$ and $20$, as well as those below or above $N/Z=50$ and $82$.
}
\label{shell}
\end{figure}

It is particularly interesting to study orbitals with relatively high $j$ values, as illustrated in Fig.~\ref{shell}, where a rich spectrum can arise from the coupling of just a few valence nucleons. Table \ref{tab:number} lists all possible states for systems with $n$ identical particles in a $j=9/2$ shell, which are relevant for intermediate-mass and heavy nuclei involving the $g_{9/2}$ and $h_{9/2}$ orbitals. The maximum seniority is five, considering particle-hole symmetry, as the degeneracy is $2\Omega=2j+1=10$ (or the pair degeneracy is $\Omega=5$). The largest angular momenta are $12$ for the even-$n$ systems and $25/2$ for the odd systems.

For even-$n$ systems within $j=9/2$ orbital, in addition to those shown in Fig.~\ref{scheme}, one can have a second set of states with $I^\pi=0^+-8^+$ but with seniority $v=4$ and two extra $4^+$ and $6^+$ states. In other words, the $I^\pi=4^+$ and $6^+$ states are not uniquely defined by angular momentum and seniority quantum numbers and are not “multiplicity-free,” as is often described. As a result, seniority symmetry may be broken in systems with $j=9/2$ and higher spin values for a general two-body interaction. Systems with $j\leq7/2$ are simpler in this regard, as all states can either be uniquely defined by angular momentum and seniority or or remain unmixed because of particle-hole symmetry at mid-shell, ensuring that the seniority symmetry is automatically conserved irrespective of the interaction. 

We highlight two striking features of the seniority scheme in atomic nuclei: the regular behavior of two-particle multiplets and long-lived isomeric states known as seniority isomers.

Groups of nuclei dominated by different numbers of valence nucleons in the same single-$j$ orbital show great similarity in both spectra and electromagnetic transition properties. Examples include:  
Protons in the $0g_{9/2}$ orbital in $N=50$ neutron-deficient isotones $^{92}$Mo, $^{92}$Ru, and $^{92}$Pd (Fig.~1 in Ref.~\cite{PhysRevLett.87.172501}); Neutrons in the $1g_{9/2}$ orbital in Z=82 neutron-rich Pb isotopes $^{210-216}$Pb~\cite{Gottardo2012}.  
Other classic examples involve protons or neutrons in the $0h_{11/2}$ orbitals in Sn isotopes and $N=82$ isotones, as well as systems involving $f_{7/2}$ and $h_{9/2}$ orbitals (see Chapter 21 in Talmi's textbook~\cite{talmi1993}).

\begin{figure*}[htp]
\centering
\includegraphics[width=0.85\textwidth]{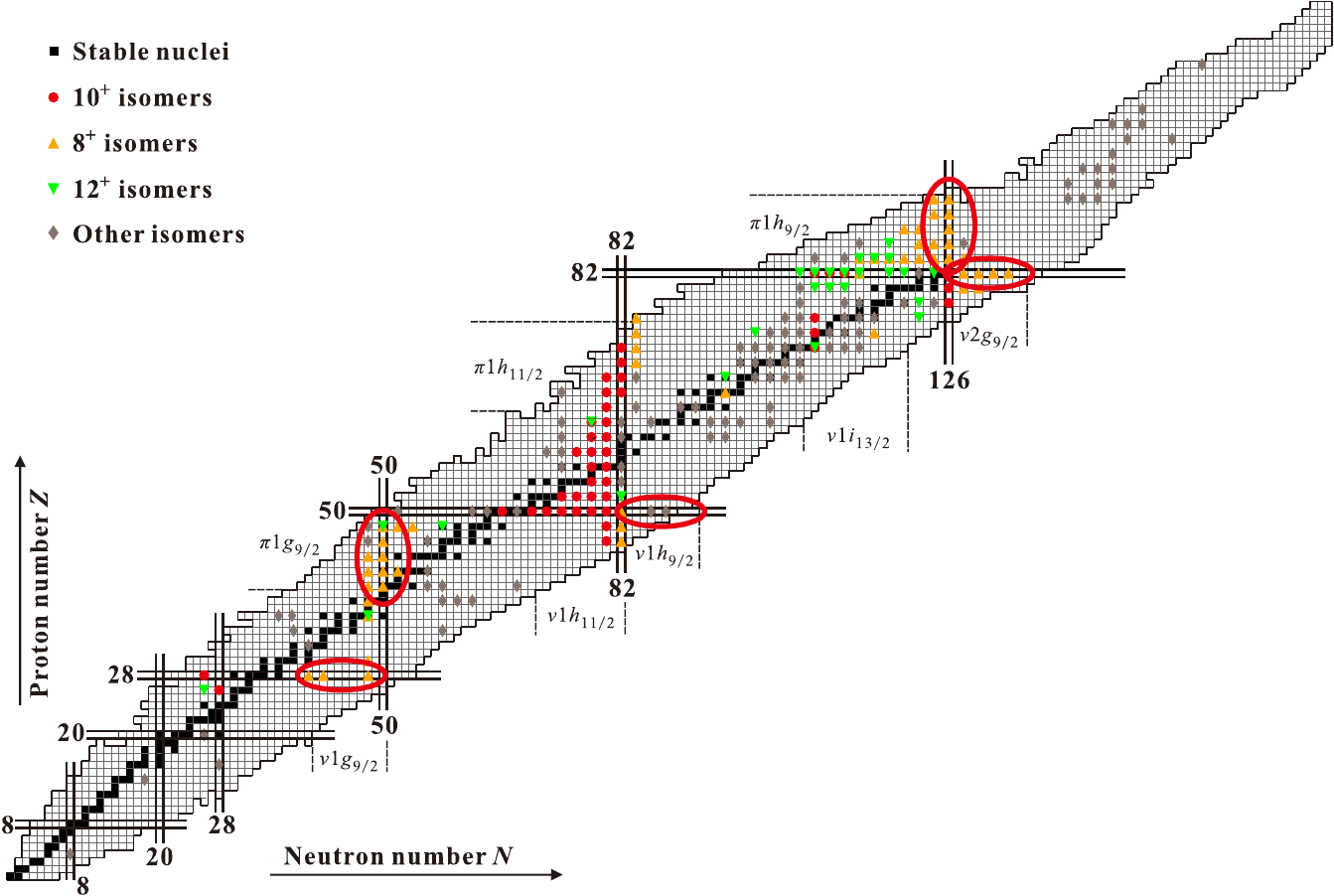}
\caption{Systematics of isomers with different angular momenta in even-even nuclei across the nuclear chart. Most $I=8^+$, $10^+$, and $12^+$ states are seniority isomers arising from the coupling of nucleons in the $j=9/2$ ($g/h$), $11/2$ ($h/i$), and $13/2$ ($i/j$) orbitals. Lower-spin seniority isomers with $I=6$ may occur in neutron-deficient Sn isotopes above $N=50$, neutron-rich Sn isotopes above $N=82$, and in $N=28$ isotones. Red circles highlight regions dominated by the coupling of nucleons in various $j=9/2$ orbitals, including Pb and Ni isotopes and the $N=50$ isotones, which have been extensively studied recently.  
Courtesy of Wenqiang Zhang.}
\label{isomer}
\end{figure*}

Another characteristic pattern in the seniority model is the presence of seniority isomers~\cite{sym14122680,Walker2024,VanIsacker2024,Watanabe_2024} with $v=2$ and $I=2j-1$, as illustrated in Fig.~\ref{scheme}. These states are typically very long lived mostly due to the facts that: i) the energy gap between the seniority isomers and the lower-lying $I=2j-3$ state is relatively small and ii) the $B(E2;I=2j-1 \rightarrow I=2j-3)$ value is very low and even vanishes at mid-shell. The latter is related to the fact that, in seniority coupling in general, matrix elements of even-tensor operators with $\Delta v=0$ vanish at mid-shell $n=(2j+1)/2$. We will explain that in more detail in Sec.~\ref{transition}. In Fig.~\ref{isomer} we marked the presence of isomers with various angular momentum values across the nuclear chart.
More comprehensive studies on the nuclear isomers may be found in Refs.~\cite{Dracoulis_2016,Jain2021,GARG2023101546}. The occurrence of seniority isomers is summarized in Ref.~\cite{sym14122680,VanIsacker2024}. Fig.~7 in Ref.~\cite{physics4030048} shows some typical seniority isomers involving $j = 9/2$ orbitals.

Giulio Racah's seminal series of papers~\cite{racah1942a,racah1943b,racah1949c} applied group theory to atomic spectra, where seniority was introduced in connection with the so-called coefficient of fractional parentage (CFP). The same framework can be applied to atomic nuclei, where an elegant description of the nuclear spectrum can be obtained, as illustrated in the left-most panel in Fig.~\ref{fig:spectra} and in Fig.~\ref{scheme}, and will be explained in more detail later.
We would like to point out that this early connection to group theory shows both the seniority model's inherent mathematical elegance and its potential for providing deep insights into the underlying symmetries of quantum systems (such as quasi-spin SU(2) or orthogonal/symplectic algebras), which is crucial for simplifying complex many-body problems and finding analytical or semi-analytical solutions. One can state that the power of seniority lies not just in its definition of unpaired particles, but in its connection to fundamental algebraic structures that simplify the many-body problem. It laid the groundwork for a systematic approach to classifying complex quantum states. 

\subsection{Building the wave function}
The wave function for a two-particle pair coupled to total angular momentum $J=0$ (resulting in seniority $v=0$) can be written as
\begin{eqnarray}
|J=0;M=0\rangle=\sum_m \langle jm, j{-}m \mid 00\rangle a_m^{\dagger} a_{-m}^{\dagger} |-\rangle\nonumber\\
=\frac{1}{\sqrt{2j+1}} \sum_m (-1)^{j-m} a_m^{\dagger} a_{-m}^{\dagger} |-\rangle,
\end{eqnarray}
where $a_m^\dagger$ creates a particle in the single-particle state $|jm\rangle$, on top of the vacuum or inert core $|-\rangle$, and the Clebsch-Gordan coefficients ensure proper angular momentum coupling. This state forms the building block of the $v=0$ paired configuration, as only time-reversed pairs $(m,-m)$ contribute, reflecting the essence of nuclear pairing.

The corresponding pair creation operator for the $j$-shell is defined as
\begin{equation}\label{pair-oper}
P_j^\dagger = \frac{1}{\sqrt{2j+1}}\sum_m (-1)^{j-m} a_m^\dagger a_{-m}^\dagger.
\end{equation}
In much of the literature, the pair creation operator is also introduced in an unnormalized form by omitting the factor $1/\sqrt{2j+1}$, in which case the resulting state must be normalized separately.

For a system with $n$ even particles, the (unnormalized) ground-state wave function with all particles paired and the low-lying excited states with two unpaired nucleons can be expressed as
\begin{equation}\label{wave-function}
\begin{aligned}
& |{\rm g.s.}\rangle = |v=0; J=0\rangle = \left(P_j^{\dagger}\right)^{n/2} |-\rangle, \\
& |v=2; JM\rangle = \left(P_j^{\dagger}\right)^{(n-2)/2} P^{\dagger}_{JM}(j^2) |-\rangle,
\end{aligned}
\end{equation}
where $P^{\dagger}_{JM}(j^2)$ creates a two-particle excitation with total angular momentum $J$ defined as
\begin{eqnarray}
	P^\dagger_{JM} = \frac{1}{\sqrt{2}} \sum_{m_1,m_2} \langle j m_1, j m_2 | J M \rangle \, a^\dagger_{m_1} a^\dagger_{m_2},
\end{eqnarray}
which reduces to $P_j^\dagger$ for $J=0$. States with $v=4,6,\dots$ are obtained by breaking additional pairs. 

Eq.~\eqref{wave-function} shows that seniority provides a natural hierarchy in constructing nuclear states: from the fully paired ground state ($v=0$), to configurations with a few broken pairs ($v=2,4,\dots$), each with well-defined angular momentum. This forms a simple yet remarkably predictive framework for the structure of nuclei dominated by pairing interactions.

\subsection{The monopole pairing interaction}
It is often argued that the strong attractive pairing interaction in nuclei arises from the short-range part of the nucleon-nucleon force and the large probability amplitude at short relative distances for the
$J=0$ two-body wave function.  For a system of $n$ identical fermions confined to a single-$j$ orbital and interacting through a pairing force, seniority is a conserved quantum number. This symmetry holds not only for pure pairing but also for a broad class of short-range nucleon–nucleon interactions. 

In a system of $n$ identical fermions confined to a single-$j$ orbital, the monopole (isovector) pairing interaction acts only in the two-body channel with total angular momentum $J=0$. In second-quantized form, the pairing Hamiltonian can be written as
\begin{eqnarray}
H_{\rm pair} =& -\,G \sum_{m>0} 
a^\dagger_{jm} a^\dagger_{j\bar m} a_{j\bar m'} a_{jm'}, 
\qquad \bar m \equiv -m\nonumber\\
=&V_0P_j^{\dagger}P_j,
\end{eqnarray}
where $G>0$ is the pairing strength.
Equivalently, in the two-body matrix element (TBME) representation, the interaction is nonzero only for pairs coupled to $J=0$:
\begin{equation}
V_J = \langle j^2; J | \hat V | j^2; J \rangle
=
\begin{cases}
-\frac{2j+1}{2} G, & J=0,\\
0, & J\neq 0.
\end{cases}
\end{equation}

Racah originally introduced this simplified interaction in the context of electron configurations, which allowed for analytical solutions. For a system of $n$ identical particles in a single-$j$ shell, the eigenenergies of the pairing Hamiltonian depend on the seniority quantum number $v$ and particle number $n$ are given by
\begin{equation}\label{eq:pair_energy}
\begin{aligned}
E(n) &= -G \frac{n-v}{4} (2j + 3 - n - v) \\
     &= \frac{n(n-1)}{4} G - \frac{v(v-1)}{4} G - \frac{1}{2} (n-v)(j+1) G.
\end{aligned}
\end{equation}
This formula shows that states with the same seniority $v$ are degenerate, independent of their total angular momentum $I$, reflecting the seniority-conserving nature of the monopole pairing interaction, as illustrated in Fig.~\ref{scheme}.

As shown in Figs.~21.1 \& 21.2 in Ref.~\cite{talmi1993}, the above formula can reproduce nicely the binding energies of a given isotopic or isotonic chain of semi-magic nuclei and their odd-even staggering behavior.
The staggering for neutron can be extracted via the three-point mass filter formula written as
\begin{multline}
	\label{eq:3pointC}
	\Delta^ {(3)}_{C}(N)
	=\frac{1}{2}\left[B(N,Z)+B(N-2,Z)-2B(N-1,Z)\right] \\
	=\frac{1}{2}[S_{2n}(N,Z)-2S_n(N-1,Z)],
\end{multline} 
where $N$ and $Z$ denote the total neutron and proton numbers of the system.
If one assumes  $v=0$ for the ground state of even-even system and $v=1$ for that of an odd system, the expression above can be simplified as~\cite{changizi2015empirical,changizi2016odd}
\begin{eqnarray}
	\label{eqs}
	E(n) &=& \frac{n(n-1)}{4}G - \left\lfloor \tfrac{n}{2} \right\rfloor (j+1)G, \\
	&=& \left\lfloor \tfrac{n}{2} \right\rfloor \left( \left\lfloor \tfrac{n}{2} \right\rfloor -1 \right) G 
	+ \delta_{v,1}\left\lfloor \tfrac{n}{2} \right\rfloor G
	+ \left\lfloor \tfrac{n}{2} \right\rfloor E_2. \nonumber
\end{eqnarray}
where $\left\lfloor n/2\right\rfloor$ denotes the largest integer not exceeding n/2 and corresponds to the total number of $v=0$ pairs.  The symbol $\left\lfloor f\right\rfloor$ (sometimes called the floor function)
means that the greatest integer less than or equal to the real value of function $f$.
The nonlinear term  $\left\lfloor \tfrac{n}{2} \right\rfloor \left( \left\lfloor \tfrac{n}{2} \right\rfloor - 1 \right)$ is related to the energy loss due to the Pauli blocking effect. The $\delta_{v,1}$ term in the above equation indicates the energy loss in the $v=1$ odd system due to the unpaired particle which blocks the scattering of other pairs to its own level. 
$E_2$ is the energy of a single pair, which determines the theoretical odd-even staggering for a single-$j$ system:
\begin{equation}
	\Delta^ {(3)}_{C}(n)=-\frac{1}{2}E_2 = \frac{1}{2}(j+\frac{1}{2})G.
\end{equation}

\subsubsection{Seniority scheme for multi-$j$ systems}
The seniority scheme not only provides exact analytical solutions to the pairing Hamiltonian in a single-$j$ shell, but it can also be extended to multi-$j$ systems, where it either yields exact solutions or allows accurate approximations via the generalized seniority scheme that preserves particle number.
For two particles in a non-degenerate system with the same constant pairing as above, the energy can be evaluated through 
the dispersion relation,
\begin{equation}
	\label{dren}
	G\sum_{i}\frac{2j_i+1}{2\varepsilon_i-E_2}=2.
\end{equation}
The corresponding wave function amplitudes are given by 
\begin{equation}
	X_i=N_n\frac{2j+1}{2\varepsilon_i-E_2}
\end{equation}
where
$N_n$ is the normalization constant and $X_i$ are the amplitudes which have the same phase for the ground state for an attractive pairing interaction. 

For a larger system with $n$ particles,
the total energy for such a system follows closely a relation similar to Eq. (\ref{eqs}) as~\cite{changizi2015empirical}
\begin{eqnarray}
	E(n)
	&\simeq& \left\lfloor \tfrac{n}{2} \right\rfloor \left( \left\lfloor \tfrac{n}{2} \right\rfloor - 1 \right)\mathcal{G} 
	+ \delta_{v,1}(\varepsilon_b+\delta)\nonumber\\
	&&+ \left\lfloor \tfrac{n}{2} \right\rfloor E_2,
\end{eqnarray}
where $\varepsilon_b$ is the single-particle energy of the unpaired orbital in an odd system, and $\delta$ is the energy loss due to the associated blocking effect. $\mathcal{G}$ is a coefficient that is related to the pairing strength $G$ and single-particle gaps.

The exact solutions of the pairing Hamiltonian for a large multi-$j$ system can be derived by diagonalizing the Hamiltonian in the seniority-zero subspace which spans a tiny portion of the total shell model space~\cite{liu2021pairdiag,LIU2021107897}.

\subsection{Conservation of seniority symmetry for a general effective interaction}

The seniority scheme and the pairing Hamiltonian approximation are built on the observation that the $J=0$ pairing force dominates the low-energy 
structure of nuclei near closed shells. While residual 
interactions with $J \neq 0$ are weaker, they can still mix states of different seniority and thus 
destroy the simplicity of the pure pairing model. The two-body Hamiltonian in a single-$j$ shell is completely determined by the interaction between pairs of nucleons, each pair being coupled to a definite angular momentum from which we can calculate the energy of any 
$n$-particle state using CFPs. The general Hamiltonian for a single-$j$ system can be expressed as
\begin{eqnarray}
	H =  \sum_{J} \sum_{M=-J}^{J} V_J \, P^\dagger_{JM} P_{JM},~~~~~~
\end{eqnarray}
where the single-particle energy is neglected and 
\begin{eqnarray}
	V_J = \langle j^2; J | V | j^2; J \rangle
\end{eqnarray}
 is the TBME of the interaction as mentioned earlier.
 
For a broad 
range of $j$ values, most two-body interactions still conserve seniority to a large extent: for $j \leq 7/2$ all 
interactions preserve seniority, for $j \geq 9/2$ seniority 
seems to remain a good approximation as well.
It turns out that the rotationally invariant interaction TBMEs must satisfy $\left\lfloor (2j-3)/6 \right\rfloor$
linear constraints in order to conserve seniority:
\begin{itemize}
	\item 
	Seniority is conserved for systems with $j\leq 7/2$ as $(2j-3)/6<1$ and $\left\lfloor (2j-3)/6 \right\rfloor=0$.
	\item The TBMEs must satisfy one condition to conserve seniority for orbitals $j=9/2,11/2$ and 13/2, as $1\leq (2j-3)/6<2$ and $\left\lfloor (2j-3)/6 \right\rfloor=1$. 
	\item For $j = 15/2$, two constraints are required.
This may correspond to the highest single-particle orbital of interest in nuclear physics.
\end{itemize}
These constraints constitute the necessary and sufficient conditions for the exact preservation of seniority under a given effective interaction.
It can be stated that the number of constraints is much smaller than the number of TBMEs, so the constraints are not particularly restrictive.  And it turns out that seniority is approximately preserved by many realistic interactions.

\subsection{Solvability of the single-$j$ system} \label{solvable}
Solvability refers to the ability to find exact, closed-form solutions for the system's eigenvalues and eigenfunctions. For many-body systems, such exact solutions are rare due to the complexity introduced by interactions between particles. But there are many interesting examples in single-$j$ systems.
Ref.~\cite{PhysRevLett.87.172501}, titled ``Partially Solvable Pair-Coupling Models with Seniority-Conserving Interactions", presented an algebraic framework for constructing solvable 
and partially solvable shell-model Hamiltonian based on the conservation of seniority in a 
single-$j$ shell. As explained at the beginning of the paper, a model is often referred to be solvable if all its energy levels can
be determined analytically. A system is partially solvable and is having a
partial dynamical symmetry~\cite{Alhassid1992,Leviatan1996} if some
of its energy levels can be determined analytically.
Ref.~\cite{PhysRevLett.87.172501} employs a quasi-spin tensor decomposition of the two-nucleon interaction, building on the quasi-spin formalism originally introduced by Kerman~\cite{Kerman1961} and further developed by Helmers~\cite{Helmers1961}, from which the necessary algebraic conditions for seniority conservation listed above are derived.

The authors employ the $SU(2)$ quasi-spin algebra to classify operators as quasi-spin tensors and states by their quasi-spin ($S,S_0$), where $S$ is the total quasi-spin quantum number, analogous to angular momentum and $S_0$
is the projection (analogous to magnetic quantum number 
$m$ in spin), and angular momentum ($J,M$) quantum numbers. The quasi-spin quantum numbers are related to seniority $\nu$ and the number of particles $n$ by 
\begin{equation}
    S=1/2(\Omega-\nu),
\end{equation}
and
\begin{equation}
S_0=1/2(n-\nu).\end{equation}
The maximum quasi-spin occurs when all particles are paired with \(v = 0\) (\(S = \frac{\Omega}{2}\)), which decreases by 1/2 if one add an unpaired particle, and \(S_0\) counts the number of pairs.
 A general Hamiltonian with seniority-conserving two-body interactions can be written in the 
compact form~\cite{talmi1993}
\begin{equation}
    H = \epsilon \hat{n} - G \, \hat{S}_1 \hat{S}_2 + V_0 ,
\end{equation}
where $\hat{S}_i$ are SU(2) quasi-spin generators, $\hat{n}$ is the number operator, and 
$V_0$ is a linear combination of scalar operators $\left( C_J \otimes C_J \right)^{(0)}$. 
The quasi-spin and symplectic algebras form a dual pair, ensuring that states can be labeled 
by particle number $n$, seniority $v$, and angular momentum $J$.
By deriving the algebraic conditions, the authors show that seniority is conserved for a wide range of two-body interactions for larger values of $j$. This makes it possible to construct realistic solvable and partially solvable single-shell models with eigenstates classified by a spectrum generating algebra. The paper provides explicit expressions for the seniority-two states and for the states of largest and next largest angular momentum for any seniority. These states are referred to as ``multiplicity-free'' because they are uniquely defined by seniority and angular momentum quantum numbers. For those multiplicity-free states, explicit analytic 
energies can be obtained. For example, in the $j=9/2$ shell with even particle number $N$,
\begin{equation}
    E_{v,J} = E_{N0} + a J(J+1) + b \, v (2j+3-v) + c \, Z_{v,J},
\end{equation}
where $Z_{v,J}$ are eigenvalues of scalar operators such as $(C_3 \otimes C_3)^{(0)}$.

Applications to the $N=50$ isotones demonstrate that seniority-conserving interactions reproduce 
observed spectra and that the framework provides both conceptual insight into 
partial dynamical symmetries and a practical tool for shell-model studies of medium-mass nuclei.

The two-body interaction within a $j$-shell can decomposed into irreducible tensor components under the quasi-spin SU(2) algebra. Based on that it can be shown that most components conserve seniority.
A two-body interaction in a single-$j$ shell is expanded as
\begin{equation}
    V = \frac{1}{4} \sum_J (2J+1) V_J \, (A^\dagger_J \otimes B_J)^{(0)} ,
\end{equation}
where the operator set $(A,B,C)$ form components of quasi-spin tensors defined as
\begin{align}
    A^\dagger_{JM} &= \frac{1}{\sqrt{2}} \sum_{mn} (jm, jn | JM) \, a^\dagger_{jn} a^\dagger_{jm}, \\
    B_{JM} &= \frac{1}{\sqrt{2}} \sum_{mn} (jm, jn | JM) \, a_{jm} a_{jn}, \\
    C_{JM} &= (a^\dagger_j \otimes a_j)_{JM}, \qquad (J \neq 0), \\
    C_{00} &= \tfrac{1}{2}\left[(a^\dagger_j \otimes a_j)_{00} + (a_j \otimes a^\dagger_j)_{00}\right] 
           = \sqrt{\tfrac{1}{2j+1}} \, \hat{S}_0 ,
\end{align}
where $a^\dagger_{jm}, a_{jm}$ are fermion creation and annihilation operators.  To classify interactions by quasi-spin rank, the bilinear products are normal-ordered. The 
relations
\begin{align}
    (B_J \otimes A^\dagger_J)^{(0)} &= (A^\dagger_J \otimes B_J)^{(0)} 
      + \frac{2J+1}{V} (\hat{n} - V), \label{eq:12} \\
    (C_J \otimes C_J)^{(0)} &= 
      \begin{cases}
        \tfrac{1}{2V} (\hat{n}-V)^2 , & J=0 , \\
        \sum_g M^V_{Jg} (A^\dagger_g \otimes B_g)^{(0)} 
          + \tfrac{2J+1}{2V} \hat{n}, & J \neq 0 ,
      \end{cases} \label{eq:13}
\end{align}
introduce the matrix $M^V_{Jg}$. It is given by
\begin{equation}
    M^V_{Jg} = 2(2J+1)(2g+1)
    \begin{Bmatrix} j & j & g \\ j & j & J \end{Bmatrix}, \label{eq:14}
\end{equation}
where the curly bracket denotes a $6j$ symbol.
By diagonalizing the corresponding matrix $M^V_{Jg}$, Rowe and Rosensteel showed that all eigenvalues are either -1 or 2, which give linearly independent combinations of quasi-spin scalar and rank-two quasi-spin tensors, respectively. The number of eigenstates with eigenvalue 2 determines the  linear combination constraint conditions of interactions  in order that it should conserve seniority.

\subsection{Seniority and the Purity of Single-$j$ Configurations}

The seniority scheme can be embedded within large-scale shell-model and \textit{ab initio} frameworks, yet its conceptual elegance is most clearly seen in single-$j$ or other simplified systems. One may naturally question the validity of the seniority scheme itself and the purity of the single-$j$ shell approximation. Even after more than 75 years since the discovery of the nuclear shell model, the interpretation of single-particle orbitals remains somewhat controversial~\cite{doi:10.1142/S0218301305003570,physics4030048}. From a purely theoretical standpoint, one might even challenge the very concept of nuclear single-particle levels—the foundation of seniority coupling—since, in principle, these levels are not directly observable quantities~\cite{Duguet2015}.

For instance, the low-lying states in $^{209}$Pb, traditionally interpreted as single-particle excitations outside the doubly magic $^{208}$Pb core, are in reality 209-particle states. Similarly, the states in $^{210}$Pb with seniority $v=0,2$ correspond to fully interacting 210-particle systems. Even when employing large-scale shell-model or \textit{ab initio} no-core shell-model calculations, which are often regarded as the most faithful representations of the nucleus, the resulting wave functions still depend on the chosen single-particle basis and on the truncation of the model space.

This dependence has important practical implications. As discussed in Sec.~\ref{expt-p}, comparisons between shell-model calculations performed in different model spaces require caution. The choice of model space strongly influences both the effective interaction and the resulting wave functions. Calculations restricted to a single major shell may emphasize seniority-conserving properties, whereas extended model spaces that include cross-shell excitations can "dilute" such patterns by introducing additional correlations. Apparent discrepancies between results from different model spaces may therefore reflect differences in representation rather than genuine differences in nuclear structure.

At the same time, one should not overlook the simple and remarkably regular patterns displayed by complex nuclei, as revealed by seniority systematics and collective behavior. Regardless of the nuances of the shell model, the seniority framework remains a powerful and convenient tool: it captures the essential features of complex nuclear wave functions and provides clear, physically meaningful interpretations of the structural similarities observed among neighboring nuclei. In this sense, seniority serves both as a practical computational device and as a conceptual lens for understanding the emergent simplicity underlying the apparent complexity of atomic nuclei.

\section{Partial seniority conservation}\label{partial}

\begin{figure}[ht]
	\centering
	\includegraphics[width=0.5\textwidth]{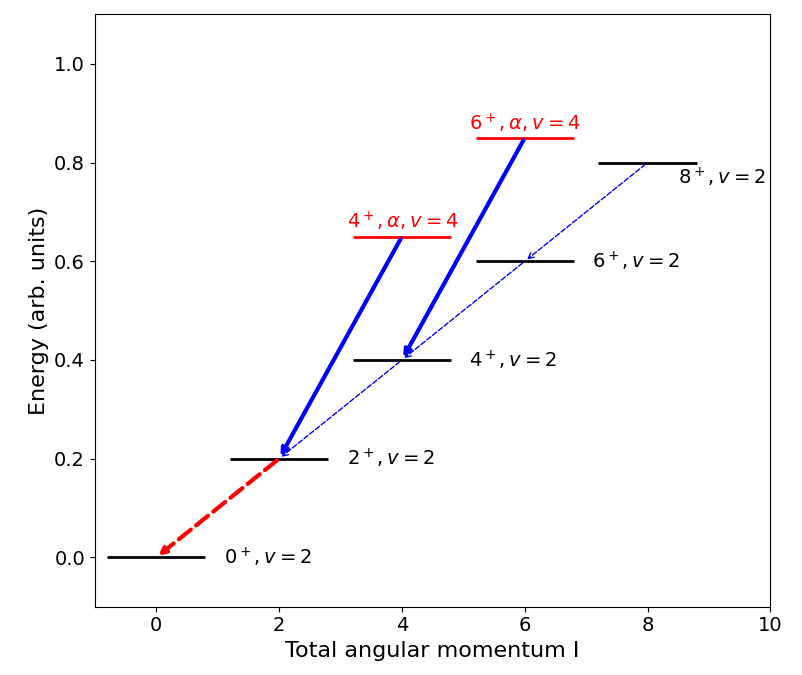}
	\caption{Illustration of the low-lying states for an $n=4$ system in a $j=9/2$ orbital and the associated E2 cascade. In the seniority scheme, transitions with $\Delta v = 0$ are typically weak, whereas the three $\Delta v = 2$ transitions can be relatively strong. The two $v = 4$ states, labeled as $\alpha$, exhibit partial seniority conservation. These states are expected to be low-lying, though their relative positions depend on the effective interaction and the nuclear region.
In addition, there are another set of $v = 4$, $I=4,6$ states which are expected to lie at substantially higher energy.}
	\label{alpha}
\end{figure}

Understanding the concept of solvability discussed in Sec. \ref{solvable} is crucial as both a theoretical tool and a conceptual guide because it provides insights into the structure and symmetry behavior of quantum systems and allows for deriving analytical solutions for energy levels, transition rates, and other properties, offering exact insights that numerical diagonalizations, however precise, cannot easily reveal.
Preserving seniority, however, does not guarantee that the system is exactly solvable nor does not ensure that explicit algebraic formulas for its eigenvalues and eigen functions exist. In Ref.~\cite{Isacker1} seniority is interpreted as a form of partial dynamical symmetry.
The solvability of the single-$j$ system may be grouped into three categories~\cite{qian2018partial}:

{\it Type 1:} Solvability arises in its simplest form, when a given angular momentum $J$ corresponds to a single state within the configuration space. In such cases, the state is automatically solvable: the uniqueness of the state ensures that its eigenvalue can always be expressed as a fixed linear combination of TBMEs, independent of the interaction details. This type of solvability is a direct consequence of angular momentum uniqueness and has been recognized since the earliest developments of the shell model.

{\it Type 2:} Solvability represents a far more subtle and surprising phenomenon that emerges when multiple states share the same angular momentum $J$. Ordinarily, such degeneracy leads to configuration mixing under a general two-body Hamiltonian, and the energies must be obtained through explicit diagonalization, depending on the details of the interaction and seemingly precluding analytic solutions. However, remarkable exceptions exist.  In the classic case of the $(9/2)^4$ configuration, two states defy this expectation: they remain exactly solvable despite the presence of degeneracy. Their wave functions retain fixed analytic forms independent of the interaction, a phenomenon which we refer to as partial seniority conservation. This discovery highlights the presence of hidden symmetries that act selectively within the Hilbert space. These states are concrete realizations of a partial dynamical symmetry, where exact algebraic structure persists for a restricted subset of states even though the full system is not solvable.  So far, only two such states are firmly established, as illustrated in Fig.~\ref{alpha}. Determining whether these are isolated occurrences or part of a broader systematic pattern remains an open question.

{\it Type 3:} 
Solvability arises when a state is uniquely defined by its total angular momentum $I$ and seniority $v$. 
This ensures exact solvability for a \emph{seniority-conserving} interaction since no other states with the same $I$ and $v$ exist to mix with it. 
Classic examples include mid-shell systems such as $(9/2)^5$ listed in Table~\ref{tab:number}, corresponding to $^{213}$Pb with five neutrons in the $1g_{9/2}$ shell (Sec.~\ref{208Pb}) and intermediate-mass nucleus like $^{95}$Rh (Sec.~\ref{95Rh}). However, the situation is richer than this simple picture suggests. 
In these cases, the off-diagonal Hamiltonian matrix elements connecting states of seniority differing by two vanish identically, ensuring that \emph{certain states} are analytically solvable even for \emph{seniority-non-conserving} interactions. 
For higher-$j$ configurations, for example $(11/2)^6$, additional solvable or partially solvable states exist even when $J$ and $v$ do not uniquely define a single state. 
Examples include the $v=6, J=0$ and $v=4, J=3$ states, which remain solvable for any interaction. 

Moreover, there exist partially solvable states of a different kind. 
These states possess well-defined seniority and angular momentum for all interactions, 
yet they can mix among themselves because they are not uniquely determined by $v$ and $I$. 
As a result, their internal structure depends on the specific interaction. 
Such cases represent a distinct and more general form of partial solvability, extending beyond the strict Type~2 classification.

From a broader perspective, these three categories of solvability illustrate the layered structure of symmetries in nuclear models. Type 1 solvability reflects the simplest consequences of angular momentum uniqueness. Type 3 solvability shows how seniority conservation can protect states from mixing. Type 2 solvability, meanwhile, reveals the most intriguing possibility: the emergence of exact solutions in situations where standard symmetry arguments would predict none. Taken together, these cases demonstrate that solvability is not an binary (all-or-nothing) concept, but rather a continuous spectrum shaped by the interplay of angular momentum coupling, seniority, and hidden dynamical features.

So far partial seniority conservation of Type-2 has been predicted only in systems with four particles in  $j= 9/2$ shell. The property was first discovered through  numerical experiments in Ref.~\cite{Escu} and has since been studied in various theoretical approaches~\cite{Escu,Zamick, Isacker1,Isacker2,qi2011partial,qi2017partial,PhysRevC.98.061303,Qi3}. Furthermore, as mentioned, there are some  experimental evidences indicating its presence in nuclear systems~\cite{das2022nature,PhysRevResearch.6.L022038}.

There are three $I=4$ (or 6) states for four particles in  $j= 9/2$ shell, among which one has $v=2$ and two others with $v=4$. Those two $v=4$ states were not expected to be uniquely defined. Any linear combination of them provides a new $v=4$ state. However, it was first found in Ref.~\cite{Escu} that special $I=4$ (and 6) states exist which stand out and remain unmixed with the others under any interaction. In other words, the matrix elements between that state and the two other states vanish, and the latter remain orthogonal to it, even when seniority is not conserved by the interaction. Those are the type-2 partial seniority conserved states that we are referring to. As illustrated in Fig.~\ref{alpha}, these two states are not expected to lie far from the yrast $v=0,2$ states while the ``third" $I=4$ (and 6) states may locate at rather high excitation energy and only mix slightly with the $v=2$ state.

\subsection{Analytical proof for the existence of partial seniority conserved states}
As introduced in the section above, we can write the Hamiltonian matrix elements as the combinations
of the interaction terms $V_J$ as,
\begin{equation}
	H^I_{ij}=n(n-1)/2\sum_JM^I_{ij}(J)V_J,
\end{equation}
where $i,j$ denote the basis states which run over the three $v=2,4$ states in the case of concern.
If a state $|\alpha_1\rangle$ is an eigenvector of any Hamiltonian, it must also be a common eigenvector of all matrices $M^I$:
\begin{equation}
	M^I(J)|\alpha_1\rangle=m_I(J)|\alpha_1\rangle
\end{equation}
where we use $m$ to denote the eigen value of the matrix. This provides a sufficient condition for the vanishing of the non-diagonal Hamiltonian matrix elements.
As the partial seniority conserved state $\alpha_1$ has $v=4$, we can assume it is of the form
\begin{equation}\label{29}
	|j^4,v=4,\alpha_1,I\rangle=c_1|j^4,v=4,\beta_1,I\rangle+c_2|j^4,v=4,\beta_2,I\rangle,
\end{equation}
where $\beta_{1,2}$ denote any set of $v=4$ bases and their amplitudes are denoted by  $c$.
For such a state, immediately we can have,
\begin{align}\label{30}
\frac{H^I_{2\beta_1}}{H^I_{2\beta_2}}
&= \frac{M^I_{2\beta_1}(J)}{M^I_{2\beta_2}(J)} \nonumber\\
&= \frac{\displaystyle [\, j^3 (v_3=3, I_3=j) j I \mid\} j^4, v=4, \beta_1, I \,]}%
       {\displaystyle [\, j^3 (v_3=3, I_3=j) j I \mid\} j^4, v=4, \beta_2, I \,]},
\end{align}
and
\begin{align}
\frac{H^I_{\beta_1\beta_1}-H^I_{\beta_2\beta_2}}{H^I_{\beta_1\beta_2}}
&= \frac{M^I_{\beta_1\beta_1}(J)-M^I_{\beta_2\beta_2}(J)}{M^I_{\beta_1\beta_2}(J)}.
\end{align}

The vanishing of non-diagonal matrix elements between the two states
can be attributed to to a special relation among certain one-particle CFPs~\cite{Zamick}:
\begin{eqnarray}
	\nonumber \frac{[j^4(\alpha_1,v=4,I)jI_5|\}j^5,v=3,I_5=j]}{[j^4(\alpha_2,v=4,I)jI_5|\}j^5,v=3,I_5=j]} \\
	=\frac{[j^4(\alpha_1,v=4,I)jI_5|\}j^5,v=5,I_5=j]}{[j^4(\alpha_2,v=4,I)jI_5|\}j^5,v=5,I_5=j]},
\end{eqnarray}
where the states $|j^5,v=5,I_5=j\rangle$ and $|j^5,v=3,I_3=j\rangle$ can be uniquely specified. 

The above relation can be proven by examining the recursion relation of the CFPs~\cite{Qi3}.
The $v\rightarrow v-1$ one-particle CFPs can be factorized as
\begin{eqnarray}\label{sun}
	[j^{n-1}(v-1,\alpha_1,I_1)jI|\}j^{n} v,\alpha, I]\nonumber \\
	=\sqrt{\frac{v(2j+3-n-v)}{n(2j+3-2v)}}
	R(j,v-1,\alpha_1I_1;v\alpha I).
\end{eqnarray}
where the factor $R$ is independent of the particle number $n$.
Table~\ref{cfp-46} presents the one-particle CFPs \([j^3(v_3 I_3) j I \,\|\, j^4, \alpha, I=4,6]\) for two \(v=4\) states: 
the partially seniority-conserved state \(\alpha_1\), which retains the seniority structure 
with minimal mixing, and the state \(\alpha_2\), which is orthogonal to \(\alpha_1\) 
and exhibits a complementary configuration. These CFPs illustrate how the seniority 
quantum number is partially preserved in the four-particle system.

\begin{table}	\caption{One-particle CFPs $[j^3(v_3I_3)jI|\}j^4,\alpha,I=4,6]$ for the partial seniority conserved state $\alpha_1$ and the state  $\alpha_2$ that is orthogonal to it. Both states have $v=4$. \label{cfp-46}}
	\centering
		\begin{tabular}{ccc}
			\hline
			$I_3$&$\alpha_1$&$\alpha_2$\\
			\hline
		$	I=4$\\
			\hline
			3/2&    $\displaystyle-\sqrt{\frac{1547}{1727*60}} $   &   $\displaystyle\frac{59}{3\sqrt{1727}}$   \\
			5/2&      $\displaystyle\frac{8\sqrt{17}}{\sqrt{210*1727}}$  &  $\displaystyle-\frac{19\sqrt{13}}{3\sqrt{2*1727}}$         \\
			7/2&    $\displaystyle-\frac{19\sqrt{51}}{2\sqrt{7*1727}}$    &   $\displaystyle-\frac{\sqrt{65}}{3\sqrt{1727}}$    \\
			9/2&    0    &$\displaystyle\frac{\sqrt{1727}}{33\sqrt{13}}$   \\
			11/2&  $\displaystyle\frac{11\sqrt{21}}{3\sqrt{1727}}$     &    $\displaystyle\frac{16\sqrt{85}}{3\sqrt{13*1727}}$          \\
			13/2&  $\displaystyle-\frac{8\sqrt{255}}{5\sqrt{1727}}$       &    $\displaystyle\frac{23\sqrt{7}}{2\sqrt{13*1727}}$     \\
			15/2&  $\displaystyle\frac{13\sqrt{133}}{\sqrt{510*1727}}$     &   $\displaystyle\frac{76\sqrt{19}}{3\sqrt{26*1727}}$         \\
			17/2&   $\displaystyle\frac{12\sqrt{7}}{\sqrt{17*1727}}$        &  $\displaystyle-\frac{5\sqrt{65}}{2\sqrt{3*1727}}$      \\
\hline
		$	I=6$\\
\hline
			3/2&    $\displaystyle \sqrt{\frac{2261}{715*281}} $   &   $\displaystyle -\frac{46\sqrt{3}}{3\sqrt{143*281}}$  \\
			5/2&      $\displaystyle -\frac{4\sqrt{646}}{\sqrt{385*281}}$  &  $\displaystyle-\frac{43}{\sqrt{66*281}}$      \\
			7/2&    $\displaystyle -\frac{45\sqrt{323}}{\sqrt{6006*281}}$    &   $\displaystyle\frac{2\sqrt{10}}{3\sqrt{143*281}}$   \\
			9/2&    0    & $\displaystyle\frac{\sqrt{281}}{3\sqrt{286}}$     \\
			11/2&  $\displaystyle -\frac{55\sqrt{19}}{\sqrt{4862*281}}$     &    $\displaystyle\frac{13\sqrt{210}}{3\sqrt{143*281}}$            \\
			13/2&  $\displaystyle-\frac{42\sqrt{19}}{5\sqrt{715*281}}$       &    $\displaystyle\frac{13\sqrt{119}}{4\sqrt{429*281}}$        \\
			15/2&  $\displaystyle\frac{507\sqrt{3}}{\sqrt{230945*281}}$     &   $\displaystyle-\frac{43\sqrt{7}}{\sqrt{143*281}}$          \\
			17/2&   $\displaystyle\frac{70\sqrt{21}}{\sqrt{2431*281}}$        &  $\displaystyle\frac{177\sqrt{19}}{4\sqrt{715*281}}$      \\
			21/2&    $\displaystyle\frac{1320}{\sqrt{46189*281}}$   &  $\displaystyle-\frac{77\sqrt{7}}{\sqrt{8580*281}}$    \\
            \hline
		\end{tabular}
\end{table}

\subsection{The pair content}
Following the CFPs as introduced,
for four identical nucleons in a single-$j$ shell, the state can be written as the tensor product of two-particle states as $|j^2(J_{\alpha})j^2(J_{\beta});I\rangle$ where $J_{\alpha}$ and $J_{\beta}$ are the so-called principal parents. 
The corresponding two-particle CFPs can be constructed within the principal parent scheme and be expressed in closed forms in terms of $9j$ symbols. 
 The overlap between such overcomplete paired states is~\cite{Qi3}
\begin{eqnarray}
	\nonumber A^j_I(J_{\alpha}J_{\beta};J_{\alpha}'J_{\beta}')&=&\langle j^2(J_{\alpha})j^2(J_{\beta});I|j^2(J'_{\alpha})j^2(J'_{\beta});I\rangle\\
	\nonumber &=&\delta_{J_{\alpha}J_{\alpha}'}\delta_{J_{\beta}J_{\beta}'}+(-1)^{I}\delta_{J_{\alpha}J_{\beta}'}\delta_{J_{\beta}J_{\alpha}'}\nonumber\\
&&	-4\hat{J}_{\alpha}\hat{J}_{\beta}\hat{J}_{\alpha}'\hat{J}_{\beta}'
	\left\{
	\begin{array}{ccc}
		j&j&J_{\alpha}\\
		j&j&J_{\beta}\\
		J_{\alpha}'&J_{\beta}'&I
	\end{array}
	\right\},
\end{eqnarray}

In particular, for $I=4$, one can construct the $v=2$ state as $|j^4[J_{\alpha}=0J_{\beta}=4],v=2,I\rangle$ and the first $v=4$ state as $|j^4[J_{\alpha}=2J'_{\beta}=2],v=4,I\rangle$.
Similarly, for $I=6$, one can construct the $v=2$ state as $|j^4[J_{\alpha}=0J_{\beta}=6],v=2,I\rangle$ and the first $v=4$ state as $|j^4[J_{\alpha}=2J'_{\beta}=4],v=4,I\rangle$.
Although the special $v=4$ states $|a\rangle$ cannot be constructed directly from the one-particle or two-particle CFPs, it is found that these $I=4$ and 6 states have very large overlap with the states $|j^4[J_{\alpha}=2J_{\beta}=2],v=4,I\rangle$ and $|j^4[J_{\alpha}=2J_{\beta}=2],v=4,I\rangle$, respectively. The corresponding overlaps are calculated to be~\cite{Isacker1,Qi3}
\begin{eqnarray}
\langle j^4\alpha_1,v=4,I=4\rangle|j^4[J_{\alpha}=2J_{\beta}=2],v=4,I\rangle\nonumber\\=10\sqrt{255}/\sqrt{25591}\approx0.998220
\end{eqnarray}
and
\begin{eqnarray}
\langle j^4\alpha_1,v=4,I=6\rangle|j^4[J_{\alpha}=2J_{\beta}=4],v=4,I\rangle\nonumber\\=2\sqrt{6783}/\sqrt{27257}\approx 0.9974
\end{eqnarray}
for the $I$=4 and 6 states, respectively.

For completeness, the recursion relation for the matrix element between states with $\Delta v=2$ is
\begin{align}
	&\langle j^nv\alpha JM|V|j^nv-2,\alpha'JM\rangle\nonumber\\&=\frac{2j+1-2n}{2j+1-2v} \sqrt{\frac{(n-v+2)(2j+3-n-v)}
		{2(2j+3-2v)}}\nonumber\\&\times\langle j^{v}v\alpha JM|V|j^{v}v-2,\alpha'JM\rangle,\label{rec}.
\end{align}
As a result, the 
matrix element of $V$ between states with seniorities $v$ and $v-2$ always vanish in the middle of one single $j$ orbit,
\begin{equation}
	\langle j^nv\alpha JM|V|j^nv-2,\alpha'JM\rangle=0, \ \ \ \text{when}\ n=(2j+1)/2.\label{mid}
\end{equation}
which shows the particle-hole conjugate symmetry as will be discussed below.

\section{Quest of the wave function}\label{sec4}

\subsection{Symbolic shell model}
There has been increasing interest in developing symbolic modeling techniques for studying many-body systems in an exact manner. For simple systems with $j\leq 7/2$, the eigenvalues of all the states can be expressed in  a closed-form linear combinations of the TBMEs of the nuclear shell-model interaction since, as mentioned above, all of them are either uniquely defined by the angular momentum and seniority quantum numbers, or they do not mix due to particle-hole symmetry. The expansion coefficients are positive rational numbers and are independent of actual values of the TBME. Sometimes those systems are described as exactly ``solvable"~\cite{PhysRevLett.87.172501,Talmi2010}. For larger $j$ values the systems can become rather involved and may not be solvable except in some special cases.

A symbolic shell model approach was presented in Ref.~\cite{qian2018partial} starting from the so-called $M$-scheme representation of the shell model. It is worth noting that, in practice, in practice, all large-scale configuration-interaction shell model calculations in large model spaces are performed in the $M$-scheme today where Hamiltonian matrices up to dimension of $2\times 10^{10}$ can be evaluated on a modest-sized supercomputer~\cite{qi2016shell,qi2016large}, as most angular momentum projection or coupling algorithms and the evaluation of the Hamiltonian matrix elements required for $jj$ coupling scheme calculations are very time-consuming and difficult to parallelize.
On the other hand, the Hamiltonian matrix elements between  $M$-scheme bases can be simply expressed as linear combinations of the TBME.

It can readily be seen that the eigenvalue takes a simple closed form for states uniquely defined by their angular momentum. In general, however, the eigenvalues obtained from the symbolic shell-model calculations above can still be complex, nonlinear functions of the TBMEs. As examples, Table~\ref{tab1} presents the energy expressions for selected $v=3$ states listed in Table~\ref{tab:number}.

\begin{table}
\caption{\label{tab1} The closed-form expression obtained from the symbolic shell-model calculation for the energy of selected states uniquely specified by the total angular momentum $I$ and seniority $v=3$ for three and five particles in a single $j=9/2$ shell.}
\begin{tabular}{lccc}
\hline
Configuration& I
& $v$ & Energy
\\ \hline
$(9/2)^3$ & 3/2 & 3 & $\frac{24 V_4}{11}+\frac{9 V_6}{11}$ \\
  & 5/2 & 3 & $\frac{5 V_2}{6}+\frac{13 V_4}{22}+\frac{52 V_6}{33}$ \\
  & 7/2 & 3 & $\frac{52 V_2}{33}+\frac{60 V_4}{143}+\frac{V_6}{165}+\frac{714 V_8}{715}$ \\
  & 11/2 & 3 & $\frac{17 V_2}{33}+\frac{170 V_4}{143}+\frac{56 V_6}{165}+\frac{684 V_8}{715}$ \\
  & 13/2 & 3 & $\frac{10 V_2}{11}+\frac{27 V_4}{143}+\frac{17 V_6}{22}+\frac{323 V_8}{286}$ \\
  & 15/2 & 3  & $\frac{57 V_4}{143}+\frac{21 V_6}{11}+\frac{9 V_8}{13}$ \\
  & 17/2  & 3  & $\frac{125 V_4}{143}+\frac{57 V_6}{110}+\frac{209 V_8}{130}$ \\
  \hline
  			$(9/2)^5$ & 5/2 & 3 & $\frac{2}{5}V_0+\frac{85}{66}V_2+\frac{371}{110}V_4+\frac{559}{330}V_6+\frac{357}{110}V_8$ \\
& 7/2 & 3 & $\frac{2}{5}V_0+\frac{155}{66}V_2+\frac{2373}{1430}V_4+\frac{346}{165}V_6+\frac{2499}{715}V_8$ \\
& 9/2 & 3 & $\frac{2}{5}V_0+\frac{28}{33}V_2+\frac{1884}{715}V_4+\frac{562}{165}V_6+\frac{1938}{715}V_8$ \\
			& 11/2 & 3  & $\frac{2}{5 }V_0+\frac{50}{33}V_2+\frac{73}{55}V_4+\frac{1267}{330 }V_6+\frac{321 }{110 }V_8$ \\
			& 13/2 & 3 & $\frac{2}{5}V_0+\frac{50}{33}V_2+\frac{123}{55}V_4+\frac{302}{165 }V_6+\frac{221}{55}V_8$ \\
			& 15/2 & 3 & $\frac{2}{5}V_0+\frac{5}{11}V_2+\frac{2274}{715}V_4+\frac{223}{110}V_6+\frac{5631}{1430}V_8$ \\
			& 17/2 & 3 & $\frac{2}{5}V_0+\frac{32}{33}V_2+\frac{829}{715}V_4+\frac{632}{165}V_6+\frac{2603}{715}V_8$ \\
  \hline\end{tabular}
\end{table}

The analytical wave functions and eigenvalues for the two partially seniority conserved states (labeled as $\alpha_1$) and the other states that are orthogonal to them in the $(9/2)^4$ configuration can also be obtained via symbolic calculation. The energy of the $4^+$ $\alpha_1$ states reads
\begin{small}
\begin{align}
E_{4^+}[(9/2)^4,v=4,\alpha_1]=\frac{68}{33}V_2+V_4+\frac{13}{15}V_6+\frac{114}{55}V_8
\end{align}
\end{small}

The $v=2$ and $\alpha_2$ states can mix for a general seniority-non-conserving interaction which can be evaluated from diagonalizing the corresponding $2\times2$ symmetric matrix. The two diagonal matrix elements are
\begin{small}
\begin{align}
V_{4^+}[(9/2)^4,v=4,\alpha_2]=\frac{47}{99} V_2+\frac{300}{143} V_4+\frac{1214}{495} V_6+\frac{697 }{715}V_8,\nonumber\\
V_{4^+}[(9/2)^4,v=2]=\frac{3}{5} V_0+\frac{67}{99} V_2+\frac{746}{715} V_4+\frac{1186}{495} V_6+\frac{918 }{715}V_8.
\end{align}
\end{small}
for the $v=2$ and $\alpha_2$ states, respectively.
The non-diagonal matrix element between the two states is
\begin{equation}\label{key}
V_{v=2,\alpha_2;4^+}=\frac{\sqrt{314} }{6435}\left(65 V_2-315 V_4+403 V_6-153 V_8\right).
\end{equation}

The wave functions of those two $\alpha_1$ states acquire the simple forms
\begin{eqnarray}
|\alpha_1,4^+\rangle &=& \frac{1}{\sqrt{146795}} \Big\{
-24 \sqrt{7},\; 12 \sqrt{42},\; 12 \sqrt{7},\; 3 \sqrt{7},\nonumber\\&&
-14 \sqrt{42}, -14 \sqrt{42},\; 6 \sqrt{175},-82 \sqrt{3},\; 28 \sqrt{7},\nonumber\\
&&   18 \sqrt{7},
-112 \sqrt{3}, 52 \sqrt{7},\; -26 \sqrt{42} 
\Big\}.
\end{eqnarray}and
\begin{eqnarray}
|\alpha_1,6^+\rangle &=& \frac{1}{\sqrt{90763}} \Big\{
-120 \sqrt{2},\; 120 \sqrt{\dfrac{6}{7}},\; \dfrac{-340}{\sqrt{7}} ,\; 50 \sqrt{\dfrac{6}{7}},\nonumber\\&&  
64 \sqrt{3}, 5 \sqrt{3},\; -60 \sqrt{2},\; -10 \sqrt{42},\; 60 \sqrt{2} 
\Big\}.
\end{eqnarray}
within the $M$-scheme basis, arranged as defined in \ref{Mscheme}, with the detailed symbolic calculation procedure provided, and subject to the restriction $M=I$.

In addition to symbolic calculations with angular momentum projection operation as described above (which in principle also works for large systems), for single-$j$ systems of concern in this review, one can simply construct the matrix for the angular momentum operator $I^2$. One can still obtain a set of coupled states with conserved angular momentum from the symbolic diagonalization of that matrix. The states thus derived will be random if there are more than one states for a given angular momentum. As above, one can solve it together with a pairing Hamiltonian which will automatically differentiate states with different seniority.

\subsection{Quest for the ``magic" generator}
Although one can obtain the wave functions of those partially seniority conserving states with the help of the above symbolic calculations or other analytical techniques, there is no simple and straightforward way to generate those states directly so far. In other words, we still do not have an operator like that in Eq.~(\ref{proj2}) that can project out those states directly. An interesting attempt along that direction was presented in Ref.~\cite{PhysRevC.106.024308}. Neerg\aa{}rd started also with the $M$-scheme bases and constructed the unique $v=0$ (and $I=0$) state from the $M=0$ bases which can be written as
\begin{equation}
\begin{aligned}
|\phi_0\rangle &=
\left| 9/2, 7/2, -7/2, -9/2 \right\rangle
-
\left| 9/2, 5/2, -5/2, -9/2 \right\rangle \\
&+
\left| 9/2, 3/2, -3/2, -9/2 \right\rangle
-
\left| 9/2, 1/2, -1/2, -9/2 \right\rangle \\
&+
\left| 7/2, 5/2, -5/2, -7/2 \right\rangle
-
\left| 7/2, 3/2, -3/2, -7/2 \right\rangle \\
&+
\left| 7/2, 1/2, -1/2, -7/2 \right\rangle
+
\left| 5/2, 3/2, -3/2, -5/2 \right\rangle\\
&-
\left| 5/2, 1/2, -1/2, -5/2 \right\rangle 
+
\left| 3/2, 1/2, -1/2, -3/2 \right\rangle
\end{aligned}
\end{equation}
which spans 10 out of the in total 18 $M=0$ states. It may be useful to mention that the total number of $M=0$ bases equals to the total number of
states with different spins for the $n=4$ system as listed in Table \ref{tab:number}.
That $I=0$ state can be generated from the symbolic shell-model calculation as described above or simply through the $J=0$ pair creation operator $P^\dagger_j$ defined in Eq.~(\ref{pair-oper})
which creates a pair of fermions coupled to total angular momentum $J=0$ and seniority $v=0$.
To obtain a system with two pairs, one can apply the pair creation operator twice which generates exactly the above state.
By construction, the total seniority must be zero because the operation involves only two pairs coupled to spin $J=0$.

One can further verify the total angular momentum of the state with the help of the spin operators:
\[
\mathbf{I} = (I_x,I_y,I_z),
\]
and
\[
I_0 \equiv I_z,\qquad I_\pm \equiv I_x \pm i I_y.
\]
which obey the standard \({SU}(2)\) commutation relations
\begin{equation}\label{eq:su2}
[I_0,I_\pm] = \pm I_\pm,\qquad [I_+,I_-] = 2 I_0.
\end{equation}
\begin{equation}\label{eq:Ialpha}
[I_0,\alpha^\dagger_m] = m\,\alpha^\dagger_m,\qquad
[I_+,\alpha^\dagger_m] = \begin{cases}
\alpha^\dagger_{m+1}, & m<j,\\[4pt]
0, & m=j.
\end{cases}
\end{equation}
where $\alpha^\dagger_m$ are the \emph{unnormalized} creation operators:
\[
\alpha^\dagger_m \equiv \sqrt{\frac{(j+m)!}{(j-m)!}}\, a^\dagger_m.
\]
One has for the ladder operators
\begin{equation}
  I_\mu (P^\dagger)^2|0\rangle 
  = [I_\mu,(P^\dagger)^2]|0\rangle 
  = 2 P^\dagger [I_\mu,P^\dagger]|0\rangle = 0,
\end{equation}
where \(\mu=0,\pm\).
Therefore $(P^\dagger)^2|0\rangle$ is annihilated by the total angular momentum operators $I_\mu$ and lies in the $I=0$ subspace.  
It can thus be shown that  $|\phi_0\rangle $ satisfies
\[
I_+|\phi_0\rangle = 0,\qquad I_0|\phi_0\rangle=0.
\]

Ref.~\cite{PhysRevC.106.024308} then introduced (essentially by hand) another spin $I=0$ vector of the form (in the unnormalized single-particle representation)
\begin{equation}
\begin{aligned}
|\phi_1\rangle =&
-5\left| 9/2, 7/2, -7/2, -9/2 \right\rangle
+5\left| 9/2, 5/2, -5/2, -9/2 \right\rangle \\
&+
\left| 9/2, 3/2, -3/2, -9/2 \right\rangle
-7\left| 9/2, 1/2, -1/2, -9/2 \right\rangle \\
&+
9\left| 7/2, 5/2, -5/2, -7/2 \right\rangle
-3\left| 7/2, 3/2, -3/2, -7/2 \right\rangle \\
&-
9\left| 7/2, 1/2, -1/2, -7/2 \right\rangle
-6\left| 5/2, 3/2, -3/2, -5/2 \right\rangle \\
&-
14\left| 9/2, 3/2, -5/2, -7/2 \right\rangle
-6\left| 7/2, 5/2, -3/2, -9/2 \right\rangle \\
&+
16\left| 9/2, 1/2, -3/2, -7/2 \right\rangle
+6\left| 7/2, 3/2, -1/2, -9/2 \right\rangle \\
&-
25\left| 9/2, -1/2, -3/2, -5/2 \right\rangle
-6\left| 5/2, 3/2, 1/2, -9/2 \right\rangle \\
&+
9\left| 7/2, 1/2, -3/2, -5/2 \right\rangle
+6\left| 5/2, 3/2, -1/2, -7/2 \right\rangle
\end{aligned}
\end{equation}
Although somewhat tedious, one can verify that  $|\phi_1\rangle $ also satisfies
\[
I_+|\phi_1\rangle = 0
\]
and also has spin $I=0$. The state $|\phi_1\rangle $ is not unique and is not orthogonal to $|\phi_0\rangle $ where one has inner products
\[
\langle\phi_0,\phi_0\rangle = 10,
\qquad
\langle\phi_0,\phi_1\rangle = -5.
\]
One can therefore construct the unique $v=4$, $I=0$ state as 
\[
\Phi' = \phi_1 + \tfrac{1}{2}\phi_0
\]
We know it has $v=4$ because there are only $I=0$ states available.

For completeness, we would like to mention that $|\phi_0\rangle $ would be identical in the representation of the \emph{normalized single-particle states} as the normalization factor for the paired single-particle orbitals cancel each other while $|\phi_1\rangle $ becomes
{\footnotesize
\begin{equation}
\begin{aligned}
|\phi_1\rangle &=
-5\left| 9/2, 7/2, -7/2, -9/2 \right\rangle
+5\left| 9/2, 5/2, -5/2, -9/2 \right\rangle \\
&+
\left| 9/2, 3/2, -3/2, -9/2 \right\rangle
-7\left| 9/2, 1/2, -1/2, -9/2 \right\rangle \\
&+
9\left| 7/2, 5/2, -5/2, -7/2 \right\rangle
-3\left| 7/2, 3/2, -3/2, -7/2 \right\rangle \\
&-
9\left| 7/2, 1/2, -1/2, -7/2 \right\rangle
-6\left| 5/2, 3/2, -3/2, -5/2 \right\rangle \\
&-
2\sqrt{21}\left| 9/2, 3/2, -5/2, -7/2 \right\rangle
-2\sqrt{21}\left| 7/2, 5/2, -3/2, -9/2 \right\rangle \\
&+
4\sqrt{6}\left| 9/2, 1/2, -3/2, -7/2 \right\rangle
+4\sqrt{6}\left| 7/2, 3/2, -1/2, -9/2 \right\rangle \\
&-
5\sqrt{6}\left| 9/2, -1/2, -3/2, -5/2 \right\rangle
-5\sqrt{6}\left| 5/2, 3/2, 1/2, -9/2 \right\rangle \\
&+
3\sqrt{6}\left| 7/2, 1/2, -3/2, -5/2 \right\rangle
+3\sqrt{6}\left| 5/2, 3/2, -1/2, -7/2 \right\rangle
\end{aligned}
\end{equation}}
which leads to a total normalization factor
\[
\langle\phi_1,\phi_1\rangle = 1075.
\]

Consider the operation
\begin{equation}
\Phi_0 = \operatorname{span} \{ n_m \phi_{0,1} \},
\end{equation}
where $n_m = a_m^{\dagger} a_m$ is the number operator. It was observed that the span operation, $\operatorname{span} \{\}$, defines the set of all linear combinations of the states $n_m \phi_{0,1}$ and forms a subspace of 14 linearly independent states, denoted $[\Phi_0]$. Of these, 5 states are derived from $\phi_0$ and 9 from $\phi_1$.
In this way, one can divide the full 18 bases into two subgroups: The subspace generated by the number operator with dimension 14 and the leftover subspace with dimension 4 (denoted as $[\Phi_{1}]$). The author claimed that the two partial seniority conserved $\alpha_1$ states are reserved in the later small space $[\Phi_{1}]$. He proved that point by showing the two subspaces are invariant under the two-body interaction, meaning that no interaction can mix states from the two subspaces.

However, one must be cautious with such a unitary transformation of the original $M$-scheme bases, because the two-body interaction is rotationally invariant and cannot mix states with different spin values. Therefore one can easily construct an arbitrary subspace that is invariant under the two-body interaction.
Our analyses show that the subspace $[\Phi_{0}]$ spans all the $v=0$ and 2 states and 9 out of the 13 $v=4$ states shown in Table \ref{tab:number}. In fact, among the 13 states in $[\Phi_{0}]$ as defined in Ref.~\cite{PhysRevC.106.024308}, four are solvable states that are uniquely defined by the total angular momentum.
The subspace $[\Phi_{1}]$ contains the rest four $v=4$ states, all of which are solvable as well. 
However, the basis vectors in subspace $[\Phi_{1}]$  are not uniquely defined. They do not necessarily conserve the angular momentum quantum number either, although the seniority number is conserved, as all these states share the same value. We are therefore unable to verify the $[\Phi_{1}]$ basis vectors provided in Ref.~\cite{PhysRevC.106.024308}. But fortunately, with the help of the symbolic calculation, we can construct directly states with good angular momentum restricted to that subspace.
We did not investigate the reason in detail, but it can be confirmed that this subspace includes the unique $I=10$ and 12 states in addition to the partially seniority conserved $\alpha_1$ states.
Their eigenvectors are provided in Table \ref{phi1} as an expansion of the $M$-scheme bases. Unlike in the previous subsection, here the basis is taken as $M=0$ as in Ref.~\cite{PhysRevC.106.024308} but we have renormalized the single-particle wave functions at the end to simplify their application for other purposes.

\begin{table}[h!]
\caption{Integer eigenvectors for $I=12, 10, 6, 4$ in the $M$-scheme basis ($M=0$) for the four states contained in the $[\Phi_{1}]$ subspace. All states have $v=4$. The first two states are uniquely determined by their angular momenta, while the remaining $I=4$ and $6$ states correspond to the partially seniority-conserved $\alpha_1$ states. All four states span the full set of 18 $M$-scheme basis vectors. The eigen-functions are not normalized, but they are expressed in the normalized single-particle representation.
\label{phi1}}
\centering
\renewcommand{\arraystretch}{1.1}
\begin{tabular}{l|rrrr}
\hline
Basis & $I=12$ & $I=10$ & $I=6$ & $I=4$ \\ \hline
$|9/2, 7/2, -7/2, -9/2\rangle$ & 12 & -840 & 3024 & -5040 \\
$|9/2, 5/2, -5/2, -9/2\rangle$ & 105 & -4935 & 4914 & -1260 \\
$|9/2, 3/2, -3/2, -9/2\rangle$ & 243 & -7695 & 1386 & 1980 \\
$|9/2, 3/2, -5/2, -7/2\rangle$ & 280 & -6720 & -3654 & 3780 \\
$|9/2, 1/2, -1/2, -9/2\rangle$ & 150 & -3600 & -504 & -1800 \\
$|9/2, 1/2, -3/2, -7/2\rangle$ & 512 & -6400 & -1848 & -2200 \\
$|9/2, -1/2, -3/2, -5/2\rangle$ & 175 & -175 & 4200 & -400 \\
$|7/2, 5/2, -3/2, -9/2\rangle$ & 120 & -2880 & -1566 & 1620 \\
$|7/2, 5/2, -5/2, -7/2\rangle$ & 135 & -2205 & -3486 & 420 \\
$|7/2, 3/2, -1/2, -9/2\rangle$ & 192 & -2400 & -693 & -825 \\
$|7/2, 3/2, -3/2, -7/2\rangle$ & 525 & -525 & -966 & -2900 \\
$|7/2, 1/2, -1/2, -7/2\rangle$ & 378 & 2520 & -504 & 1720 \\
$|7/2, 1/2, -3/2, -5/2\rangle$ & 405 & 5805 & 4704 & 2280 \\
$|5/2, 3/2, 1/2, -9/2\rangle$ & 42 & -42 & 1008 & -96 \\
$|5/2, 3/2, -1/2, -7/2\rangle$ & 270 & 3870 & 3136 & 1520 \\
$|5/2, 3/2, -3/2, -5/2\rangle$ & 240 & 5280 & 3024 & -1520 \\
$|5/2, 1/2, -1/2, -5/2\rangle$ & 270 & 8010 & -5376 & 160 \\
$|3/2, 1/2, -1/2, -3/2\rangle$ & 42 & 1890 & -5376 & -3360 \\
\hline
\end{tabular}
\end{table}

In addition to the four vectors above, one may be interested in deriving all states in the coupled scheme. For that we have evaluated explicitly the matrix elements of angular momentum operator $I^2$. One can diagonalize the matrix to obtain all eigen values of $I^2$, their corresponding eigen vectors and the multiplicity for each angular momentum.
We will not list all the eigenvectors of $I^2$ for simplicity. But they are available as a simple python script upon request.

\section{The electromagnetic transition properties}\label{transition}

One of the most well-known features of the seniority scheme is seniority isomerism. In particular, long-lived isomers tend to appear near the half-filled valence shell, where the \( B(E2) \) values are nearly vanishing.
 This suppression originates from the direct dependence of the matrix elements of the 
electric multiple tensor operator on the factor \((\Omega - n)\). At mid-shell, this factor approaches zero, 
resulting in strongly hindered transition rates and reduced electromagnetic moments. 
The E2 transition matrix elements between states
with the same seniority are related to each other as~\cite{talmi1993}
\begin{equation}
    \langle j^n vI ||E2||j^n vI'\rangle=\frac{\Omega-n}{\Omega-\nu}\langle j^v vI ||E2||j^v vI'\rangle.
\end{equation}
As a result, the E2 transitions between $v=2$ states along the yrast cascade are mostly observed to be weak.
One can also get the reduced transition matrix elements for states
~with the seniority quantum numbers differ by two, which is expressed as~\cite{talmi1993}
\begin{align}
&	\langle j^n v I \,||\, E2 \,||\, j^n v \pm 2 I' \rangle\nonumber \\
	&= \sqrt{\frac{(n-v+2)(2 \Omega + 2 - n - v)}{2 (2 \Omega + 2 - 2 v)}} 
	\langle j^v v I \,||\, E2 \,||\, j^{v \pm 2} v I' \rangle \nonumber \\
	&= \sqrt{\frac{n (2 \Omega - n)}{2 \Omega - 2}} 
	\langle j^v v I \,||\, E2 \,||\, j^{v \pm 2} v I' \rangle 
	\quad (v=2 \rightarrow v=0)
\end{align}

The reduced transition probability, as we will discuss below, is related to the above matrix elements by
\begin{equation}
  B(E2;I\rightarrow I') =\frac{1}{2I+1} \langle j^n vI ||E2||j^n v'I'\rangle^2.
\end{equation}

\begin{figure}[ht]
\centering
\includegraphics[width=0.5\textwidth]{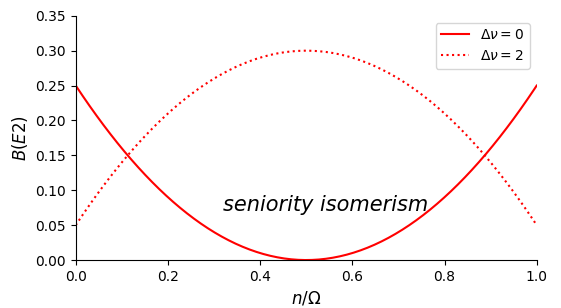}
\caption{Illustration on the evolution of the $B(E2)$ values as a function of the occupation of the orbital for $\Delta v=2$ and 0 transitions, concerning for examples the commonly studied $2^+\rightarrow 0^+_{gs}$ and $I=2j-1\rightarrow 2j-3$ transitions. The absolute $B(E2)$ value shown in the plot is random and has no special physical meaning.}
\label{fig:s-isomer}
\end{figure}

\begin{figure*}[htp]
\centering
\includegraphics[width=0.95\textwidth]{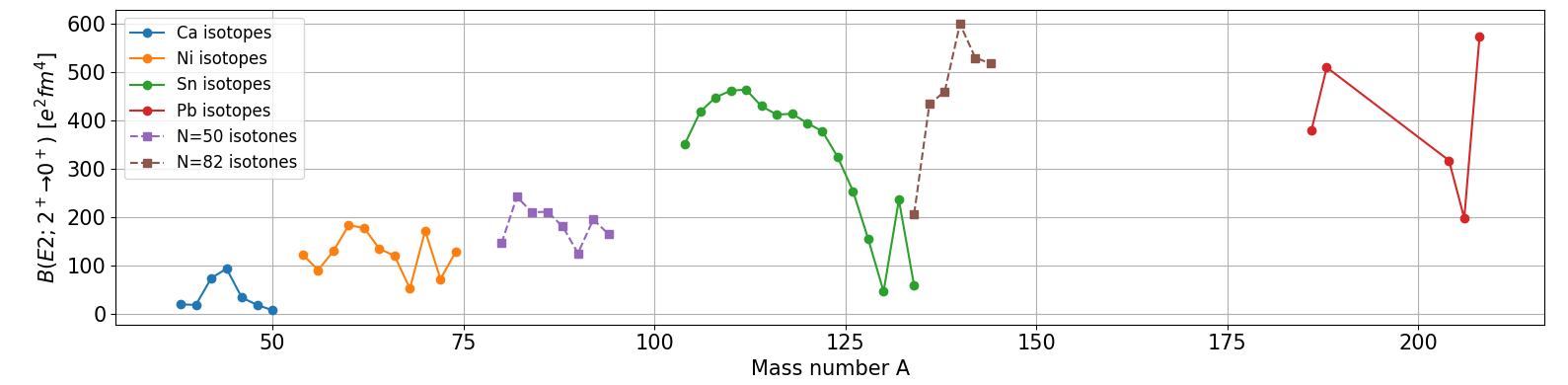}
\caption{Evolution of the experimental $B(E2)$ values as a function of mass number for known semi-magic isotopic and isotonic chains. Data taken from~\cite{Pritychenko_2016,PhysRevLett.116.122502,PhysRevLett.129.112501}.}
\label{2plus}
\end{figure*}

In Fig.~\ref{fig:s-isomer} we show a schematic plot on how the $B(E2)$ values evolve as a function of the occupancy. The transitions for $2^+_1$ to the ground state with $\Delta v=2$ follow a parabolic behavior with maximum at mid-shell. Known experimental data are plotted in Fig.~\ref{2plus}.
We focus on single-$j$ systems in the present work, but parabolic behavior is well known for multi-$j$ systems that can be described by the generalized seniority scheme. One of the most typical and most heavily studied systems is the Sn isotopic chain (see, for example, Refs.~\cite{Morales_2011,Back2013}.)
For the $I=2j-1$ isomeric state, one may expect all transitions to be weak. They become even more strongly suppressed, if not vanishing, near mid-shell.

\begin{figure}[ht]
\centering
\includegraphics[width=0.5\textwidth]{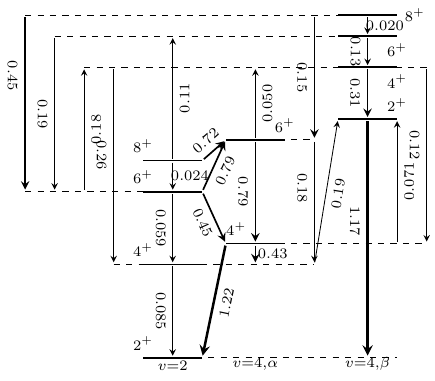}
\caption{Illustration of the E2 cascade and relative E2 transition strengths (normalized to $B(E2; 2_{1}^{+} \rightarrow 0_{1}^{+})$, denoted as $B_{20}$) for a system with four particles (holes) in the $j=9/2$ shell within the seniority scheme. 
The two partially seniority-conserved $v=4$, $\alpha$ states do not mix with other states for any $g_{9/2}$ interaction, as explained in the main text. Their exact energies depend on the effective interaction; in the plot shown, the interaction is adjusted to nuclei near $N=Z=50$ (see Sec.~\ref{RuPd}). The second $6^+$ state can lie below the yrast $8^+$ state in neutron-rich Ni isotopes, as discussed in Sec.~\ref{Nineutron}, but the E2 strengths remain unchanged. 
\\
Some transitions are drawn pointing upwards for convenience. The arrow widths are roughly proportional to the $B(E2)$ strengths, not the transition rates as is usually done. Adapted from Ref.~\cite{qi2017partial}.
}
\label{fig:The-E2-transition}
\end{figure}

The existence of partial conservation
of seniority in $j=9/2$ shells plays an essential role in our understanding
of the electric quadrupole transitions of the nuclei involved. That was studied in details in Refs.~\cite{qi2017partial,PhysRevC.108.064313,PhysRevC.110.034320}.
In Fig.~\ref{fig:The-E2-transition} a detailed calculation is given
on the relative E2 transition strengths for a $(9/2)^{4}$ system
calculated with a seniority-conserving interaction tuned for the $^{100}$Sn region. The transition strength doesn't depend on the interaction for a seniority conserved system as plotted.

\begin{figure}
\includegraphics[width=0.49\textwidth]{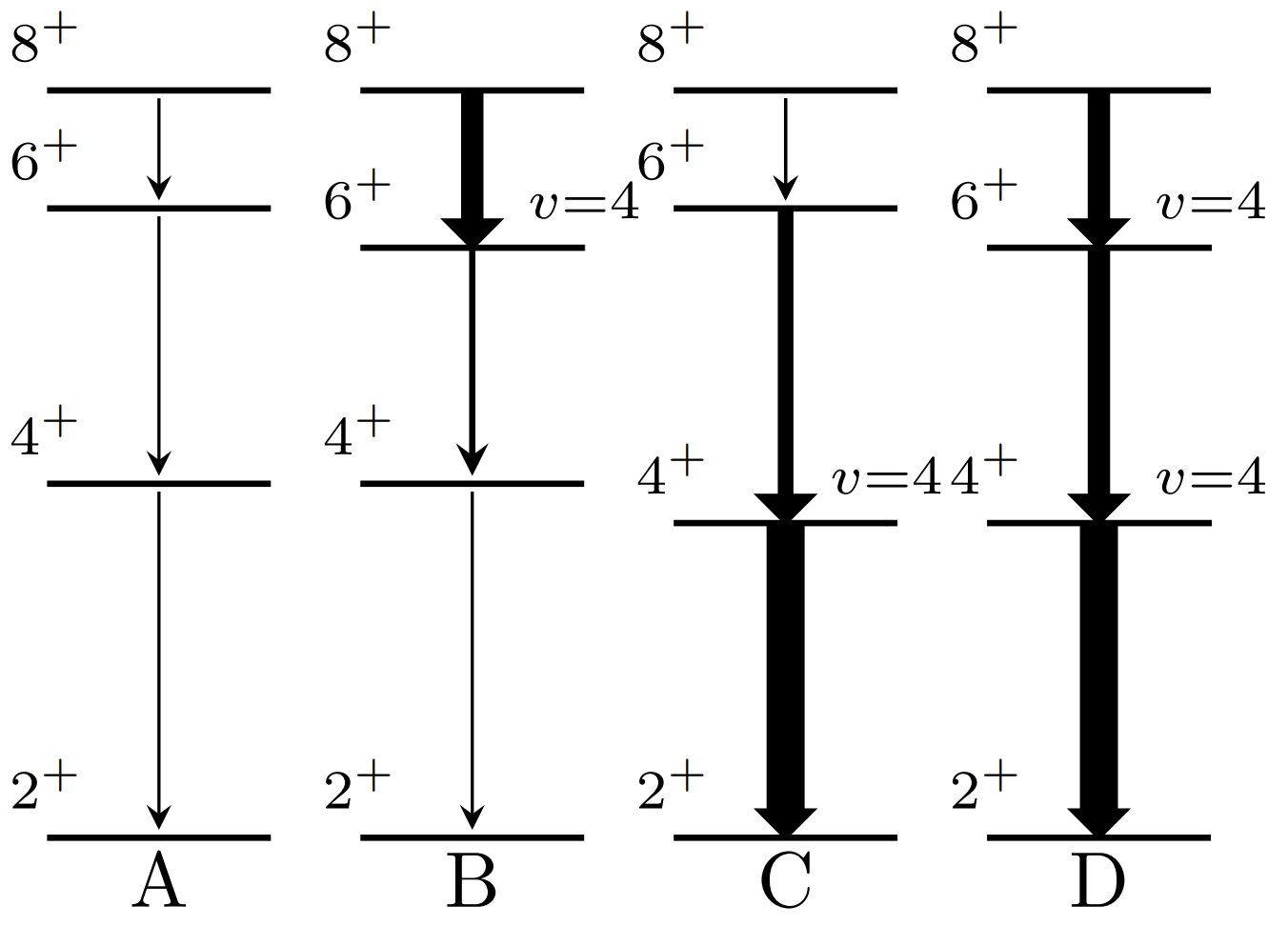}\caption{Schematic illustration of E2 decay patterns for yrast states 
in a $(9/2)^4$ configuration (based on Fig.~\ref{fig:The-E2-transition}): 
(A) The lowest excited levels are mainly seniority $v=2$, leading to weak 
E2 links among them; 
(B) A particular seniority $v=4$, $6^{+}$ state becomes yrast, producing a 
strong $B(E2;8_{1}^{+}\!\rightarrow6_{1}^{+})$, so the $8_{1}^{+}$ level 
is not isomeric; 
(C) Similar to case~B, but the special seniority-$v=4$, $4^{+}$ state is 
yrast instead; 
(D) Both the $v=4$, $4^{+}$ and $6^{+}$ levels are yrast, resulting in a collective-band-like sequence with enhanced in-band E2 strengths. Adapted from Ref.~\cite{qi2017partial}.\label{fig:E2-transitions-for}}
\end{figure}

As indicated in Fig.~\ref{fig:The-E2-transition}, the E2 transitions between the two special
$v=4,\alpha$ states and between those states are strong and are proportional to $B(E2;2^+_1\rightarrow 0^+_1)$. The transitions between those $v=4$ states and the $v=2$ states are also strong. 
A schematic plot for the influence of the relative positions of low-lying
states on the yrast E2 transition properties are shown in Fig.~\ref{fig:E2-transitions-for}.
As one can imagine, several E2 cascade scenarios may emerge in different nuclear systems:
\begin{itemize}
    \item[(A)] The lowest excited levels are mainly seniority-$v=2$, leading to weak E2 links among them, combined with a relatively strong transition to the ground state. This is the most common feature.
    \item[(B)] A special seniority-$v=4$, $6^{+}$ state becomes yrast, producing a strong $B(E2;8_{1}^{+}\!\rightarrow6_{1}^{+})$  or the yrast $6^{+}$ state lies sufficiently below the $8^{+}$ level to open a fast decay path. As a consequence, the isomerism of the $8_{1}^{+}$ state is quenched. As will be explained in Sec. \ref{Nineutron}, the $8_{1}^{+}$ states in $^{72,74}$Ni are not expected
to be isomeric as the $v=4$, $6^{+}$ becomes lower. 
    \item[(C)] Similar to case~(B), but the special seniority-$v=4$, $4^{+}$ state is yrast instead.
    \item[(D)] Both the $v=4$, $4^{+}$ and $6^{+}$ levels are yrast, resulting in a band-like sequence with enhanced in-band E2 strengths.
      \item[(E)] The existence of partial seniority conservation prevents the low-lying $v=2$ and 4 states from mixing. However, mixing can still occur, leading to tremendously enhanced or quenched E2 transitions. That has been observed to happen in $N=50$ isotones below $^{100}$Sn. The phenomenon remains controversial, as different experiments report varying results. See Sec. \ref{RuPd}.
\end{itemize}

The presence of the uniquely defined seniority $v=4$ ($\alpha$) states allows for a particularly transparent explanation of the small $B(E2;8_{1}^{+}\!\rightarrow6_{1}^{+})$ value.  That concerns in particular
the nucleus $^{94}$Ru which exhibits an $8^{+}$ isomer at $E_x=2.644$~MeV with a half-life of $71~\mu\text{s}$. Its isomeric nature stems from a strongly hindered $E2$ decay to the $6^{+}$ level and the small energy spacing between these two yrast states. Because these special $v=4$, $|\alpha_1\rangle$ states do not mix with others, the physical $6^{+}$ and $8^{+}$ yrast wave functions can be written, to an excellent approximation, as
\[
|{j^{4};I}_{1}\rangle=c_{2}^{I}\,|{j^{4},v=2;I}\rangle +c_{4}^{I}\,|{j^{4},v=4;I}\rangle,
\]
with amplitudes $c_{v}^{I}$. This is because, for $I=8$, there is only one $v=2$ state and one $v=4$ state, whereas for $I=6$, there are two $v=4$ states; however, only one (the $|\alpha_2\rangle$ state) is included, since the $|\alpha_1\rangle$ state does not mix in, as mentioned above.
 Defining $M_{v_{1}v_{2}}\equiv M(E2;8^{+}(v_{1})\!\rightarrow\!6^{+}(v_{2}))$, the reduced matrix element becomes
\[
M(E2;8_{1}^{+}\!\rightarrow\!6_{1}^{+})
= c_{2}^{8}c_{2}^{6}M_{22}
+\bigl(c_{4}^{8}c_{2}^{6}M_{42}+c_{2}^{8}c_{4}^{6}M_{24}\bigr)
+c_{4}^{8}c_{4}^{6}M_{44}.
\]
In practice, $\lvert c_{2}^{I}\rvert\gg \lvert c_{4}^{I}\rvert$ because the above $v=4$ configurations lie at comparatively high excitation energies (see Fig.~\ref{fig:The-E2-transition}). Moreover, $\lvert M_{22}\rvert$ and $\lvert M_{44}\rvert$ are significantly smaller than the mixed-seniority terms $\lvert M_{24}\rvert$ and $\lvert M_{42}\rvert$. With the usual phase convention $M_{22}$ is positive whereas the others carry the opposite sign. Consequently, the dominant suppression of $M(E2;8_{1}^{+}\!\rightarrow\!6_{1}^{+})$ arises from a cancellation between the leading $c_{2}^{8}c_{2}^{6}M_{22}$ term and the mixed-seniority contributions in brackets, which is favored when $c_{4}^{I}$ shares the sign of $c_{2}^{I}$.

A peculiar feature of the partial seniority conservation observed in Ref.~\cite{qi2017partial} is that:
\begin{itemize}
    \item Unlike in a pure single-$j$ system, the $v=2$ and $v=4$ states can mix when the model space is extended to multi-orbitals. That doesn't necessarily mean that cross-orbital excitations become dominant or critical but that they serve as mediators for the two supposedly pure configurations to mix.
    \item The diagonal TBMEs do not contribute to state mixing, which is instead driven by the cross-orbital non-diagonal TBMEs. From the perspective of a perturbative expansion, these higher-order terms become significant precisely because the leading-order contribution vanishes.

    \item Consequently, calculations of $B(E2)$ are highly sensitive to the non-diagonal TBMEs, as illustrated in Fig.~5 of Ref.~\cite{qi2017partial}. Unfortunately, these TBMEs are often poorly constrained, because they may not strongly affect the energies, which are typically the primary criterion in the optimization of effective interactions.
\end{itemize}

\begin{figure}[ht]
\centering
\includegraphics[width=0.48\textwidth]{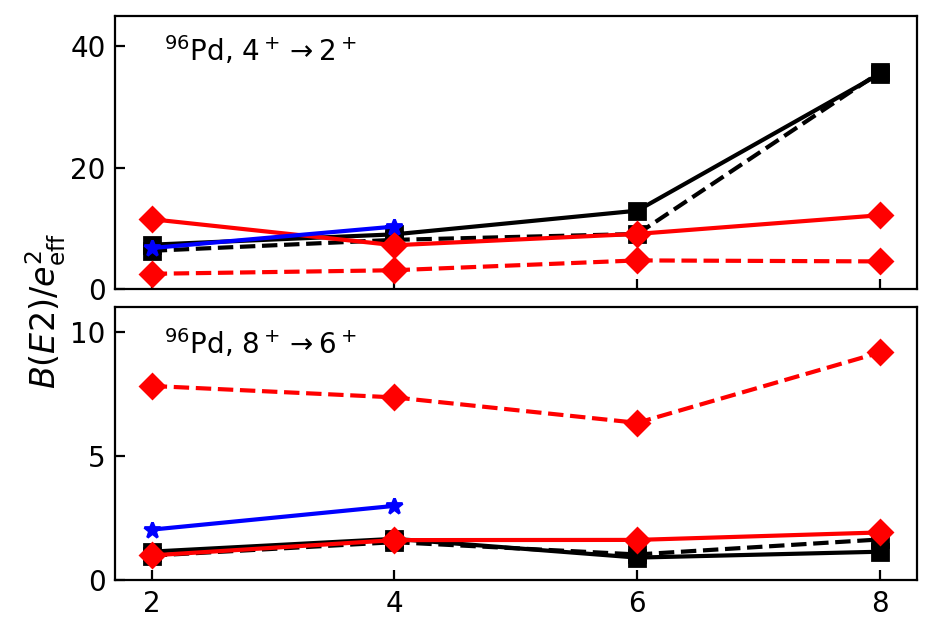}
\caption{Illustration of the different $B(E2)$ results (in $\text{e}^2\text{fm}^4$ but divided by the effective charge $\text{e}_{\text{eff}}$) obtained in various model spaces using different effective interactions (indicated by different symbols, with details omitted for simplicity). The $8^+ \rightarrow 6^+$ and $4^+ \rightarrow 2^+$ transitions in the yrast cascade of the $^{96}\mathrm{Pd}$ nucleus are taken as examples. Adapted from Ref.~\cite{qi2017partial}.}
\label{96Pd}
\end{figure}

We can illustrate this effect of configuration mixing in a matrix form. The original $I=4$ or 6 states form a $3\times 3$ Hamiltonian matrix of the  symmetric form
\[
A_{3\times3}=
\begin{pmatrix}
\lambda & 0 & 0\\
0 & \alpha & \beta\\
0 & \beta & \gamma
\end{pmatrix},
\]
the vector \(e_1=(1,0,0)^{\mathsf T}\) is an eigenvector (no mixing) with eigenvalue \(\lambda\) because the couplings \(A_{12}=A_{13}=0\) vanish, as in the case of the partial seniority conservation. Extending to a higher dimension in the presence of other orbitals (as one increases the model space) to the simplest $4\times 4$ matrix, for example:
\[
A_{4\times4}=
\begin{pmatrix}
\lambda & 0 & 0 & a\\
0 & \alpha & \beta & b\\
0 & \beta & \gamma & c\\
a & b & c & d
\end{pmatrix},
\]
Any nonzero \(a\) (as induced by cross-orbital non-diagonal TBMEs) destroys that isolation: the eigenvector that continuously “evolves” from \(e_1\) acquires components from the other basis vectors. If \(x\) is its eigenvalue and $x_{1,2,3,4}$ the components of its eigenvector, one convenient parametrization (fixing the overall scale by setting \(x_1=1\)) is
\[
x_4=-\frac{\lambda - x}{a},\qquad
\begin{pmatrix}x_2\\ x_3\end{pmatrix}
= -\frac{x_4}{\Delta(x)}\,
\begin{pmatrix}
\gamma - x & -\beta\\[2pt]
-\beta & \alpha - x
\end{pmatrix}
\begin{pmatrix} b\\ c\end{pmatrix},
\]
where \(\Delta(x)=(\alpha - x)(\gamma - x)-\beta^2\). Thus, for generic nonzero \(a,b,c\), one obtains \(x_2,x_3,x_4\neq 0\): the previously unmixed \(e_1\) mixes with the other three states in the \(4\times4\) system.
That is exactly the behavior of the partial seniority conserved configurations in the presence of other orbitals.

That was partly motivated by  numerical experiments showing that, as is often see in literature, the calculated E2 strength can be very sensitive to the model space and interaction used in large-scale multi-j shell model calculations, as illustrated in Fig.~\ref{96Pd}. One has to carefully examine the wave functions before jumping to the conclusion that the cross-orbital excitations are important.

\section{Experimental progresses and future opportunities}\label{expt-p}
While seniority coupling, particularly seniority isomerism, appears in many regions across the nuclear chart, as illustrated earlier in Fig.~\ref{isomer}, the partially seniority-conserved $j=9/2$ systems are found primarily in the following regions:
\begin{itemize}
  \item $N=126$ isotones ($\pi0h_{9/2}$) above $^{208}$Pb

  \item Pb isotopes ($\nu1g_{9/2}$) above $^{208}$Pb
    \item Neutron-rich Ni isotopes ($\nu0g_{9/2}$) below $^{78}$Ni

  \item $N=50$ isotones ($\pi 0g_{9/2}$) below $^{100}$Sn
  \item Neutron-rich $N=82$ isotones ($\pi 0h_{9/2}$) below $^{132}$Sn

    \item Neutron-rich Sn isotopes ($\nu0h_{9/2}$)

\end{itemize}
There is no experimental data yet on the $\nu0h_{9/2}$ coupling in the low-lying structure of the neutron-rich Sn isotopes, which may be more mixed and dominated by the $1f_{7/2}$ orbital instead. In the $^{208}$Pb/ $N=126$ region, one may also expect seniority coupling involving $\pi h_{11/2}$ and $\nu i_{13/2,11/2}$.

One can expect rapid progress in experimental studies of those regions in connection with the availability of  new and future large-scale facilities (FRIB-MSU~\cite{Brown_2025}, RIBF-RIKEN~\cite{PhysRevLett.133.072501}, FAIR-GSI, HIAF-IMP~\cite{Zhou2022}, RAON, as well as the $N=126$ factory at ANL). 
In Asia, the construction of the HIAF (High Intensity Heavy-ion Accelerator Facility) at IMP, China, is near completion~\cite{Zhou2022}. Beam commissioning of that facility has been carried out in autumn 2025. It has the capability of intense pulsed heavy-ion and radioactive beams and large energy range (MeV/u to ~GeV/u). The IMP-DRAGON High-Purity Germanium multi-detector array has been developed for 
$\gamma$-ray spectroscopy~\cite{ LI2025170804}. The  $^{208}$Pb region could be one of the most prioritized regions in their first experiments. The HIAF facility's high-energy radioactive beamline, HIRIBL, will be able to generate and characterize a range of drip-line nuclei with $Z = 60-90$. In addition, 
the RIKEN RIBF facility in Tokyo has been running very successfully in past decades~\cite{Sakurai2018-ge,Watanabe2019-ij} and is undergoing
ongoing upgrades. There is also active development in new detector array comprising a set of Clovers and the DEGAS detectors.

\subsection{Nuclei in the $^{208}$Pb / $N=126$ region}\label{208Pb}
Ref.~~\cite{Kiss2024} provides a state-of-the-art review of experimental methods, recent results and future prospects for neutron-rich nuclei with $55 \leq Z \leq 92$, with emphasis on structure phenomena (shell evolution, shape transitions, isomerism) and astrophysical relevance (r-process peaks). The $^{208}$Pb / N=126 region is identified as a priority in relation to the fact that experimental coverage is still sparse and $N=126$ controls the r-process A$\sim$195 peak. 

Speculating about the near future, there may be dedicated campaigns to reach $^{208}$Pb / N=126 nuclei via fragmentation, fission, or multi-neutron transfer reactions~\cite{Colovic2019-zn} at upcoming facilities, including the future FAIR~\cite{Aumann_2024} facility. These campaigns would be combined with Coulomb excitation and lifetime $(B(E\lambda))$ measurements with large-scale arrays like AGATA~\cite{Bracco_2021}. There have also been theoretical efforts in refining shell-model calculations for those heavy nuclei~\cite{qi2016large,yoshinaga2021large,PhysRevC.106.044314,PhysRevC.105.024315,PhysRevC.103.054303}, aiming for both nuclear structure and their $\beta$ decay properties that could be relevant for the r-process. There has been renewed interest in the nature of $^{208}$Pb itself which was expected for long to be the best doubly magic nucleus~\cite{henderson2025deformation}.

Many isomeric states (including the $8^+$ $\nu g_{9/2}$ seniority isomers and higher-spin $\alpha$-decaying spin traps) are abundant in the Pb/Hg/Po/Tl region. As mentioned earlier, the even-even $^{210\textit{,}212\textit{,}214\textit{,}216}$Pb nuclei all exhibit $8^+$ seniority isomer.  That was also observed in $^{212}$Po and $^{214}$Po, although the known $8^+$ states in $^{216,218}$Pb have not been confirmed to be isomeric. Fig.~15 in Ref.~~\cite{Kiss2024} compared the E2 transition strengths of $Z=80,81,82$ isomers with those of the $N=50$ isotone, which show great similarity for both the even-even and odd-A cases.

The seniority coupling and its partial conservation may also play a role at higher-lying states. There could also be states where both $j=9/2$ $\nu g_{9/2}$ and $\pi h_{9/2}$ are active. Several \(\alpha\)-decaying high-spin isomeric states with simple structure are known, including, for example, the $25 / 2^{+}$ state in ${ }^{211} \mathrm{Po}$ and $18^{+}$ state in ${ }^{212}$Po, which have $\pi h_{11 / 2}^2 v g_{9 / 2}$ and $\pi h_{9 / 2}^2 v (g_{9 / 2} h_{11 / 2})^{10^+}$ configurations, respectively.  One may build new isomeric states on top of those like the newly observed $23^+$ and $21^-$ states in Ref.~\cite{ZAGO2022137457}.  Similar isomeric states have been observed at both GSI and ISOLDE.

There have also been excited states  identified in some neutron-rich nuclei around $^{208}$Pb involving the cross-shell excitations of single-neutron and/or single-proton orbitals. Those excitations may be coupled to configurations with multi-neutrons in Ref.~\cite{PhysRevC.98.024324}.

In a recent paper entitled ``Manifestation of the Berry phase in the atomic nucleus $^{213}$Pb''~\cite{VALIENTEDOBON2021136183}, the authors successfully identified the neutron-rich isotope $^{213}$Pb, which was produced at GSI via the fragmentation of a relativistic $^{238}$U beam on a 2.5 g/cm$^2$ Be target. The reaction fragments were separated in the FRS spectrometer and implanted into a double-sided silicon-strip detector (DSSSD). For detection, the RISING setup was used, consisting of 105 high-purity Ge detectors arranged in 15 clusters around the implantation point. 
From that highlighted experiment, delayed coincidence spectra revealed six transitions with energies: 772, 488, 369, 311, 190, and 176 keV. The measured half-life of the isomeric state was 
    \[
      t_{1/2} = 0.26(2)\ \mu\text{s}.
    \]
 Among those, the cascade $772 \to 369 \to 190$ keV was assigned to the sequence $21/2^+ \to 17/2^+ \to 13/2^+ \to 9/2^+$ (ground state).
A second decay branch involving the $488$, $311$, and $176$ keV transitions, requiring an unobserved $71$ keV $E2$ transition. This places a second $17/2^+$ state ($17/2^+_2$) at 1260 keV, which also decays via an intermediate $15/2^+$ state.
The newly observed isomer and its $\gamma$-decay properties provide solid evidence of seniority conservation. It was argued to be driven by a Berry phase associated with particle-hole conjugation, which, if true, would be the first clear manifestation of such a geometric phase in nuclear physics. Irrespective of that, the experiment has successfully determined the reduced transition probabilities for the decay from the $21/2^+$ isomeric states which are
    \[
      B(E2; 21/2^+ \to 17/2^+_1) = 1.1(4)\ e^2\ \text{fm}^4, 
    \]
       \[
      B(E2; 21/2^+ \to 17/2^+_2) = 32(5)\ e^2\ \text{fm}^4.
    \]

In the cases of  $^{213}$Pb and $^{95}$Rh (that will be discussed below in Sec.~\ref{95Rh}), with five neutrons filling the $1g_{9/2}$ subshell and the five protons filling the $0g_{9/2}$ subshell, respectively,
the valence neutrons and protons form a limited set of angular-momentum and seniority
combinations. There are 20 states in total: one  with $v=1$, nine with $v=3$, and ten with $v=5$. All of these can be uniquely defined by the angular momentum and seniority quantum numbers. As listed in Table~\ref{tab:number} and Fig.~3 in Ref.~\cite{VALIENTEDOBON2021136183}, those states include three different $I^\pi = 9/2^+$ states with
$v = 1, 3, 5$, two $13/2^+$ states with $v = 3, 5$, and two
$17/2^+$ states with $v = 3, 5$ and the $21/2^+$ with $v = 3$.

The particle-hole conjugation~\cite{VanIsacker2024} is related to the operation that
    transforms an $n$-fermion configuration in a single-$j$ shell into a $(2\Omega - n)$-fermion configuration with $\Omega$ being the pair degeneracy as introduced earlier. The transformation takes the form
    $$
\Gamma\left|j^n v J\right\rangle=(-)^{(n-v) / 2}\left|j^{2 j+1-n} v J\right\rangle,
$$with $\Gamma$ being the transform operator.
    At half-filling ($n=\Omega$), states map onto themselves, but acquire a geometric phase.
    This phase, normally invisible, becomes observable at mid-shell because it connects particle-hole conjugation to seniority conservation.
    
    A distinctive feature of midshell nuclei is that a two-body interaction restricted to a single-$j$ shell can only connect states
whose seniority differ by $\Delta v = 4$.  Consequently, the two
$13/2^+$ and the two $17/2^+$ levels, associated with $v = 3$ and
$v = 5$, cannot mix through such interactions.  Under these conditions,
seniority remains an exactly conserved quantity for all states except the $I=9/2$ states. 
A detailed calculation in single-$j$ shell on the E2 transition properties of all above states can be found in Table V in Ref.~\cite{PhysRevC.108.064313}.

In Fig.~3 of Ref.~\cite{VALIENTEDOBON2021136183}, the calculated complete energy spectrum of a system with five nucleons in the $1g_{9/2}$ orbital has been derived using an empirical set of TBMEs, which agree remarkably well with the few known experimental data.
The spectrum confirms that, for mid-shell   $^{213}$Pb with five neutrons in the $1g_{9/2}$ shell,  all states conserve seniority, except for possible mixing between $J^\pi=9/2^+$ states of $v=1$ and $v=5$, as a consequence of the particle-hole conjugation symmetry.
The shell-model calculation can reproduce well the asymmetric $B(E2)$ values listed above. It is suggested that  the $17/2^+_1$ state has predominantly $v=3$ character, leading to suppression of the $B(E2)$ value as it carries the same seniority number as the $21/2^+$ while the $17/2^+_2$ state has predominantly $v=5$ character.  
    The large $B(E2)$ strength towards $17/2^+_2$ and the suppression towards $17/2^+_1$ are direct consequences of seniority conservation at mid-shell.

\begin{figure*}[htp]
\centering
\includegraphics[width=0.7\textwidth]{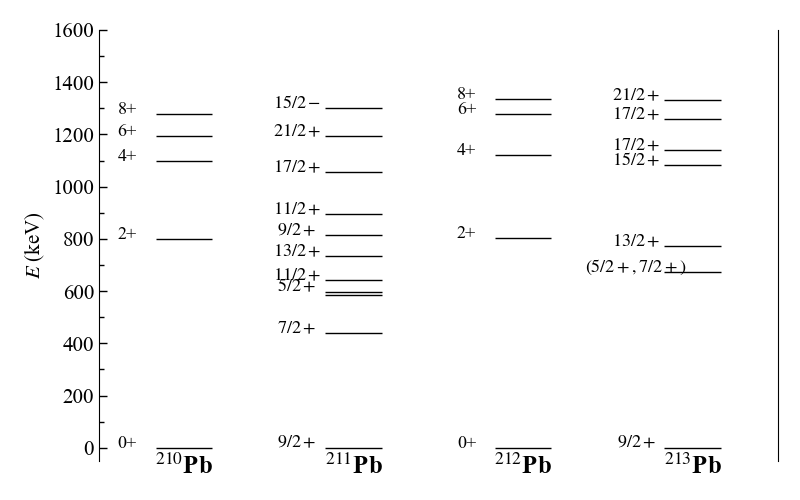}
\caption{Comparison between the low-lying spectra of odd-A $^{211,213}$Pb from Refs.~\cite{VALIENTEDOBON2021136183,LANE200534} and those of the neighboring even-even $^{210,212}$Pb taken from the NNDC database.}
\label{Pb-isomer}
\end{figure*}

In Fig.~\ref{Pb-isomer} we compare the low-lying spectra of odd-A $^{211,213}$Pb and those of $^{210,212}$Pb. The $21/2^+$ isomeric state in 
$^{211}$Pb was observed in Ref.~\cite{LANE200534} in experiment at Argonne National Laboratory with a $^{208}$Pb ion beam of 1360 MeV  from the ATLAS accelerator. There is currently no spectroscopic information available for the heavier isotope $^{215}$Pb. However, Ref.~\cite{PhysRevC.87.067303} successfully produced a $^{215}$Pb beam using resonant laser ionization, followed by mass separation at the ISOLDE-CERN on-line mass separator, and may have identified $\gamma$ rays from the transition of $^{215}$Pb and/or $^{215}$Bi.

\begin{figure}[htp]
\centering
\includegraphics[width=0.35\textwidth]{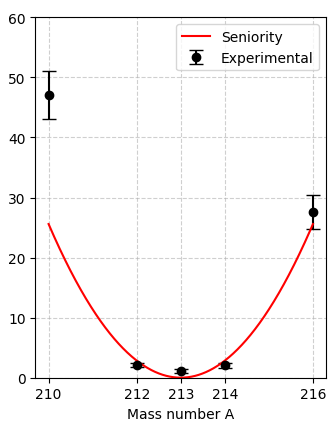}
\caption{$B\left(E2; 8^{+} \rightarrow 6^{+}\right)$ transition strengths for even-even Pb isotopes from Ref.~\cite{PhysRevLett.109.162502}, compared with predictions from the seniority model, and the analogous $B(E2; 21/2^+ \to 17/2^+_1)$ for mid-shell $^{213}$Pb~\cite{VALIENTEDOBON2021136183}, where the $B(E2)$ value is expected to vanish in the ideal seniority scheme. 
\\
For completeness, the experimental half-lives of the $8^+$ isomeric states are $T_{1/2}=201(17)$~ns, $6.0(8)$~$\mu$s, $6.2(3)$~$\mu$s, and $0.40(1)$~$\mu$s for $^{210,212,214,216}$Pb, respectively, compared with $T_{1/2} \approx 260$~ns for the $21/2^+$ state in $^{213}$Pb.
}
\label{Pb-isomer-2}
\end{figure}

Regarding the even-even Pb isotopes of concern, spectroscopic studies of the isomeric states in neutron-rich isotopes up to $^{216}$Pb  were performed via the fragmentation of a high-energy $^{238}$U beam at the FRS-RISING setup at GSI~\cite{PhysRevLett.109.162502,Gottardo2014-ea}. In the experiment, projectile fragments were separated with the FRS spectrometer and implanted in silicon-strip detectors while coincident $\gamma$ rays were detected with the RISING array.  Isomeric decays in $^{214}$Pb, and $^{216}$Pb were observed for the first time, and the extracted half-lives ranged from microseconds to sub-microseconds. A systematic comparison between the experimental spectra of  $^{210,212,214,216}$Pb and those from the shell-model calculations with the Kuo-Herling interaction was presented in Fig.~2 in Ref.~\cite{PhysRevLett.109.162502}. There is little further spectroscopic information available for the $n=4$ systems $^{212,214}$Pb beyond the $8^{+}$ isomeric states shown in Fig.~\ref{Pb-isomer}.

The experimental $B\left(E 2 ; 8^{+} \rightarrow 6^{+}\right)$ transition strengths for even-even Pb isotopes and $^{213}$Pb are plotted in Fig.~\ref{Pb-isomer-2}. In particular, the values for $^{212,214}$Pb are observed to be quite similar to each other. However, it was noted in Ref.~\cite{PhysRevLett.109.162502} that, while excitation energies were reproduced within about 100 keV using the Kuo-Herling interaction, the predicted $B(E2)$ strength for the $8^+ \rightarrow 6^+$ transition in $^{212}$Pb was overestimated by factors of two to five, which leads to an artificial asymmetry around mid-shell. Adjustments to single-particle energies improved the agreement only partially, and other modern realistic interactions such as CD-Bonn gave similar inconsistencies, indicating that the discrepancies are not interaction-dependent. The authors then explored the explicit inclusion of particle-hole excitations from the $^{208}$Pb core, which simulate effective three-body forces and two-body operators. Those corrections seem to improve the agreement with experiment by restoring midshell symmetry, and yielding consistent quadrupole moment signs. It was thus speculated that the effective three-body interaction may play a central role in describing electromagnetic transition rates in heavy neutron-rich systems. 
 Traditionally, the shell model uses two-body effective interactions within a limited valence space, often neglecting explicit higher-order effective terms that arise from renormalization, including effective three-body forces and two-body transition operators~\cite{Derbali2018-vs}. In intermediate-mass and heavy nuclei, those kinds of effective three-body forces contribute mostly at the mean field level and rarely lead to genuine three-body effects.
In addition, one must consider the facts that: The uncertainty in shell-model studies  (or any microscopic modeling) of the nuclear lifetimes can be quite large and the observed discrepancy may not be significant enough to judge the quality of the model calculations; And that, near the mid-shell, the $B(E2)$ calculation can be very sensitive to tiny changes in the wave function and the occupation of the neutron $1g_{9/2} $ orbital while most interactions are optimized accordingly to energy criterion which show much less sensitivity.  
Therefore, for now, such speculation on the role of three-body interactions may not yet be fully convincing.

In addition to seniority model and large-scale shell model studies, there were extensive studies by Jan Blomqvist and collaborators within the multistep shell model approach~\cite{BLOMQVIST199345} and quasi-particle multistep shell model method~\cite{POMAR1990381} on both low-lying and core-excited states in the Pb region.  The calculations were done with the Kuo-Herling interaction~\cite{kuo1971a,HERLING1972113} which was further adjusted to fit experimental data~\cite{blomqvist1984a}. That is the interaction that we have been continuing refining~\cite{qi2016large,Qi2025-uc}.
Three-quasiparticle states analysis in odd-mass lead isotopes with the modified Kuo-Herling interaction were shown in Ref.~\cite{PhysRevC.47.554}.

We hope the large-scale facilities like the already-running RIBF and FRIB, the soon-to-be-online HIAF, and the future FAIR will offer many physics opportunities in the Pb region including in particular
precision lifetimes and $B\left(E 2 ; 8^{+} \rightarrow 6^{+}\right)$ values across Pb, Hg, Po isotopic chains
through fast timing, recoil-distance, or isomer-decay spectroscopy in combination of the advent of new fast-timing scintillators~\cite{PhysRevResearch.6.L022038,Li2025a,Zhang2025a,Ballan_2023}. The detailed knowledge on the $B(E2)$ trend will be a sensitive probe of seniority conservation, energy evolution of the $1g_{9/2}$ orbital and mixing effects from core-excited configurations. The search for the missing yrare $4^+$ and $6^+$ states in $^{212,214}$Pb and neighboring isotones will reveal unique information on the role played by the partial seniority conservation in that region and benchmark large-scale microscopic models to be developed which is not only important for the understanding of nuclear structure effects but also for determining $\beta$ and first-forbidden $\beta$ decay properties that are needed for nucleosynthesis simulations.

\subsection{The neutron-rich Ni isotopes}\label{Nineutron}
The neutron-rich Ni isotopes are the lightest nuclei that involve a $j=9/2$ orbital where seniority may not be conserved. There has been a long history of studying the neutron-rich Ni isotopes and neighboring nuclei~\cite{Sahin_2019,Hagen_2019,Taniuchi_2025,Canete_2024,Sun_2024,Giraud_2022,Ballan_2023,PhysRevC.103.064328} addressing in particular key questions including: The magicity of the $N=40$ and 50 shell closures and the isomerism of the $8^+$ states for isotopes in between.
In Ref.~\cite{Taniuchi_2019} it was concluded that the neutron-rich nucleus \textsuperscript{78}Ni is indeed doubly magic, exhibiting enhanced stability and a spherical ground-state structure. This conclusion is based on the observation of its first excited $2^{+}$ state at about 2.6\,MeV, characteristic of a strong shell closure (see Fig.~4 in Ref.~\cite{Taniuchi_2019}). The experiment was carried out at RIBF, RIKEN using the MINOS device (developed by CEA-IRFU, France) together with the DALI2 spectrometer to detect emitted $\gamma$ rays. The robustness of the $N=40$ shell closure, which is not expected to be as strong as the spin-orbit shell closure $N=50$ in any case, may be inferred from the low quadrupole collectivity~\cite{PhysRevLett.88.092501}.
Theoretical studies on the structure of those nuclei with the large-scale shell model or similar approaches could be found in Refs.~\cite{Isacker_2011,Lisetskiy_2005,PhysRevC.70.044314,PhysRevC.67.044314,PhysRevC.110.034316,PhysRevC.111.044308,Tichai_2024,Li_2023,Hu_2024,xu2013shell,Sidorov_2022}.

The isomeric state in $^{70}$Ni has half-life 0.232~$\mu$s. The $6^+$ state is also pretty long-lived with $T_{1/2}=1.049$~ns for a 448~keV E2 transition. In comparison, the isomeric state in $^{76}$Ni has half-life
$T_{1/2}=547.8$~ns and transition energy 142.58~keV. One striking feature of these neutron-rich isotopes is the missing isomerism in $^{72,74}$Ni. That was seen in various shell-model calculations and is attributed to the possible scenario that the $v=4$, $6^+$ state may come lower than the yrast $8^+$ (and $v=2$ $6^+$) state (see for examples, Fig.~1 in Ref.~\cite{Isacker_2011} and Fig.~4 in Ref.~\cite{PhysRevC.70.044314}).  This can effectively reduce its lifetime  by several orders of magnitude as the E2 transition strength between the two states is expected to be strong. 

\begin{figure}[ht]
\centering
\includegraphics[width=0.5\textwidth]{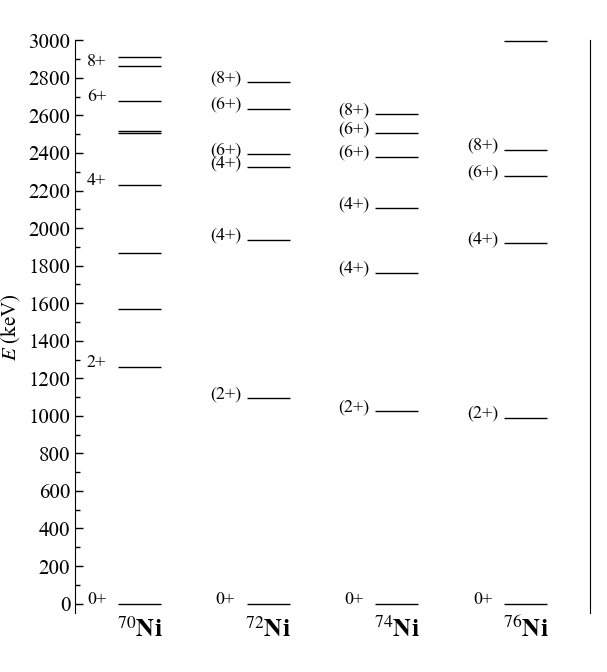}
\caption{Low-lying spectra of even-even $^{70-76}$Ni. Data are taken from Refs.~\cite{PhysRevC.93.034328,MORALES2018706} and NNDC database.}
\label{Ni-isomer-2}
\end{figure}

Refs.~\cite{PhysRevC.93.034328,MORALES2018706} present significant achievement on the investigation of the low-lying structures of the neutron-rich nickel isotopes $^{72}$Ni and $^{74}$Ni conducted within the EURICA campaign at the RIKEN RIBF facility via the $\beta$ decay of $^{72}$Co and $^{74}$Co. Even though most  spin-parity assignments are still tentative, 
their observation of previously unknown low-lying states in $^{72}$Ni and $^{74}$Ni is providing a near complete picture of the seniority scheme up to the first $8^+$ states for even-even isotopes $^{72}$Ni.
Some key experimental information we can see from that experiment~\cite{MORALES2018706} include:
\begin{itemize}
    \item Identification of two $\beta$-decaying isomers in $^{74}$Co (high-spin and low-spin states) feeding distinct level structures in $^{74}$Ni.
    \item First measurement of the half-life of the $6_1^+$ state in $^{72}$Ni using in-flight $\beta$-delayed fast-timing spectroscopy, yielding $t_{1/2} = 860(60)$~ps.
    \item Extraction of reduced transition probabilities, $B(E2)$, for transitions between yrast states in $^{72}$Ni, showing a smooth downward trend with increasing spin.
    \item Observation of different decay patterns for the $6_2^+$ state: in $^{72}$Ni it decays predominantly to the $4_1^+$ state, while in $^{74}$Ni it feeds the $4_1^+$ and $4_2^+$ states with equal intensity.
    \item  As mentioned, the disappearance of the $8^+$ seniority isomerism in $^{72,74}$Ni is explained by the predicted lowering of the seniority-$v=4$ $6^+$ state below the seniority-$v=2$ $6^+$ state.
    \item  The different decay patterns of the $6_2^+$ state in $^{72}$Ni and $^{74}$Ni suggest the involvement of seniority-nonconserving interactions.
\end{itemize}
The experimental results are compared with four different shell-model calculations to assess the conservation of seniority in the $\nu g_{9/2}$ shell.
In Fig.~\ref{Ni-isomer-2}, we compare the low-lying spectra of even-even isotopes $^{70-76}$Ni, which are expected to be dominated by the coupling of neutrons in the $0g_{9/2}$ orbital.
The observed $2^+$, $4^+$ and both $6^+$ states in $^{72}$Ni are confirmed by neutron and proton knockout reactions in the SEASTAR campaign at RIBF~\cite{Angelini2021}. The low-lying spectra of neighboring odd-A $^{73,75}$Ni have also been studied~\cite{PhysRevC.102.044331}. The transition strengths for decays from the presumed $5/2^+$, $13/2^+$, and $11/2^+$ states of those two isotopes are reported in Ref.~\cite{PhysRevC.102.014323}, and they agree reasonably well with shell-model calculations.

The reduced transition probability for  $^{74}$Ni, $B(E2;2^+\rightarrow 0^+_{gs})=128_{-45}^{+43}\, e^2\mathrm{fm}^4
$,  has been measured 
in an intermediate energy Coulomb excitation experiment~\cite{PhysRevLett.113.182501}. 
The lifetimes of the first excited $2^+$ and $4^+$ states in $^{72}\mathrm{Ni}$  were measured in Ref.\cite{PhysRevLett.116.122502} using the recoil-distance Doppler-shift method. The excited states were populated via a one-proton knockout reaction and $\gamma$-ray–recoil coincidences were detected with the GRETINA array. The measured reduced transition probabilities in $^{72}\mathrm{Ni}$ are $B(E2;2^+_1 \!\to\! 0^+) = 74(10)\,e^2\mathrm{fm}^4$ and $B(E2;4^+_1 \!\to\! 2^+_1) = 50(9)\,e^2\mathrm{fm}^4$, corresponding to lifetimes of $\tau(2^+_1)=7(1)$\,ps and $\tau(4^+_1)=38(9)$\,ps, respectively.
An interesting systematic comparison for the $2^+ $energies and $B(E2)$ values of Ni isotopes are presented in Fig.~4 of that paper.
 The results are mostly consistent with other experiments and theoretical calculations, except large $B(E2)$ values in $^{70}\mathrm{Ni}$ from Coulomb excitation measurements. The measured lifetime of the $4^+$ state supports a seniority $v=4$ decay which, as mentioned above, is consistent with the disappearance of the $8^+$ isomer in $^{72}\mathrm{Ni}$.

\subsection{$N=50$ isotones ($\pi 0g_{9/2}$) below $^{100}$Sn}\label{RuPd}
The region surrounding 
$^{100}$Sn has become one of the most intensively explored domains in contemporary nuclear physics, both experimentally and theoretically~\cite{Faestermann_2013,G_rska_2022}.
The robustness of the $N=Z=50$ shell closures in $^{100}$Sn was questioned until recently. $^{100}$Sn is now considered to be doubly magic, as supported by significant experimental evidence, including, for examples:
The half-life and $Q_{\beta}$ measurement for super-allowed Gamow–Teller decay of $^{100}$Sn at GSI~\cite{Hinke_2012},
which showed an exceptionally large GT strength concentrated in a single transition of single-particle nature; studies of alpha decay to $^{101}$Sn~\cite{Darby_2010}, indicating dominant single-particle character; direct 
$B(E2)$ measurement on neutron-deficient Sn isotopes and neighboring nuclei~\cite{Guastalla_2013,Back2013}; as well as precise measurements of the masses~\cite{Mougeot_2021,PhysRevLett.133.132503} and electromagnetic-moment and charge-radius systematics~\cite{ Karthein_2024} on $N\sim Z$ In and Ag isotopes.

\begin{figure*}[htp]
\centering
\includegraphics[width=0.95\textwidth]{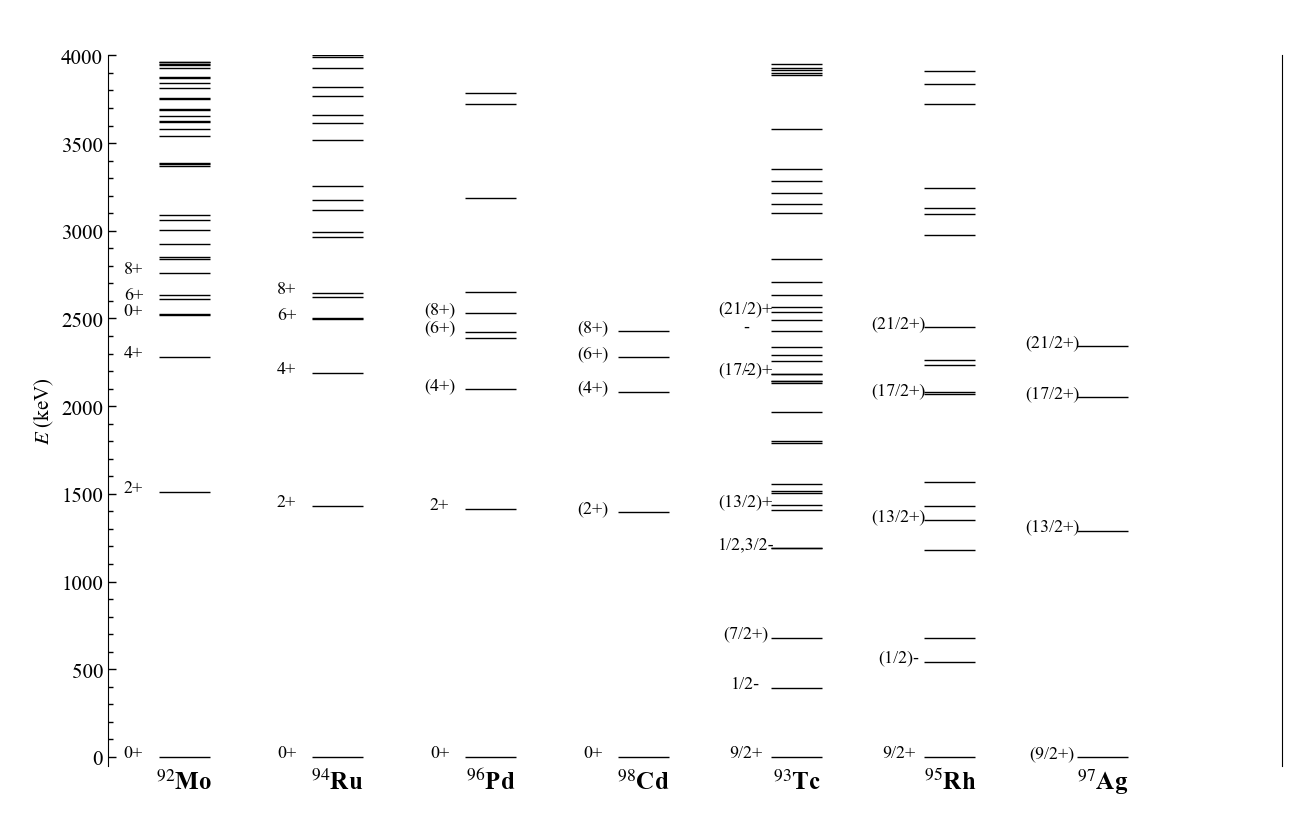}
\caption{Low-lying spectra of even-even and odd-$A$ $N=50$ isotones. Data are taken from the NNDC database. Some spin assignments are omitted for simplicity. There is currently no clear indication of possible candidates for the second $4^+$ and $6^+$ states in $^{94}$Ru and $^{96}$Pd, which are crucial for understanding partial seniority conservation.}
\label{n50}
\end{figure*}

One should safely expect many of the low-lying states in Mo-In $N=50$ isotones to be dominated by the coupling of valence protons in $\pi 0g_{9/2}$ orbital~\cite{Ertoprak_2018,PhysRevC.86.014318} and by the seniority scheme. That is well supported by the known data on the level structure, $8^+$ isomer half-lives as well as the emerging $B(E2;2_1^+\!\to\!0_1^+)$. In Fig.~\ref{n50} we compare the low-lying spectra of the $N=50$ isotones. Detailed comparison between experimental data and theory on the seniority structure in those even-even nuclei can be found in Refs.~\cite{PhysRevLett.87.172501,Isacker1,Isacker_2011}. The half-life of the $8^+$ isomers in $^{94}$Ru and $^{96}$Pd have been measured long ago~\cite{H_usser_1977,Grawe_1983}. The half-life of the yrast $8^+$ state in $^{94}$Pd was recently measured at FAIR-0, GSI~\cite{YANEVA2024138805}, yielding a reduced transition probability of $B(E2; 8^+ \rightarrow 6^+) = 205~\mathrm{e}^2\mathrm{fm}^4$, which indicates that the seniority scheme is largely preserved. The $N=Z$ nucleus $^{92}$Pd is, however, known to be dominated by neutron-proton coupling rather than seniority coupling~\cite{cederwall2011evidence}. For the heavier $N=50$ isotone $^{98}$Cd, both the $8^+$ and $6^+$ states are isomeric, with half-lives measured in Ref.~\cite{PhysRevC.96.044311}.
One may even expect some similarity in the structure between $N=50$ and $N=82$ isotones for proton systems in the $\pi 0g_{9/2}$ orbital (see, for example, Fig.~4 in Ref.~\cite{PhysRevLett.113.042502}) and between $N=50$ isotones and the neutron-rich Ni isotopes (Fig.~3 in Ref.~\cite{Isacker_2011}).

\subsubsection{The $n=4$ systems $^{94}$Ru and $^{96}$Pd}
The pair of nuclei $^{94}$Ru and $^{96}$Pd provide an ideal ground to test the partial seniority conservation. The properties of their low-lying states should be fairly well described by the coupling of four proton particles or four proton holes in the orbital $0g_{9/2}$. Their structure should be exactly the same if $0g_{9/2}$ is completely isolated and the interaction remains the same.
These nuclei have been heavily studied recently, leading to controversial results from both experimental and theoretical perspectives.

In the work of Mach \emph{et al.}~\cite{Mach2003,PhysRevC.95.014313}, low-lying yrast states in $^{94}$Ru and $^{96}$Pd were populated using fusion--evaporation reactions with a 145\,MeV $^{36}$Ar beam delivered by the Cologne FN Tandem accelerator impinging on enriched Ni targets ($^{58}$Ni for $^{94}$Ru and $^{60}$Ni for $^{96}$Pd). Prompt $\gamma$ rays were detected with an array of HPGe detectors for high-resolution spectroscopy in coincidence with LaBr$_3$(Ce) scintillators that provided sub-nanosecond timing (an older generation of fast-timing detectors). Lifetimes of the $2^+_1$, $4^+_1$, and $6^+_1$ states were extracted with the advanced time-delayed $\gamma$--$\gamma(t)$ fast-timing method, employing both centroid-shift and de-convolution analyses. This setup allowed the determination of very short lifetimes, most notably of the $4^+$ states. 

Ref.~\cite{das2022nature} reported results from recent experiment conducted as part of the FAIR-0 campaign at GSI/FAIR using the DESPEC setup, composed of the AIDA implantation detector array, HPGe detectors, and the FAst TIMing Array (FATIMA) of LaBr$_3$(Ce) scintillators.
Excited states in $^{94}$Ru were populated through $\beta$-delayed proton emission from $^{95}$Pd. The latter was produced via projectile fragmentation of an 850 MeV/nucleon \(^{124}\)Xe beam impinging on a 4 g/cm$^2$ $^9$Be target.
They used $\gamma\text{–}\gamma$ coincidences with fast timing from LaBr$_3$(Ce) detectors to determine the lifetimes of low-lying yrast states in \(^{94}\)Ru. The $\gamma$--$\gamma$ fast-timing technique enabled extraction of sub-ns lifetimes, yielding values of $\tau(2^+_1) \approx 15(10)$\,ps and $\tau(4^+_1)=32(11)$\,ps.

In Ref.~\cite{PhysRevLett.129.112501}, excited states in $^{90}$Zr, $^{92}$Mo, and $^{94}$Ru were populated via multi-nucleon transfer using a 717\,MeV $^{92}$Mo beam incident on an enriched $^{92}$Mo target. Lifetimes were determined with the recoil-distance Doppler-shift (RDDS) technique employing a high-precision plunger device at the IKP Cologne. Prompt $\gamma$ rays were detected with a partial AGATA array, while reaction products were identified in the VAMOS$^{++}$ magnetic spectrometer, enabling clean ion selection and event-by-event Doppler correction. Distances between the target and degrader foils ranged from 19 to 4000\,$\mu$m, corresponding to recoil velocities of $\beta \approx 0.12$ before and $\beta \approx 0.11$ after the degrader. Analysis combined the decay-curve method (DCM) and the differential decay-curve method (DDCM), with careful treatment of long-lived feeders and gating on the total kinetic energy loss (TKEL) to isolate specific reaction channels. This setup provided lifetimes in the ps–ns range with high precision for yrast $2^+$, $4^+$, and $6^+$ states along the $N=50$ isotonic chain.

There is another very recent experiment~\cite{PhysRevC.108.064313} that reported fast-timing lifetime measurements of key excited states in 
$^{92}$Mo using a hybrid array of HPGe and LaBr$_3$(Ce) detectors at the Cologne FN Tandem accelerator. Their measured lifetimes differ noticeably from values obtained using the RDDS technique~\cite{PhysRevLett.129.112501}, highlighting a dependence on the experimental method for determining lifetimes and derived B(E2) strengths. In their following up paper~\cite{PhysRevC.110.034320}, lifetimes for the low-lying excited states in $^{93}$Tc and $^{94}$Ru were reported and compared with calculations  in single-$j$ and $p_{1/2}g_{9/2}$ model space where, as one would expect, the seniority symmetry is roughly conserved and $\Delta v=0$ transitions are  predicted to be rather weak. 
There has been another recent experiment measuring the $B(E2,4^+\rightarrow2^+)$ value in $^{94}$Ru at HIRFL, IMP. The data is still being analyzed~\cite{Zhang2025a}. 
Neighboring nuclei like $^{95}$Rh may also be detectable with  IMP-DRAGON on their existing HIRFL beamline and at the BTANL/BRIF (ISOL) facility in Beijing~\cite{NAN2025104188}.

The key experimental results from above four experiments are summarized in Table \ref{tab:all_results}. 
Ref.~\cite{PhysRevLett.129.112501} explicitly reported a discrepancy with Ref.~\cite{das2022nature} and presented robustness checks (TKEL gating and feeder subtraction) that still yielded consistent limits $\tau\simeq 72$--95\,ps for the $4_1^+$ state, thereby supporting their longer lifetime data.

The two experiments from Refs.~\cite{das2022nature,PhysRevLett.129.112501} present conflicting conclusions, which is causing considerable confusion as well around the \(4^+ \to 2^+\) transition in \(^{94}\)Ru. Ref.~\cite{PhysRevLett.129.112501} interpreted their \emph{smaller} $B(E2)$) as indication for restoring the expected hindrance at mid-filling and aligning with large-space shell-model predictions that favor largely $\nu{=}2$ yrast structures and over all seniority conservation with only mild configuration mixing, consistent with the observed  and calculated global $E2$ patterns.
On the other hand, the anomalous pattern—enhanced \(4^+ \to 2^+\), suppressed \(6^+ \to 4^+\) observed in Ref.~\cite{das2022nature} is interpreted as evidence for existing of partial seniority conserved states and seniority-symmetry breaking due to subtle interference between \(v=2\) and \(v=4\) states.

\begin{table*}[ht]
\caption{Half-lives and reduced transition probabilities $B(E2)$ reported for low-lying states in $N=50$ isotones from  recent papers~\cite{PhysRevC.95.014313,das2022nature,PhysRevLett.129.112501,PhysRevC.108.064313} using different experimental methods as briefly explained in the main text. Results for $^{94}$Ru from Refs.~\cite{PhysRevC.95.014313,das2022nature} seem to agree but Ref.~\cite{PhysRevLett.129.112501} gives a $B(E2;4^+_1 \to 2^+_1)$ value that is three times smaller. The most recent result from Ref.~\cite{PhysRevC.110.034320} falls in between the two values. There is also noticeable mismatch between Ref.~\cite{PhysRevLett.129.112501,PhysRevC.108.064313} on $^{92}$Mo measurement. }
\label{tab:all_results}
\centering
\renewcommand{\arraystretch}{1.3}
\begin{tabular}{lcccc}
\hline
\textbf{Experiment} & \textbf{Nucleus / Transition} & \textbf{Half-life $\tau$ (ps)} & \textbf{$B(E2)$ ($e^2$fm$^4$)} \\
\hline
Mach \emph{et al.} & $^{94}$Ru $2^+_1 \to 0^+_1$ &$\le 0.0144$ & $\ge 9.5$ \\
    (2017) ~\cite{PhysRevC.95.014313}                     & $^{94}$Ru $4^+_1 \to 2^+_1$ & $\le 0.0721$ & $\ge 46$  \\
                          & $^{96}$Pd $2^+_1 \to 0^+_1$ & $\le 0.0245$ & $\ge 6$ \\
                          & $^{96}$Pd $4^+_1 \to 2^+_1$ &$1.443(144)$ & $3.8(4)$ \\
\hline
Das \emph{et al.}   & $^{94}$Ru $2^+_1 \to 0^+_1$ & $\leq 15 $ & $\geq 10$ \\
              (2022) ~\cite{das2022nature}           & $^{94}$Ru $4^+_1 \to 2^+_1$ & $32(11)$ & $103(24)$ \\
                                     & $^{94}$Ru $6^+_1 \to 4^+_1$ & $91(3)\times 10^3$ & $3.0(2)$ \\

\hline
P\'erez-Vidal \emph{et al.}
                          & $^{90}$Zr $4^+_1 \to 2^+_1$ & $4.2(4)$ & $304(29)$ \\
        (2022) ~\cite{PhysRevLett.129.112501}                  & $^{90}$Zr $6^+_1 \to 4^+_1$ & $18^{+5}_{-8}$ & $122^{+34}_{-54}$ \\
                          & $^{92}$Mo $2^+_1 \to 0^+_1$ & $0.59(15)$ & $196(56)$ \\
                          & $^{92}$Mo $4^+_1 \to 2^+_1$ & $35.5(6)$ & $84.3(14)$ \\
                          & $^{94}$Ru $2^+_1 \to 0^+_1$ & $0.8(4)$ & $165(80)$ \\
                          & $^{94}$Ru $4^+_1 \to 2^+_1$ & $87(8)$ & $38(3)$ \\
                          \hline
Ley \emph{et al.}
                     & $^{92}$Mo $2^+_1 \to 0^+_1$ & $\leq 3$ & $\geq 35$ \\
  (2023) ~\cite{PhysRevC.108.064313}                           & $^{92}$Mo $4^+_1 \to 2^+_1$ & $22.5(11)$ & $132^{+7}_{-6}$ \\
(2024) ~\cite{PhysRevC.110.034320}       & $^{94}$Ru $2^+_1 \to 0^+_1$ & $\leq 2$ & $\geq 68$ \\
  & $^{94}$Ru $4^+_1 \to 2^+_1$ & $66(2)$ & $50 (2)$ \\
  & $^{94}$Ru $6^+_1 \to 4^+_1$ & $95.5(6)\times 10^3$ & $2.85(2)$ \\
\hline
\end{tabular}

\end{table*}

We summarize what we know so far as follows:
\begin{itemize}
  \item The measured \(B(E2; 2^+ \to 0^+)\) and \(B(E2; 8^+ \to 6^+)\) in the yrast cascade behaves as expected under seniority conservation. 
\item There is no experimental information on the location of the second $4^+$ or $6^+$ states in  $^{94}$Ru or $^{96}$Pd.
\item E2 transition strengths are known for yrast states in both nuclei. For $^{96}$Pd, the $B(E2;6_1^+\!\to\!4_1^+)$ has been known~\cite{Alber1989} and is already in  the NNDC database; $B(E2;4_1^+\!\to\!2_1^+)$ was reported in Ref.~\cite{PhysRevC.95.014313} as shown in Table~\ref{tab:all_results}. For $^{94}$Ru, there are two consistent measurements~\cite{das2022nature,PhysRevC.110.034320} of $B(E2;6_1^+\!\to\!4_1^+)$; there are three measurements of $B(E2;4_1^+\!\to\!2_1^+)$ that seem to disagree with each other but all show rather large values. 
Resolving this discrepancy is crucial for understanding the structure of \(j = 9/2\) nuclei and for testing theoretical models and nuclear interactions.
\item Despite the differences in measurements, the few known transitions in  $^{94}$Ru or $^{96}$Pd show two interesting patterns: One of the two transitions, $B(E2;6_1^+\!\to\!4_1^+)$ or $B(E2;4_1^+\!\to\!2_1^+)$, is almost completely suppressed, and the two nuclei exhibit precisely opposite patterns.
\item The anonymously suppressed $B(E2;6_1^+\!\to\!4_1^+)$ in $^{96}$Pd and $B(E2;4_1^+\!\to\!2_1^+)$ in $^{94}$Ru actually provide solid evidence for the existence of the partial seniority conserved states. That will be explained in more detail in the section below. 
\end{itemize}

So far, there is no clear indication on the existence of the $4_{2}^{+}$ and $6_{2}^{+}$ in $^{94}$Ru or $^{96}$Pd.
A tentative search for the $6_{2}^{+}$ state in $^{94}$Ru was reported in Ref.~\cite{PhysRevC.75.047302}. 
For $^{94}$Ru and $^{96}$Pd, unlike in neutron-rich Ni isotopes, the two partially seniority conserved states are expected
to lie just above the yrast $I=4$ and $I=8$ states, respectively, in most typical shell-model calculations. 
The $4_{2}^{+}$ states in $^{94}$Ru and $^{96}$Pd were also predicted to be lower in energy 
than the $6_{1}^{+}$ states in the $pg$-shell calculations, 
suggesting a strong competition between collective quadrupole and aligned configurations in this region.

\subsubsection{Theoretical interpretation of the $^{94}$Ru and $^{96}$Pd data}
Refs.~\cite{PhysRevC.95.014313,das2022nature,PhysRevLett.129.112501} not only presented different experimental results but also came to different or even seemingly contradictory conclusions. Ref.~\cite{PhysRevC.95.014313} compared three different shell-model calculations done in two different model spaces, $\pi f_{5/2}p_{1/2}g_{9/2}$ (marked SMCC and SMLB) and $gds$ (SDGN).
As presented in Table~1 in that paper, the calculations generally produce similar results, but notable discrepancies arise, especially with regard to
the $4^+\rightarrow2^+$ and $6^+\rightarrow4^+$ transitions in $^{94}$Ru and $^{96}$Pd. The latter calculations reproduced the enhanced $4^+\rightarrow2^+$ transition in  $^{94}$Ru and strongly suppressed $4^+\rightarrow2^+$ transition in  $^{96}$Pd (though failed to describe the suppressed $6^+\rightarrow4^+$ transition in  $^{94}$Ru). The paper interpreted the results as indication for 
``breakdown of the seniority quantum number due to particle-hole excitations across the $ N = Z = 50$ shell"~\cite{Grawe17}. But one must consider that the calculations in different model spaces were done with different interactions. It is not clear whether the difference  in the TBMEs, the model space, or the inclusion of some higher-lying orbitals that are critical for reproducing the data.
One also needs to be careful when opening up the model space. For example, any observation of non-zero transition strength between states with same seniority in a mid-shell system would indicate a breakdown of the shell-model description, since such
$\Delta v = 0$ transitions must vanish identically in the
single-$j$ seniority model, as discussed below. But from a general perspective, those contributions from configuration mixing are mostly tiny components of the wave function and play a minor role in explaining nuclear structure.

A systematic calculation on all even-even $N=50$ isotones within $\pi 0g_{9/2}$ was done in Ref.~\cite{PhysRevLett.129.112501} within the $f_{5/2}pg_{9/2}$ model space and realistic CD-Bonn potential. A remarkably good agreement is obtained for the excitation energies of the yrast  \(2^{+}\) to \(8^{+}\)  states and the overall trend of the $B(E2)$ values except for those of $4^+\rightarrow2^+$. It was concluded that seniority is largely
conserved. Some interesting observations from their calculations (as shown in Fig.~3 in that paper):
\begin{itemize}
\item The calculated $B(E2;2_1^+\!\to\!0_1^+)$ exhibits a \emph{maximum} near mid-filling of $\pi g_{9/2}$ (a $\Delta\nu{=}2$ pattern). That is quite reasonable though the limited amount of data doesn't show any clear trend yet.
\item The calculated $B(E2;4_1^+\!\to\!2_1^+)$, $B(E2;6_1^+\!\to\!4_1^+)$, and $B(E2;8_1^+\!\to\!6_1^+)$ show \emph{minima} and nearly vanishing values near mid-filling in $^{94}$Ru but not in $^{96}$Pd. That is opposite to the observed pattern for the $B(E2;4_1^+\!\to\!2_1^+)$ transitions which show vanishing value at $^{96}$Pd and a relatively speaking still quite strong value at $^{94}$Ru.
\end{itemize}

Both papers above focused on the bulk properties or the general conservation or breaking of the seniority. Neither study examined the partially conserved seniority nature of the $4^{+}$ and $6^+$ states. That was the main focus of Refs.~\cite{das2022nature,PhysRevC.110.034320}. 
It may be interesting to mention that there is a detailed single-$j$ shell calculation on the E2 transition properties presented in above paper~\cite{PhysRevC.108.064313}, which used experimental B(E2) strengths in $^{92}$Mo  to predict  B(E2) values in the heavier $N =50$ isotones from up to $^{95}$Rh.

So, what new physics can we learn from the present three rather different data sets? Along the $N{=}50$ isotones with valence protons filling $\pi g_{9/2}$, the yrast $2_1^+,4_1^+,6_1^+,8_1^+$ states are \emph{predominantly} seniority-$v{=}2$; their $E2$ systematics display the $\Delta v{=}0$ minima and $\Delta v{=}2$ maxima expected if seniority is conserved. We do not believe that any of the above experiments or calculations fundamentally challenge the observed bulk behavior that the seniority is largely conserved for the systems in the $\pi 0g_{9/2}$ orbital. What is of more interest is the role played by the partial seniority conservation in the $4^+$ and $6^+$ states. 

There is no evidence for their existence yet. However, let us assume that the partially seniority-conserved $4^+$ and $6^+$ states lie slightly higher than the yrast $4^+$ and $6^+$ states. Their existence would prevent the mixture with the yrast $v=2$ states. That occurs because partial dynamic symmetry known for $j{=}9/2$ (special $\nu{=}4$, $J{=}4,6$ solvable states) inhibits $\Delta \nu{=}2$ mixing even when $\nu{=}4$ states lie relatively close in energy.

\begin{figure}[htp]
\centering
\includegraphics[width=0.52\textwidth]{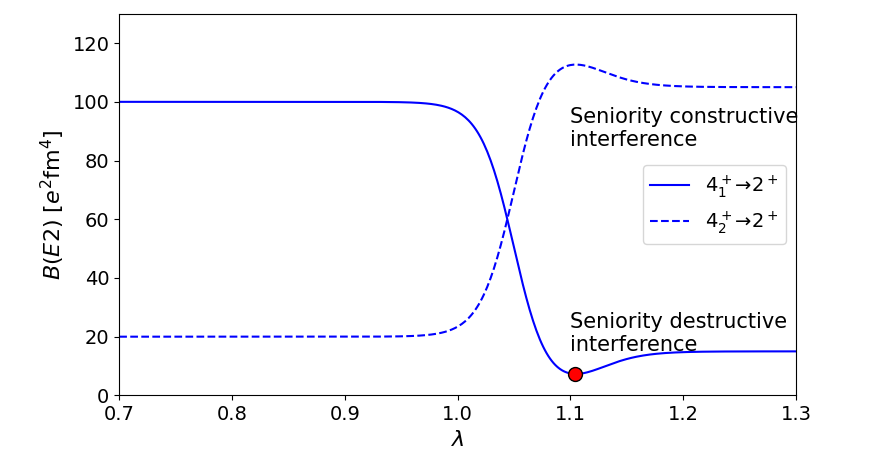}
\caption{
Schematic illustration of the behavior of $B(E2; 4^+ \rightarrow 2^+)$ as a function of the control parameter $\lambda$, which governs the mixing between the $v=2$ state and the partially seniority-conserved $v=4$ state. These two states would not mix in the absence of cross-shell non-diagonal TBMEs in the effective interaction. Their mixing, however, can be highly sensitive to the interaction strength, leading to dramatic changes over a narrow range of $\lambda$ values and to mixed-seniority states with distinct constructive or destructive interference. An anomalously suppressed $B(E2)$ value, highlighted by the red point, is predicted beyond the pure seniority-coupling limit—and is indeed observed experimentally in Both $^{96}$Pd~\cite{PhysRevC.95.014313} and $^{94}$Ru~\cite{das2022nature}.}
\label{ru94illustration}
\end{figure}

However, as explained in Ref.~\cite{qi2017partial} and also Refs.~\cite{das2022nature,PhysRevC.110.034320},
the two states with different seniority can still mix,  induced by cross-orbital non-diagonal
TBMEs of the two-body interaction, in the form
$$
\begin{aligned}
& \left|4_{1,2}^{+}\right\rangle=\alpha_4\left|4_{v=2}^{+}\right\rangle\pm\beta_4\left|4_{v=4;\alpha_1}^{+}\right\rangle, \\
& \left|6_{1,2}^{+}\right\rangle=\alpha_6\left|6_{v=2}^{+}\right\rangle\pm\beta_6\left|6_{v=4;\alpha_1}^{+}\right\rangle,
\end{aligned}
$$
which leads to constructive or destructive interference that can dramatically affect the E2 transition properties.
That is illustrated in Fig.~\ref{ru94illustration}.
The \(6^+ \to 4^+\) transition is observed to be strongly hindered in Ref.~\cite{das2022nature}, which they attribute to destructive interference between close-lying seniority-\(v = 2\) and \(v = 4\) state components.

The main property under dispute is the \(B(E2; 4^+ \to 2^+)\) transition. Ref.~\cite{das2022nature} shows a \emph{dramatic enhancement} compared to both pure seniority-model predictions and standard large-space shell-model predictions in the \(fpg\) proton-hole space relative to the doubly magic \(^{100}\)Sn. This would indicate an constructive interference effect from seniority mixing. To date, no large-scale shell-model calculation has been able to consistently reproduce the observed hindrances and enhancements in $^{96}$Pd and $^{94}$Ru.
New spectroscopy and lifetime measurements in $^{96}$Pd, and $^{94}$Ru and refined large-scale shell-model calculations in the $(f_{5/2},p_{3/2},p_{1/2},g_{9/2})$ proton valence space above a $^{56}$Ni core are required to clarify the picture and to get a more consistent description of the seniority scheme.

In Ref.~\cite{YUAN2024139018}, the $N=50$ isotones ($^{92}$Mo, $^{94}$Ru, $^{96}$Pd, and $^{98}$Cd) are calculated within the so-called Valence-Space In-Medium Similarity Renormalization Group method. This approach is essentially the same as the configuration interaction shell model calculation, done in the $gds(0g, 1d, 2s)$ model space with standard effective charges, but they replaced phenomenological interactions with \textit{ ab initio}-derived realistic nucleon-nucleon interactions via renormalization group techniques. By comparing their calculations in various truncations with experimental data, it was concluded that both proton
and neutron core excitations across the $Z =N= 50$ shell are important in reproducing the data.

One should exercise great caution when drawing conclusions from a simple comparison between model calculations and a single or limited set of experimental data.
In the past decade, there has been a significant increase in the number of both theoretical papers marketed as ``\textit{ ab initio}" theories and experimental papers comparing their data with various empirical and \textit{ ab initio} model calculations.
There is no clear consensus within the nuclear physics community on the standard theoretical framework and accuracy of \textit{ ab initio} models, as the majority of the models developed so far show large theoretical errors and can be far from the exact solution of the Hamiltonian applied.
Without considering the accuracy, one would still expect an \textit{ ab initio} theory in quantum physics to be a computational approach that derives predictions directly from solving the many-body Schrödinger equation constructed without relying on empirical parameters. This is still not fully achievable in nuclear physics because the nucleon-nucleon interaction is too complex and relies on empirical constraints from nucleon scattering data.
In fact, many of the available interactions are determined by fitting to both nucleon-nucleon and nuclear properties, leading to large uncertainties and blurring the distinction between realistic and empirical studies.

\subsubsection{The odd-$A$ mid-shell nucleus $^{95}$Rh}\label{95Rh}
The electromagnetic transition properties for the low-lying states in semi-magic nucleus $^{95}$Rh ($Z=45$, $N=50$), lying at the proton mid-shell of the $\pi g_{9/2}$ orbital, was measured recently in Ref.
\cite{PhysRevResearch.6.L022038}. These are the most puzzling results so far, which do not seem to be easily explained by the shell model or the seniority coupling scheme.

\begin{figure}[htp]
\centering
\includegraphics[width=0.5\textwidth]{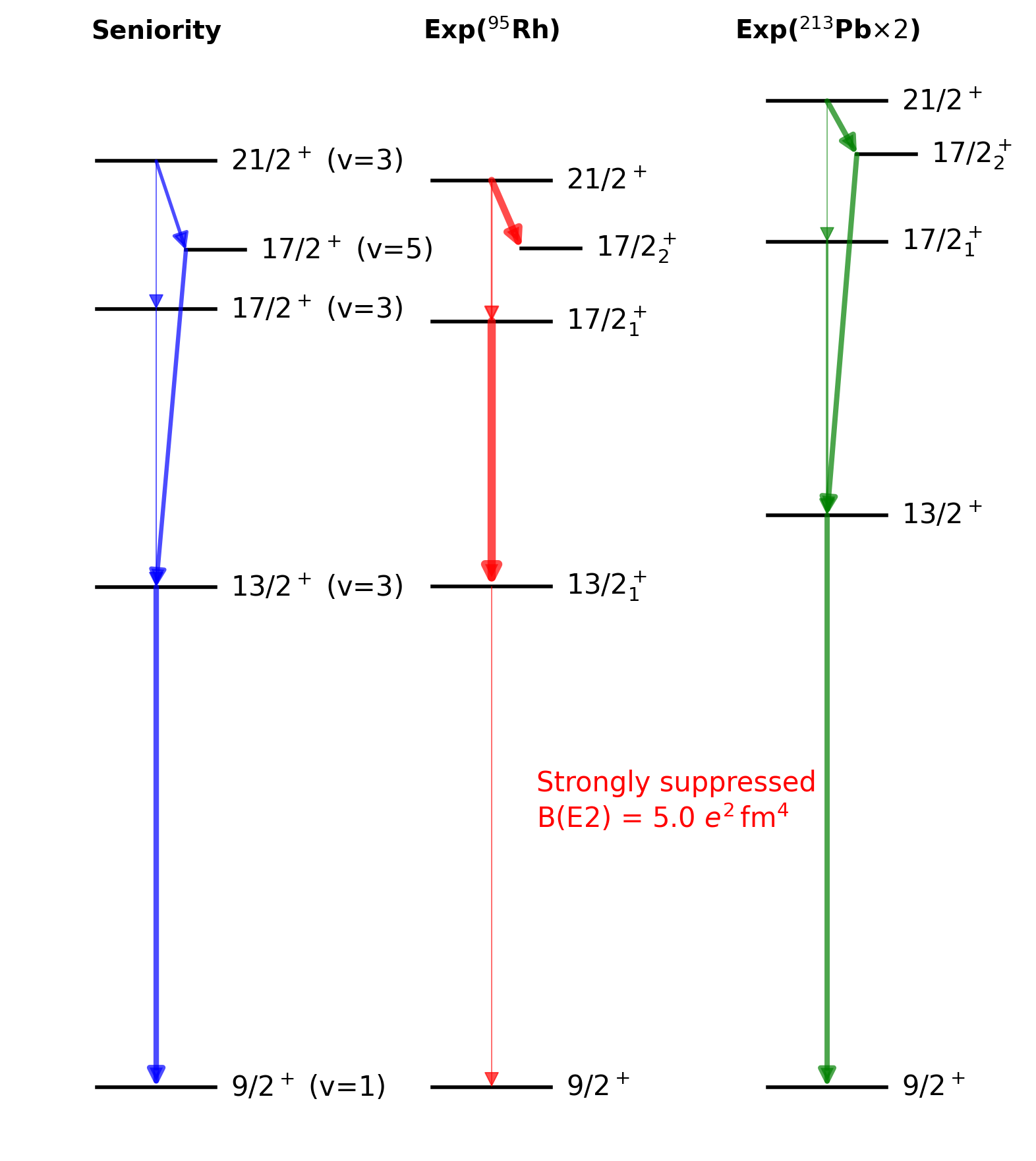}
\caption{
E2 transition properties of $^{95}$Rh compared with the typical seniority scheme and the experimental results for $^{213}$Pb, which can be considered a similar system with five neutrons in the $\nu 1g_{9/2}$ orbital. Data are taken from Refs.~\cite{PhysRevC.95.014313,PhysRevResearch.6.L022038,VALIENTEDOBON2021136183}. The level energies of $^{213}$Pb are magnified by a factor of two to facilitate comparison. 
\\
The strongly suppressed 
$B(E2; 13/2^+_1 \to 9/2^+_{\text{gs}})$ transition cannot be reproduced by any of the shell-model calculations in the various model spaces presented in Ref.~\cite{PhysRevResearch.6.L022038} or by any other structure models we have tested so far. The experiment also indicates strong $21/2^+ \to 17/2^+_{1,2}$ and $17/2^+_1 \to 13/2^+_1$ E2 transitions, which, if confirmed, suggest a highly mixed nature of the two observed $17/2^+$ states rather than the pure $v=3$ and $v=5$ seniority structures typically expected. 
\\
So far, most of our shell-model calculations with known interactions for the $17/2^+_{1,2}$ states do not support strong mixing between the two states, although the $v=5$ state can lie lower than the $v=3$ state (see also the calculations in Ref.~\cite{PhysRevC.95.014313}). The observed results are listed in Table~\ref{tab:rh95results}.
}
\label{rh95}
\end{figure}

\begin{figure}[htp]
\centering
\includegraphics[width=0.49\textwidth]{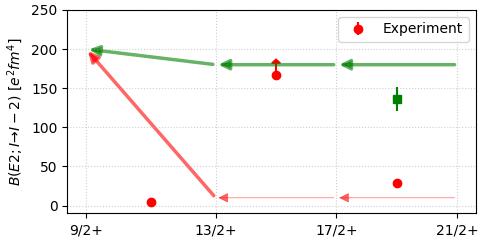}
\caption{A schematic illustration of the E2 cascades from the isomeric $21/2^+$ state in $^{95}$Rh, comparing the path through the $v=5$ $17/2^+$ state (green) and the path connecting the $v=3$ $17/2^+$ state (red) predicted by the seniority scheme and experimental data (circle and square symbols)~\cite{PhysRevResearch.6.L022038}. The arrow widths represent the relative $B(E2)$ strengths, not the transition rates.
}
\label{rh952}
\end{figure}

As for the $^{94}$Ru nucleus, the experiment reported in Ref.~\cite{das2022nature} was conducted at GSI Darmstadt using the DESPEC setup as part of the FAIR Phase-0 program. 
The $^{95}$Pd nucleus was produced via projectile fragmentation by a $^{124}$Xe beam at a kinetic energy of 850~MeV/u directed onto a $^{9}$Be target. The produced $^{95}$Pd nuclei subsequently 
underwent $\beta$ decay, populating excited states in $^{95}$Rh, including the isomeric 
$I^{\pi}=21/2^+$ state.
High-resolution $\gamma$ spectroscopy was performed using six 
triple-cluster Ge detectors, while the fast timing array, consisting of 
36 LaBr$_3$(Ce) detectors, enabled sub-100 ps resolution timing of $\gamma$ rays. 
The lifetimes of the nuclear levels were extracted using the Generalized Centroid 
Difference method, which compares $\gamma$--$\gamma$ coincidence time distributions 
and minimizes systematic uncertainties.

A clear $\beta$-delayed $\gamma$ cascade was observed:
\[
21/2^+ \xrightarrow[381~\text{keV}]{} 17/2^+_1 
\xrightarrow[716~\text{keV}]{} 13/2^+ 
\xrightarrow[1351~\text{keV}]{} 9/2^+_{\text{gs}} .
\]
The key measured lifetimes and reduced transition probabilities  $B(E2)$  are summarized in 
Table~\ref{tab:rh95results} and compared in Figs. \ref{rh95} \& \ref{rh952} with a typical seniority model and the observed patten in the heavy neutron analogue system $^{213}$Pb (five protons versus five neutrons in $ng_{9/2}$ orbital). 

\begin{table}[h!]
\caption{Experimental lifetimes and reduced transition probabilities $B(E2)$ for low-lying states in $^{95}$Rh, taken from Refs.~\cite{PhysRevResearch.6.L022038,PhysRevC.95.014313}.
}
\label{tab:rh95results}
\centering
\begin{tabular}{lcc}
\hline 
Transition & Lifetime $\tau$ [ps] & $B(E2)$ [$e^2$fm$^4$] \\\\ \hline
$21/2^+ \to 17/2^+_1$ & $3.0(4)\times 10^3$ & $29.0(4.0)$ \\\\
$21/2^+ \to 17/2^+_2$ & $3.0(4)\times 10^3$ & $136(20)$ \\\\
$17/2^+_1 \to 13/2^+_1$ & $\leq 26$ & $\geq 167$ \\\\
$13/2^+_1 \to 9/2^+_{\text{gs}}$ & $36(15)$ & $5.0^{+3.6}_{-1.6}$ \\\\ \hline 
\end{tabular}
\end{table}

The most puzzling observation for $^{95}$Rh is the strongly suppressed 
\[
B(E2; 13/2^+_1 \to 9/2^+_{\text{gs}}) = 5.0^{+3.6}_{-1.6}~e^2\text{fm}^4,
\]
which is more than a factor of 30 smaller than shell-model predictions 
($\sim 170$-$220~e^2$fm$^4$). That transition is not expected to be hindered by such a large factor in a typical shell model picture unless there are dramatic changes in nuclear structure between the two states. The observation contradicts the predictions within the seniority coupling scheme as well as all of the large-scale shell model calculations presented in that paper. We have tested various deformed mean-field and shell-model calculations, none of which predict such a feature.  Large-scale shell-model calculations
employing several commonly used effective interactions consistently
predict that the $9/2^+$ ground state is well separated from the lowest
$13/2^+$ state (dominated by $v=3$), while the two $17/2^+$ states are
usually found to lie close in excitation energy, leading to a stronger $13/2^+_1 \to 9/2^+_{\text{gs}}$ transition and a suppressed $17/2^+_1 \to 13/2^+_1$ transition.  We have also evaluated the influence of various non-diagonal TBME contributions by varying their strengths and have extended our calculations to include neutron cross-shell excitations involving the \(d_{5/2}\) and \(g_{7/2}\) configurations. No significant mixing or cross-shell excitation across the \(N=50\) shell closure has been found in our calculations so far without invoking unrealistic adjustments of the corresponding interaction TBMEs.
In general, one wouldn't expect a noticeable contribution from the cross-shell excitation to the low-lying states in $^{95}$Rh due to the large energy gap ($\sim 6$ MeV) between the $g_{9/2}$ subshell and higher-lying shells. It is also noteworthy that no evidence for significant cross-shell excitations was observed for the neutron analogue system, $^{213}$Pb, which instead shows a rather strong $B(E2; 13/2^+_1 \to 9/2^+_{\text{gs}})$. 
In that paper it was speculated that the observation provides direct evidence for broken seniority 
symmetry in $^{95}$Rh and refined theoretical models, 
potentially including three-body forces, could elucidate the mechanism behind this unexpected observation.

 In typical
calculations, the higher-lying $13/2^+_2$ state ($v=5$) appears at an
excitation energy of about $3.0$ MeV in the restricted $g$-shell
calculation and around $2.9$ MeV in the $fpg$ model space.  This state
is mainly predicted to decay into the $v=3$ $17/2^+$ state or into an
$11/2^+$ state.

Other notable features of the present experimental data include:
\begin{itemize}
\item A strong $21/2^+ \to 17/2^+_2$ transition which seems to agree well with the seniority scheme assuming $v=5$ assignment for the latter state as usually expected;
\item A relatively strong $21/2^+ \to 17/2^+_1$ transition, which is expected to vanish in the seniority scheme as $^{95}$Rh sits at the $0g_{9/2}$ mid-shell.
\item A possible strong $17/2^+_1 \to 13/2^+_1$ transition for which only the lower limit is measured in that experiment. The transition between the two states would be hindered if they have the same seniority. Shell-model calculations across different valence spaces predict approximate seniority conservation, contradicting the strong transition observed.
\end{itemize}

The wave function of a nuclear state of $n=5$ mid-shell system with angular
momentum $I$ can be expressed as a linear superposition of basis states
carrying different seniority values:
\begin{equation}
  |I,\alpha\rangle = \sum_{v=1,3,5} c_v \,
  \big|(0g_{9/2}^5,v,I)\big\rangle \;+\;
  \text{higher order terms}.
\end{equation}
As mentioned in Sec.~\ref{208Pb}, only the $I=9/2$ $v=1,5$ states can mix in the pure single-$j$ configurations. The inclusion of other neighboring orbitals may serve as mediators for the further mixture among different states.
For the $I^\pi = 9/2^+$ sequence, the three possible seniority
configurations may interact with one another with the strongest mixing allowed still being that between the $v=1$ and $v=5$ states through $\Delta v=4$
matrix elements, and with additional coupling to the intermediate $v=3$
configuration, primarily via the nearby $p_{1/2}$ orbital.  Despite this,
shell-model calculations show that the admixture of the $v=3$ and
$v=5$ components into the $v=1$ ground state is very small, on the order of
1--2\%.  This weak mixing results from the strong pairing interaction,
which elevates the higher-seniority $9/2^+$ states to excitation
energies near $1.8$ and $2.3$ MeV in large-scale shell model calculation.  The same trend can be observed in most typical
shell model effective interactions and model spaces investigated.  A
similar conclusion holds for the two $13/2^+$ states, where the predicted
mixing between the $v=3$ and $v=5$ components remains extremely
small.

\subsection{Neutron-rich $N=82$ isotones below $^{132}$Sn}
\begin{figure}[ht]
\centering
\includegraphics[width=0.5\textwidth]{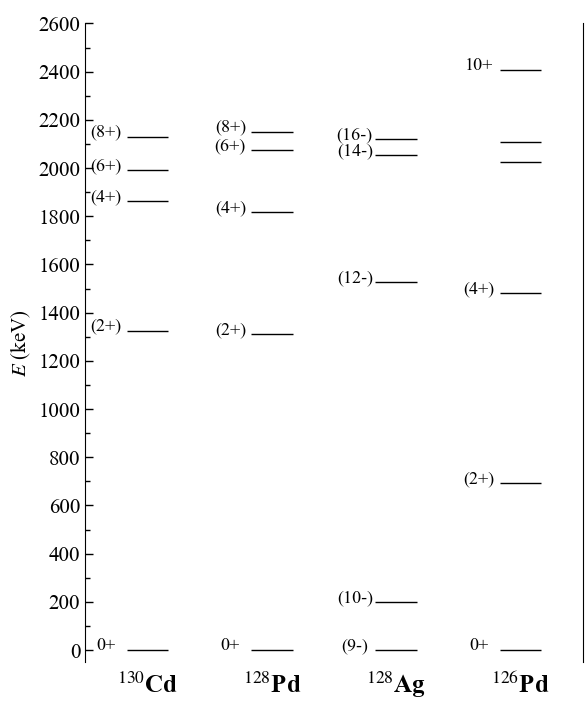}
\caption{Known spectra of even-even $N=82$ isotones $^{130}$Cd and  $^{128}$Pd in comparison with neighboring  $^{126}$Pd and  $^{128}$Ag. Data are taken from Refs.~\cite{PhysRevLett.111.152501,PhysRevLett.113.042502,wq9m-trj8} and NNDC database. 
The $8^+$ isomeric state in $^{130}$Cd has half-life 220 (30) ns while that in $^{128}$Pd has 5.8 (8) $\mu$s~\cite{PhysRevLett.111.152501}. In $^{126}$Pd there is a $10^+$ isomer with half-life 23.0 (10)~ms. The assignment of the $16^{-}$ isomer in $^{128}$Ag is tentative which has half-life 1.60(7)~$\mu$s.}
\label{Cd-isomer-2}
\end{figure}

Along the very long \(N=82\) isotonic chain, it may be possible in the near future to obtain more measurements of the spectroscopy and atomic masses of neutron-rich nuclei below \(Z=50\) with protons in the \(\pi 0g_{9/2}\) orbital. This is important not only for understanding nuclear structure~\cite{Watanabe_2019,Watanabe_2020,YUAN2016237,Watanabe_2024} but also for constraining the r-process~\cite{storbacka2024location,ctyj-ls15}, in particular the so-called weak r-process, which may affect the abundance of the nuclei with $Z \sim 40$.
The nucleus $^{131}$In has $0g_{9/2}$ hole state as its ground state with the $s_{1/2}$ hole state  being 302 keV above. The robustness of the $N=82$ shell closure is supported by magnetic dipole moment of that nucleus~\cite{Vernon_2022}. There is no spectroscopic information on  $^{129}$Ag, $^{127}$Rh or $^{126}$Ru (all of which may still be expected to be bound). The spectra of $^{130}$Cd and $^{128}$Pd are, however, already known~\cite{PhysRevLett.111.152501}.
Those spectra are plotted in Fig.~\ref{Cd-isomer-2} in comparison  with those of  neighboring nuclei.
The spectrum of the $N=81$ $^{128}$Ag nucleus has been measured recently at RIBF~\cite{wq9m-trj8}. They discovered a new seniority isomer with a spin-parity of $16^{-}$ and half-life 1.60(7) $\mu$s.
The level structure observed in these nuclei closely resembles that of $^{130}$Cd and $^{128}$Pd, suggesting that the strong $N=82$ shell closure and seniority coupling remain intact.

The most recent data for $^{128,126}$Pd shown in the figure were also from an experiment at RIBF combining EURICA with the BigRIPS separator and the WAS3ABi (DSSSD silicon detector array)~\cite{PhysRevLett.111.152501,PhysRevLett.113.042502}.
In Ref.~\cite{PhysRevLett.111.152501}, the half-life of the  $8^+$ isomeric state $^{128}$Pd was determined. The paper highlighted that the extracted electric quadrupole transition  strength $B(E2)$
  is more hindered than the corresponding transition in  
$^{130}$Cd, as can be seen in Fig.~4 of that paper, which, if true,
would agree nicely with the seniority scheme as $^{128}$Pd is expected to be in the mid-shell of the \(\pi 0g_{9/2}\) orbital.

\subsection{The Sn isotopes and $N=82$ isotones}
\begin{figure}[ht]
\centering
\includegraphics[width=0.4\textwidth]{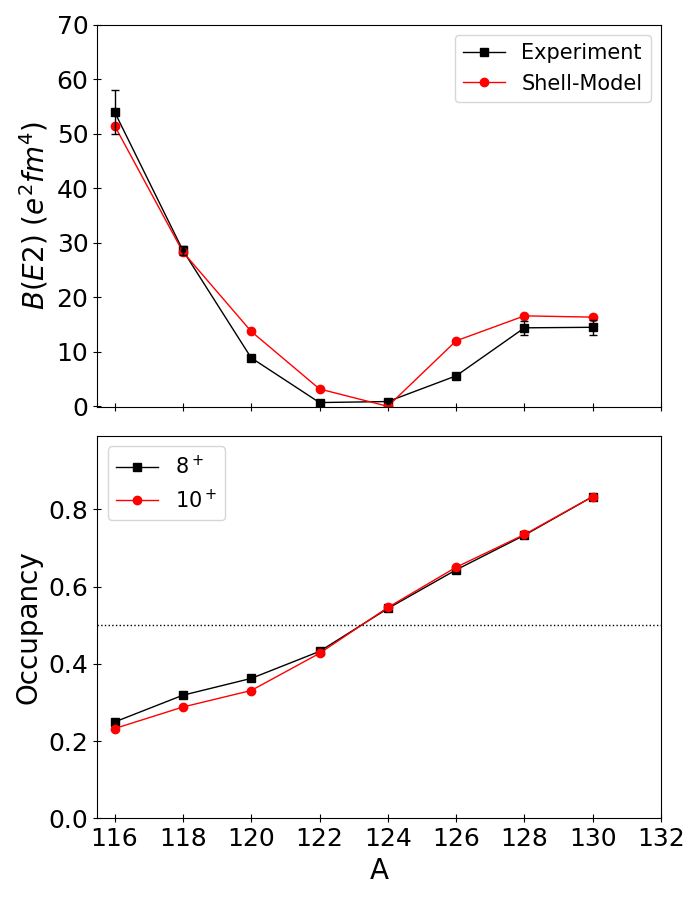}
\caption{
Upper: $B\left(E2; 10^{+} \rightarrow 8^{+}\right)$ transition strengths for even-even Sn isotopes compared with predictions from large-scale shell-model calculations~\cite{qi2012monopole}. 
\\
Lower: Calculated neutron occupation numbers for the $0h_{11/2}$ orbital in the $10^+$ and $8^+$ states. The model does not explicitly conserve seniority symmetry, and the $0h_{11/2}$ orbital is strongly mixed with neighboring $s$ and $d$ orbitals. Nevertheless, the $B\left(E2; 10^{+} \rightarrow 8^{+}\right)$ values still follow the seniority scheme, tending to vanish near mid-shell or when the $0h_{11/2}$ orbital is half-filled.}
\label{Sn-isomer-2}
\end{figure}

This region may not have a direct impact on our study of the $j=9/2$ partially conserved isomerism. However, we would like to highlight one particular feature that could be highly relevant for future studies: the emergence of 10\({}^{+}\) isomers in Sn isotopes and in the semi-magic \(N=80\) and 82 isotones, even though the proton and neutron \(h_{11/2}\) orbitals are not fully isolated~\cite{PhysRevC.99.064302}. The wave functions derived from large-scale shell model calculations show quite mixture effects between $h_{11/2}$ and neighboring orbitals~\cite{qi2012monopole,Back2013,Back2011,Cederlof2023-rj}. But the calculated properties for the $10^+$ isomers still follow simple behavior as predicted by the seniority model, as illustrated in Fig.~\ref{Sn-isomer-2}, indicating that other neighboring orbitals are playing more in the background like in the BCS pairing vacuum.
This effect has been studied by Blomqvist and collaborators since long ago~\cite{PhysRevLett.68.1671} and it appears in several regions. We think this is one of the most fascinating facts of the nuclear shell model as a ``model" which reveals simple and beautiful effects out of complex correlations.

On the proton-rich side, the low-lying negative-parity states in \(N=82\) isotones like \({ }^{149} \mathrm{Ho},{ }^{151} {\mathrm{Tm}}\) and \({ }^{153} \mathrm{Lu}\) can be explained as the proton coupling $(\pi h_{11 / 2}^{n})$~\cite{Wilson1980,McNeill1990,PhysRevLett.77.3743,PhysRevC.59.546}.
In particular, the seniority $v=3$ \(27 / 2^{-}\) isomers of \(\pi h_{11 / 2}^{n}\) character occurs at \(\sim 2.7~\mathrm{MeV}\). As $Z$ and the \(\pi h_{11 / 2}\) occupation number increase, the observed \(\mathrm{B}\left(\mathrm{E} 2,27 / 2^{-} \rightarrow 23 / 2^{-}\right.\)) values decrease sharply, reaching
a value of only \(0.45~\mathrm{e}^{2} \mathrm{fm}^{4}\) (or \(0.01 \mathrm{~W.u.}\)) in \(^{153}\mathrm{Lu}\), where the
\(\pi h_{11 / 2}\) subshell is approximately half-filled (as expected from the seniority symmetry). Yrast $0h_{11/2}^n$ states of the $N=82$ \(^{150} \mathrm{Er}\) ($\pi0h_{11/2}^4$) and \({ }^{151} \mathrm{Tm}\) ($\pi0h_{11/2}^5$) nuclei are measured in Ref.~\cite{HELPPI198211}. Results for $^{153}\mathrm{Lu}$ and $^{154}\mathrm{Hf}$ were given in Ref.~\cite{PhysRevLett.63.860}.
More information on the isomeric states in the Sn isotopes and $N=82$ isotones can be found in Ref.~\cite{sym14122680}.
Going away from the $N=82$ shell closure, the $N=83$ isotones show a nearly identical behavior as the above $N=82$ chain~\cite{McNeill1990,Broda1979,carroll2016multiparticle,wang2017reinvestigation,wang2017spectroscopic}, which may be true for $N=84$ and 80 istones as well. The $10^+$ isomer in $^{150}$Yb has been identified recently~\cite{Zhang2025}, where one notices a very long chain of isomers along the $N=80$ isotones (see, for example, Fig.~55 in Ref.~\cite{physics4030048}).

There have been extensive activities in studying the neutron-rich isotopes around $^{132}$Sn at RIBF~\cite{PhysRevC.94.051301,Wang2018-ou} as well as other labs including ISOLDE~\cite{Hoff1996-pw}, HIE-ISOLDE (see, Fig.~2 in Ref.~\cite{Nowacki2021-dh}) and Holifield RIBF at Oak Ridge National Lab~\cite{Jones2010-qu}. The ground states of neutron-rich $^{133-138}$Sn isotopes are expected to be dominated by the$1f_{7/2}$ orbital with possible strong mixing effect. The neutron $0h_{9/2}$ orbital is measured to be at 1560.9~keV above the ground state~\cite{Hoff1996-pw,Jones2010-qu}. So far there is no information on states built on top of the $0h_{9/2}$ orbital. In any case, the low-lying states in even-even Sn isotopes show quite some similarity. There are $6^+$ isomeric states identified in isotopes up to $^{138}$Sn already more than a decade ago~\cite{PhysRevLett.113.132502}. In that experiment, delayed $\gamma$-ray cascades from isomeric $6^{+}$ states in the very neutron-rich isotopes $^{136,138}$Sn was produced via projectile fission of $^{238}$U at RIBF, RIKEN. Using the BigRIPS separator, WAS3ABi silicon array, and the EURICA Ge detector system. The half-lives of $T_{1/2} = 46(7)$~ns in $^{136}$Sn and $T_{1/2} = 210(45)$~ns in $^{138}$Sn were measured. Shell-model studies on those states can be found in Refs.~\cite{PhysRevC.65.051306,PhysRevC.76.024313,PhysRevC.91.024321,YUAN2016237} which employ model spaces including both $1f_{7/2}$ and $0h_{9/2}$ as well as other neighboring orbitals.  the measured $B(E2;6^{+}\rightarrow4^{+})$ value in $^{136}$Sn, which was supposed to vanish at mid-shell with four neutrons in $1f_{7/2},$ show a relatively large strength, deviating strongly from pure seniority expectations. Shell-model calculations with realistic $V_{\text{low-}k}$ interactions reproduce level energies well but fail to account for this transition rate. It was suggested in that paper through various theoretical calculations that the $4^+$ may be a mixture between $v=2$ and 4 configurations. On the other hand, $B(E2;6^{+}\rightarrow4^{+})$ value could remain large if, as in the case of lighter Sn isotopes below $N=82$ show in Fig.~\ref{Sn-isomer-2}, the $^{136}$Sn nucleus is not located at the mid-shell due to mixing of the wave function with other orbitals.
For heavier isotopes with $Z>50$, shell-model calculations in Ref.~\cite{PhysRevC.96.034312} suggest that the wave function can be very mixed where deformation may start to play a role already at $N=86$.

\subsection{The $B_{4/2}$ anomaly and its connection to seniority symmetry}

It has long been established in nuclear-structure textbooks that one expects $B_{4/2} \ll 1$ for nuclei governed by the seniority scheme, and $B_{4/2} > 1$ for open-shell collective systems. 
However, experimental studies have revealed striking exceptions to this conventional picture. 
In $^{114}$Te~\cite{moller2005} and nearby isotopes, as well as in heavier nuclei such as $^{166}$W~\cite{saygi2017}, $^{168}$Os~\cite{grahn2016}, $^{170}$Os~\cite{goasduff2019}, and $^{172}$Pt~\cite{cederwall2018}, the observed $B_{4/2}$ values are significantly smaller than one, despite the evidently collective (rotational or vibrational) nature of their level structures. 
This unexpected behavior, commonly referred to as the \textit{$B_{4/2}$ anomaly}, remains unexplained by current microscopic approaches, including large-scale shell-model and beyond-mean-field calculations.

It has been speculated~\cite{cederwall2018} that the anomaly might be linked to a partial restoration of seniority symmetry in these open-shell nuclei. 
Nevertheless, the measured absolute $B(E2)$ values remain much larger than those typical of non-collective, seniority-dominated systems. 
Within the framework of the algebraic interacting boson model (IBM), the anomaly can be reproduced by introducing higher-order Hamiltonian terms, which generate triaxial deformation and configuration mixing~\cite{zhang2022,zhang2024,pan2024}. 
This interpretation appears to be supported by shell-model calculations employing ensembles of random interactions~\cite{Fu2025}.

At present, it seems likely that the $B_{4/2}$ anomaly reflects a distinct type of underlying physics—one not directly attributable to the breaking or restoration of seniority symmetry. 
Further theoretical and experimental investigation is needed to clarify its microscopic origin and its relation, if any, to seniority symmetry.

\section{Comparison of seniority and spin-aligned neutron-proton coupling schemes} \label{np}
Because of the large overlap between their wave functions, a strong neutron-proton (np) pair correlation has long been expected. The np interaction can strongly break the seniority symmetry, causing wave functions to contain contributions from multiple seniority components. A systematic classification of such states within the 
$jj$-coupling scheme is still not well established. The so-called stretch scheme—in which intrinsic angular momenta are maximally aligned—was introduced in the 1960s to describe rotational-like spectra in open-shell nuclei. The
np quasi-spin formalism was developed and applied in Refs.~\cite{talmi1993}. Unlike the $T=1$ neutron-neutron and proton-proton interactions which are dominated by the $J=0$ pairing interaction, the np interaction shows a much more complex structure.
 From a theoretical perspective, the np pair can couple to $J = 0$ and $T = 1$ (isovector) in exactly the same way as like-particle pairing. On the other hand, in the $T = 0$ (isoscalar) channel they can couple to different spin values from  $J=1$ to $J = J_{\rm max}=2j$, which may affect various nuclear properties. In particular it is well understood now that the $T=0$ np interaction contains a large attractive monopole channel which can affect significantly the evolution of single-particle energies as well as a large quadrupole-quadrupole correlation which can lead to large deformation in the wave function.

From a general perspective, a long-standing open question in nuclear physics is the existence of isovector and isoscalar neutron-proton pairing in atomic nuclei with $N \sim Z$ where protons and neutrons are filling in the same orbitals (see, for examples, Refs.
\cite{FRAUENDORF201424,qi2015n,le2022neutron,pan2021np,pan2020importance,liu2021evidence,VAN_ISACKER_2013} and references therein). The $J=1$ np pairing has in particular been  extensively studied within various mean field and BCS-like approaches. 
The onset of isoscalar spin-aligned np coupling scheme  is envisaged in the low-lying spectra of $N=Z$ nuclei below $^{100}$Sn~\cite{cederwall2011evidence,qi2011spin,qi2015n,cederwall2020isospin}.  When one looks into the complex wave functions of $N=Z$ nuclei like $^{92}$Pd and $^{96}$Cd (as well as odd-odd $^{94}$Ag), it is seen that indeed the low-lying states can be overwhelmingly dominated by the angular momentum coupling of $J=9$ np pairs~\cite{qi2011spin} or even quartet-like coupling in the system of four np pairs (see, table 1 in Ref.~\cite{qi2012spina}). 
The spin-aligned coupling scheme has been studied in by different groups in various approaches~\cite{xu2012multistep,WOS:000301612400004,qi2012coherence,qi2012competition,qi2012spin,PhysRevC.83.064314,PhysRevC.92.024320,PhysRevC.85.034335,PhysRevC.87.044312,PhysRevC.91.064318,PhysRevC.90.014318,PhysRevC.89.014316}. Although there is general agreement on the dominant role of the 
$0_{9/2}$
orbital in these nuclei, a frequent point of debate concerns the “purity” of the wave function and the validity of describing the system within a single-$j$ framework. As discussed earlier, addressing such questions ultimately requires deeper insights into the shell model itself, which is inherently a local theory restricted to a specific model space.

The standard shell-model approach builds its wave functions in many-particle representation which offers a straightforward and simple way of constructing the basis but the resulting wave function can be rather complex. The seniority and spin-aligned np coupling schemes offer alternative ways to express the eigen function in relatively simply manners.

\subsection{Seniority coupling scheme revisited}
\begin{figure}[ht]
	\centering
	\includegraphics[width=0.4\textwidth]{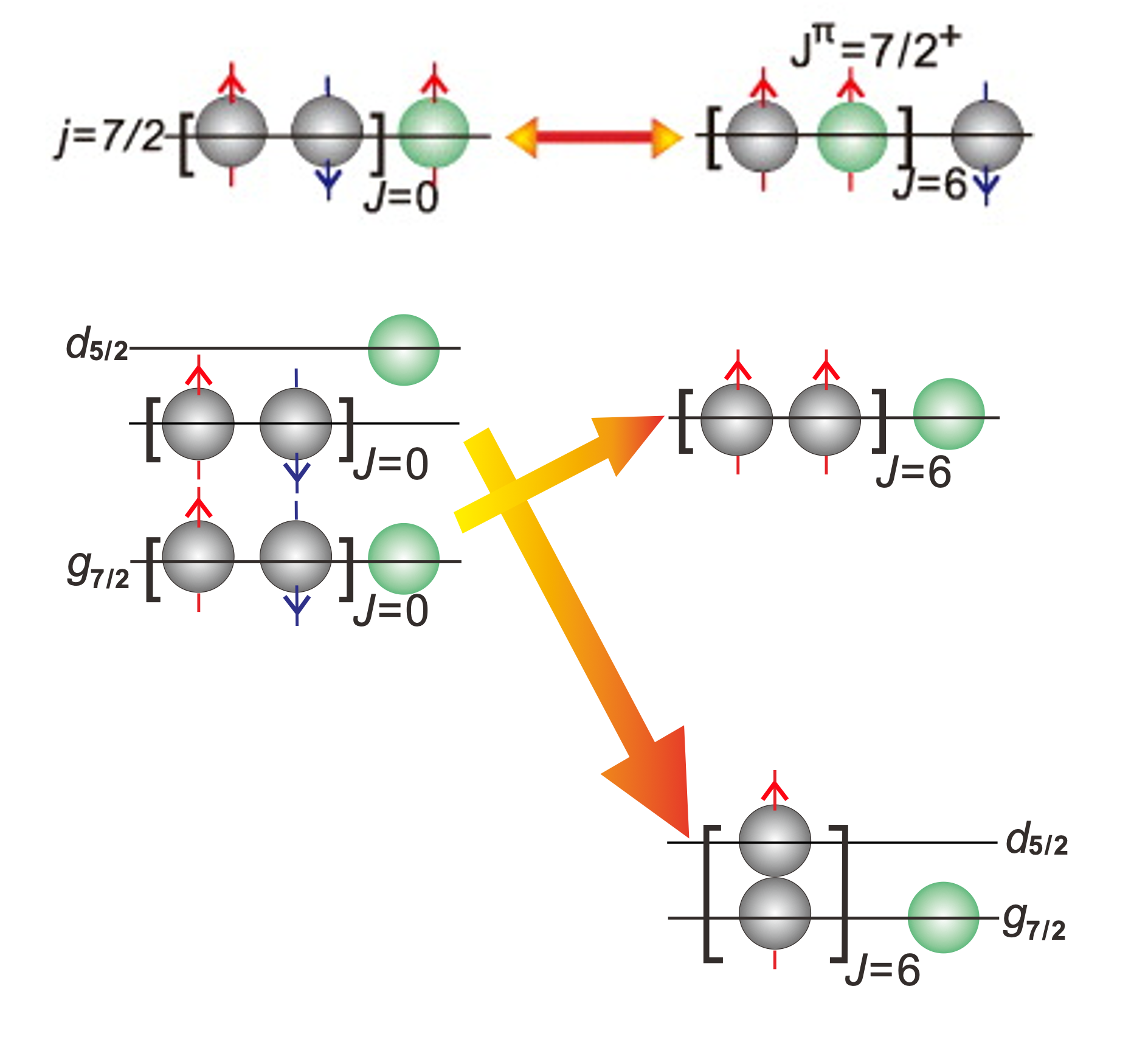}
	\caption{Upper: Illustration of the coupling of three identical particles in a single $j=7/2$ orbital. The $I=7/2$, $v=1$ state is usually described as a seniority-zero pair coupled to an odd particle. However, for $j=7/2$ or systems with lower spin $j$ values, the same state can be constructed by coupling a pair with any $J$ to the odd particle. All particles are identical, and different colors are used solely to highlight the ``unpaired'' odd particle. 
\\
Lower: Coupling of three particles in the presence of two nearly degenerate orbitals, $1d_{5/2}$ and $0g_{7/2}$. The $(0g_{7/2})^2$ $J=0$ pair can couple to either $1d_{5/2}$ or $0g_{7/2}$, giving the conventional $v=1$ configurations with spins $I=5/2$ and $7/2$. Additionally, an $I=5/2$ state can be formed by coupling an aligned $(1d_{5/2}0g_{7/2})_{J=6}$ pair, which is strongly attractive and can invert the order of the two states. This behavior was observed in $^{101,103}$Sn via $\alpha$-decay studies in Ref.~\cite{Darby_2010}.}
	\label{scheme-3}
\end{figure}

One can state that the seniority coupling is just one of the many flexible pair coupling schemes as described above.  
Here we would like to illustrate briefly a striking but unknown feature of the seniority coupling about its relation to pairs with non-zero spins:
\begin{itemize}
	\item A spin-zero ($J=0$) seniority pair is completely different from a spin $J\neq $0 pair.
	\item The coupling of a $J=0$ pair to another sub-system (particle or pairs) can, however, be equivalent to the coupling of a $J\neq 0$ pair to the other sub-system.
	\end{itemize}
Their similarity can be evaluated based on the overlap matrix we defined above.
A typical example would be the coupling of two spin-aligned np pairs with the coupling of two seniority-zero nn and pp pairs can have quite large overlap. The same maybe true for systems with identical particle. For example, the coupling of two $J=0$ pairs (resulting in obviously a seniority zero four particle system) can be exactly the same as the coupling of two aligned $J=2j-1$ pairs (or pairs with any spin values) resulting in the same spin. 	
In Fig.~\ref{scheme-3} we give another example on the coupling of three identical particles in $j=7/2$ where the coupling of a seniority zero pair to the odd-particle and the coupling of a fully aligned pair to the odd-particle give exactly the same state. If we can knock out a pair from the system, the pair can carry spin value that is nonzero. Another manifestation of the effect is that the three-particle system would have energy smaller than that of the sum of the T=0 pair plus the single-particle energy as the TBMEs with both spins contribute to the total energy of the state. It could happen under such condition that that seniority configuration may not be the ground state~\cite{qi2012monopole}. That indeed happens in the case of Sn isotopes, due to the presence of a very low-lying $d_{5/2}$ orbital, there is a spin-flip between the nuclei $^{101,103}$Sn~\cite{Darby_2010} induced by the repulsive $(0g_{7/2})^2_{J=6}$ interaction and the presence of attractive $(0g_{7/2}1d_{5/2})^2_{J=6}$ interaction.
	
	This feature exemplifies the well-known notion that``more is different". Even a simple system of three particles can reveal very rich physics not seen in the sub system of two particles.
	This feature also offers an alternate way to study the pair structure and seniority coupling of atomic nuclei.
	
\section{Summary}\label{summary}
Seniority remains one of the most powerful organizing principles in nuclear structure physics, offering analytic insight into spectra, transition rates, and isomeric behavior. At its core, the seniority quantum number counts the number of nucleons not coupled in $J=0$ pairs, thus quantifying the degree of pair breaking in a many-body system. States with $v=0$ represent ``condensates" of $J=0$ paired nucleons, while higher-seniority states correspond to broken pairs with $J\neq 0$, giving seniority a clear and physically intuitive interpretation. This definition is consistent across quantum systems, including atomic, molecular, and condensed matter contexts, highlighting its broad applicability.

In nuclear physics, seniority conservation is exact for systems of identical fermions confined to a single-$j$ orbital when $j \leq 7/2$, independent of the two-body interaction. For larger $j$ values, however, such as $j=9/2$, conservation requires the interaction to satisfy linear constraints among its TBMEs. Since realistic nuclear interactions do not generally satisfy this relation, most eigenstates in these shells are admixtures of different seniorities.

Despite this, an extraordinary phenomenon known as partial conservation of seniority has been discovered. Specifically, for four identical fermions in the $j=9/2$ shell, certain $I=4$ and $I=6$ states with seniority $v=4$ remain exact eigenstates of any two-body interaction, even when other states mix. This property arises from the algebraic structure of the CFPs, which enforce vanishing matrix elements between these special states and other configurations. As a result, these states embody a form of partial dynamical symmetry, where a subset of the spectrum retains exact symmetry even though it is broken elsewhere.
Theoretical proofs and symbolic shell-model studies confirm the robustness of this property.
From a broader perspective, the persistence of seniority in these special states provides a concrete example of partial dynamical symmetry in finite quantum systems. Unlike exact symmetries, which apply globally, partial symmetries act selectively, preserving solvability for specific eigenstates. This framework deepens the connection between algebraic models and realistic nuclear interactions, while also offering predictive power for nuclear spectroscopy.

The existence of partial seniority states and in general the approximate conservation of seniority rely on observed patterns of excitation energies, electromagnetic transition strengths, and isomeric lifetimes. We list a few regions of semi-magic nuclei and summarize the experimental progress in Sec. \ref{expt-p} which provide strong support for the seniority scheme and reveal signatures of its partial conservation in $j=9/2$ orbitals. In neutron-rich Ni isotopes, lifetimes and $B(E2)$ values up to the $8^+$ state have been measured. In $^{72}$Ni and $^{74}$Ni, the absence of long-lived $8^+$ isomers is linked to the inversion of $v=2$ and $v=4$ partial seniority conserved $6^+$ states, consistent with partial dynamical symmetry in the $\nu g_{9/2}$ shell. Among the $N=50$ isotones, $^{94}$Ru shows a classic $8^+$ seniority isomer with a hindered $E2$ decay. New Coulomb excitation and lifetime measurements reveal some strongly hindered E2 transitions in these systems which highlight strong destructive interference between $v=2$ and $v=4$ components and provide solid evidence of the existence of the partial seniority conserved states, even though their direction observation is missing. There are contradictory measurements on the enhanced E2 transition in $^{94}$Ru which we hope will be clarified in the near future.
In addition, odd-$A$ systems provide further tests: in $^{95}$Rh, precision lifetime measurements revealed a strongly suppressed $B(E2; 13/2^+ \to 9/2^+)$ transition—over 30 times smaller than shell-model predictions—pointing to a puzzle beyond the shell model interpretation.

Overall, spectroscopy and lifetime measurements confirm the persistence of seniority and its partial conservation in multiple nuclear regions. The observed anomalies—transition hindrance, isomerism, and deviations from shell-model predictions motivate continued high-precision studies at next-generation radioactive beam facilities.
Looking ahead, systematic exploration of higher-$j$ orbitals and multi-$j$ configurations will be crucial to determine whether partial conservation of seniority is unique to the $j=9/2$ case or a more general (exact or approximate) feature of quantum many-body systems. Experimental progress in measuring electromagnetic transition strengths, isomer lifetimes, and higher-seniority excitations will provide decisive tests of these theoretical predictions. The interplay between seniority-conserving configurations and competing modes, such as spin-aligned proton–neutron pairing in $N=Z$ nuclei, remains an especially promising frontier. Together, these studies will clarify the scope of seniority and np coupling schemes as both a practical organizing principle of complex nuclei.

\begin{acknowledgements}
Part of the manuscript was written during CQ’s visit at FRIB, Michigan State
University. He thanks KTH for the financial support of his sabbatical leave and
MSU theory department for their hospitality as well as support from the Olle Engkvist Foundation. The computations were enabled by resources provided by the National Academic Infrastructure for Supercomputing in Sweden (NAISS) and the Swedish National Infrastructure for Computing (SNIC) at PDC, KTH. He also thanks colleagues at KTH, including Profs. R. Liotta, R. Wyss, B. Cederwall and T. B\"ack for collaborations on the subject.

\end{acknowledgements}

\appendix

\section{General nuclear pair coupling}
One can generalize the seniority coupling to systems with pairs with non-zero spin values~\cite{ZHAO20141} as a way to build full or truncated shell-model basis in $jj$ coupled basis.
In that way one can even solve the shell-model
equation iteratively in several steps. One starts with the single-particle and two-particle states as in standard shell model. In each following step one calculates the state of a nucleus in terms
of previously calculated states of nuclei with less particles~\cite{liotta1982a,PhysRevC.41.1831,xu2012multistep}. 
The three-particle system can  be partitioned into one-
and two-particle correlated states. Thus the basis has the form
\(\left\{\left(a_{j}^{\dagger} P^{\dagger}\left(\alpha_{2}\right)\right)_{\alpha_{3}}|0\rangle\right\}\)
where we used Greek letters to label many-particle states and the number
of particles appear as a subscript. We assume in that notation that the wave function is in angular momentum coupled form and therefore the $m$ quantum number is not specified in the particle creation operator $a_{j}^{\dagger}$. The operator \(P^{\dagger}\left(\alpha_{2}\right)\) is the creation operator of the paired state.
The three-particle wave function can be written by
\begin{equation}
	\left|\alpha_{3}\right\rangle=\sum_{j,\alpha_{2}} X\left(j \alpha_{2} ; \alpha_{3}\right)\left(a_{j}^{\dagger} P^{\dagger}\left(\alpha_{2}\right)\right)_{\alpha_{3}}|0\rangle.
	\end{equation}
Unfortunately such basis is in general over-complete. But the
projection
\(F\left(\alpha_{m} \alpha_{n} ; \alpha_{s}\right)=\left\langle\alpha_{s}\left|\left(P^{\dagger}\left(\alpha_{m}\right) P^{\dagger}\left(\alpha_{n}\right)\right)_{\alpha_{s}}\right| 0\right\rangle\) is well defined.

\begin{figure}[ht]
	\centering
	\includegraphics[width=0.35\textwidth]{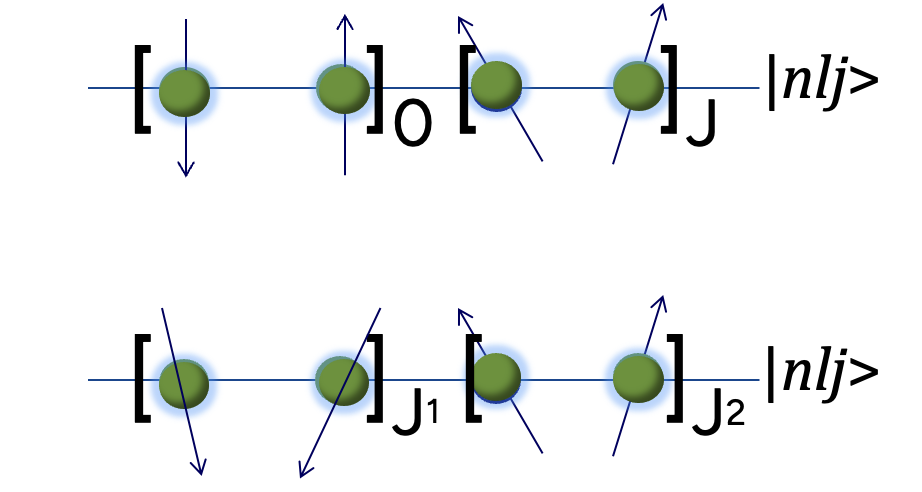}
	\caption{Illustration of the coupling of two identical pairs into a 4-particle system in a single-$j$ shell. A $v=2$ state with total spin $I$ can be constructed by coupling a $J=0$ pair with a $J=I$ pair (upper), while a $v=4$ state arises from coupling two $J\neq 0$ pairs such that $\hat{I} = \hat{J}_1 + \hat{J}_2$.
}
	\label{pair}
\end{figure}

The four-particle system can  be partitioned into the coupling of two pairs
\begin{eqnarray}\label{4pp}
	\left\{\left(P^{\dagger}\left(\alpha_{2}\right) P^{\dagger}\left(\beta_{2}\right)\right)_{\alpha_{4}}|0\rangle\right\}.
\end{eqnarray}
The wave function can be expanded as
\begin{eqnarray}
\left|\alpha_{4}\right\rangle=\frac{1}{2} \sum_{\alpha_{2},\beta_{2}} Y\left(\alpha_{2} \beta_{2} ; \alpha_{4}\right)\left(P^{+}\left(\alpha_{2}\right) P^{+}\left(\beta_{2}\right)\right)_{\alpha_{4}}|0\rangle
\end{eqnarray}
where
$$Y\left(\alpha_{2} \beta_{2} ; \alpha_{4}\right)=\left(1+\delta_{\alpha_{2} \beta_{2}}\right) X\left(\alpha_{2} \beta_{2} ; \alpha_{4}\right).$$
An illustration on the coupling of two identical pairs is given in Fig.~\ref{pair}.
Systems with higher pair numbers can be constructed in the same way step by step though it can become awkward in practice to construct bases with more than four pairs. Instead, one can express the wave function of a system with $n$ particles as the coupling of a $n-2$ subsystem and an extra pair. 
The same technique can be applied to the coupling of neutron and proton pairs, as will be shown in Sec. \ref{np}.

\subsection{One-body and two-body CFPs}
Another common way to construct the many-particle wave functions in a single-$j$ shell is to use the CFPs, which allow us to build $n$-particle antisymmetric wave functions from antisymmetric states with $n-1$ or $n-2$ particles.
The one-body CFP is denoted as
\[
\big[ j^{\,n-1}\alpha_{n-1}J_{n-1};\, j \; \big\| \; j^{\,n}\alpha_n J \big]
\]
which is the corresponding expansion coefficient for the expression of an $n$-particle  wave function in terms of an  $(n-1)$-particle wave function coupled with one additional particle:
\begin{eqnarray}
	| j^n \alpha J &= \sum_{\alpha_{n-1}, J_{n-1}} 
	\big[ j^{n-1} \alpha_{n-1} J_{n-1}; j \, || \, j^n \alpha J \big] \,\rangle \nonumber\\
	&\times | (j^{n-1} \alpha_{n-1} J_{n-1} \otimes j) J \rangle,
\end{eqnarray}
where $\alpha$ is the additional index needed to label the state. Unlike Eq. (\ref{4pp}) where the direct product gives an over-complete basis, $| j^n \alpha J \rangle$ defines a set of complete and orthonormal bases.

The two-body CFP is defined in a similar manner as
\[
\big[ j^{\,n-2}\alpha_{n-2}J_{n-2};\; (jj)J_{12}\; \big\| \; j^n \alpha J \big]
\]
It expresses an $n$-particle state in terms of an $(n-2)$-particle wave function coupled to a pair:
\begin{eqnarray}
	| j^n \alpha J \rangle = \sum_{\alpha_{n-2}, J_{n-2}, J_{12}} 
	\big[ j^{n-2} \alpha_{n-2} J_{n-2}; (jj) J_{12} \, || \, j^n \alpha J \big] \nonumber\\
	\times| ( (j^{n-2} \alpha_{n-2} J_{n-2}) \otimes (j^2 J_{12}) ) J \rangle.
\end{eqnarray}

CFPs are particularly convenient for evaluating matrix elements of one-body and two-body creation and annihilation operators. For example, one has:
\[
\langle j^{n}\alpha J \| a_j^\dagger \| j^{n-1}\alpha' J'\rangle
= \sqrt{n}\; \big[ j^{\,n-1}\alpha'J'; j \;\big\|\; j^{\,n}\alpha J \big].
\]
There can be a phase factor of $(-1)^{n-1}$
 or similar in above equation, depending on the ordering of creation operators and definition of the CFP.
For $n=3$, the one-body expansion is
\[
| j^3 \alpha J \rangle
= \sum_{\alpha_2,J_{2}}
\big[ j^{2}\alpha_{2}J_{2}; j \;\big\|\; j^{3}\alpha J \big]\;
\big| (j^{2}\alpha_{2}J_{2}) \otimes j;\; J \big\rangle.
\]
Here $\alpha_2$ is apparently redundant as the two-particle systems are uniquely defined by angular momentum $J_2$.
The diagonal two-body Hamiltonian ($H=V$) expectation value is
\[
\langle j^3\alpha J | H | j^3\alpha J \rangle
= 3 \sum_{J_{2}}
\big| \big[ j^{2}J_{2}; j\big\| j^{3}\alpha J \big]\big|^2
\; \langle j^2 J_{2} | V | j^2 J_{2}\rangle.
\]
Practical techniques for the evaluation of the CFPs can be found for example in Ref.~\cite{Deveikis_2002}.

\subsection{Seniority and CFPs}
With the introduction of seniority, the recursion relation with one-body CFPs becomes
\begin{eqnarray}
	| j^n\alpha v J \rangle
	= \sum_{\alpha',v',J'} 
	\big[ j^{n-1}\alpha', v' J'; j \,\|\, j^n\alpha v J \big]\,\nonumber\\
	\times|(j^{n-1}\alpha' v' J' \otimes j)J \rangle.
\end{eqnarray}
Similarly, one has for  two-body CFPs
\begin{eqnarray}
	| j^n \alpha v J \rangle
	= \sum_{\alpha'',v'',J'',J_{12}}
	\big[ j^{n-2} \alpha''v'' J'';\; (j^2 J_{12}) \,\|\, j^n\alpha v J \big]\,\nonumber\\
	\times\big|(j^{n-2} v'' \alpha''J'' \otimes j^2 J_{12}) J \rangle.
\end{eqnarray}
The seniority coupling scheme greatly simplifies these expansions. One-body CFPs connect $n$-particle states of seniority $v$ to $(n-1)$-particle states of $v \pm 1$, while two-body CFPs connect $n$-particle states to $(n-2)$-particle states of $v$ or $v \pm 2$. In particular, for a two-body interaction $V$:
\begin{eqnarray}
	\langle j^n \alpha v J | V | j^n\alpha' v' J \rangle
	\nonumber\\= \frac{n(n-1)}{2}\,
	\sum_{\alpha'',v'',J'',J_{2}}
	\big[ j^{n-2} \alpha'' v'' J''; (jj)J_{2} \,\|\, j^n \alpha v J \big]\,\nonumber\\
	\times\big[ j^{n-2} \alpha'' v'' J''; (jj)J_{2} \,\|\, j^n \alpha'v' J \big]\,
	\langle j^2 J_{2} | V | j^2 J_{2} \rangle.
\end{eqnarray}
As a result, the interaction matrix elements reduce to known two-particle matrix elements (TBMEs), $V_J \equiv
\langle j^2;J \,\|\, V \,\|\, j^2;J\rangle$, weighted by product of CFPs.

We take again the $n=3$ system as an example.
Its expansion in one-body form becomes
\begin{eqnarray}
| j^3, \alpha vJ \rangle
= \sum_{v_2,J_2}
\big[ j^2 v_2 J_2; j \,\|\, j^3 \alpha v J \big]\;
|(j^2 v_2 J_2 \otimes j)J\rangle.
\end{eqnarray}
Here $v_2$ is redundant. The pair takes only $v_2=0$, if the 3-particle state has $v=1$ (the $J=9/2$ ``quasi-particle''). If the 3-particle state has $v=3$, then the pair has $v_2=2$.
The two-body Hamiltonian matrix is
\begin{multline}
\langle j^3 \alpha vJ | V | j^3 \alpha' v'J \rangle \,\\= 3 \sum_{J_2} 
 [(j^2 J_2); j \,\|\, j^3\alpha vJ][(j^2 J_2); j \,\|\, j^3\alpha' v'J]
\; \langle j^2 J_2 | V | j^2 J_2 \rangle,
\end{multline}
where we dropped $v_2$ for simplicity.
For a system with $n=4$ particles, the
expansion in two-body form is
\begin{multline}
	| j^4, \alpha vJ \rangle
	= \sum_{\alpha_1 v_1,J_1,J_{2}}
	\big[ j^2 \alpha_1 v_1J_1;\, (j^2 J_{2}) \,\|\, j^4 \alpha vJ \big]\,\\
	|(j^2 \alpha_1v_1 J_2 \otimes j^2 J_{2})J\rangle.
\end{multline}
Here again the indices $\alpha_1 v_1$ is redundant which is kept for completeness. With those neglected,
the two-body Hamiltonian matrix becomes
\begin{multline}
\langle j^4 \alpha vJ | V | j^4\alpha' v'J \rangle
= 6 \sum_{J_1,J_{2}}
\big[ j^2J_1;\, (j^2 J_{2})\,\|\, j^4 \alpha vJ \big]\,\\
\big[j^2J_1;\, (j^2 J_{2})\,\|\, j^4 \alpha' v'J \big]\,
\langle j^2 J_{2} | V | j^2 J_{2} \rangle.
\end{multline}

\section{Number of pairs} \label{number-p}

A useful (but sometimes controversial) quantity in shell model studies is the number of pairs, or equivalently, the number of interaction links. This measure provides insight into how different two-body couplings contribute to the overall energy and the structure of the nuclear wave function.
For nucleons restricted to a single-$j$ orbital, the correlation energy can be expressed as~\cite{qi2010energy}
\begin{equation}
	E_I = C^I_J V_J,
\end{equation}
where $I$ denotes the total angular momentum of the system, $V_J$ are the TBMEs introduced above, and $C^I_J$ represents the number of nucleon pairs coupled to angular momentum $J$. When the two-body force $\hat{V}$ respects isospin symmetry, the set of independent TBMEs spans angular momenta $J=0$ to $2j$, yielding a total of $2j+1$ values.

The total number of pairs in a system with $n$ nucleons is given by
\begin{equation}
	\sum_{J} C^I_J = \frac{n(n-1)}{2},
\end{equation}
and the number of pairs with odd angular momentum is
\begin{equation}
	\sum_{J;{\rm odd}} C^I_J = \tfrac{1}{2}\left[\tfrac{n}{2}\left(\tfrac{n}{2}+1\right) - T(T+1)\right],
\end{equation}
where $T$ is the total isospin quantum number.
All pairs carry even spin for a system with identical particles with $T=n/2$. In general, one has
\begin{equation}
	\sum_{J;\text{even}} C^I_J
	= \frac{3}{8}n^2 - \frac{3}{4}n + \tfrac{1}{2}T(T+1)
	= \frac{3}{8}n(n-2) + \tfrac{1}{2}T(T+1).
\end{equation}

Possible confusion arises from the fact that, by definition, the number of pairs does not vanish for a completely filled single-$j$ shell. Instead, the system energy reduces to
\begin{equation}
	E_0 = \sum_{J}(2J+1)V_J,
\end{equation}
which corresponds purely to the monopole contribution, with no additional correlations. The monopole component of the interaction is defined as the weighted average of TBMEs:
\begin{eqnarray}
	\nonumber V_{jj'} &=& \frac{\sum_J (2J+1)V^J_{jj'jj'}}{\sum_J (2J+1),[1-\delta_{jj'}(-1)^J]} \\
	&=& \frac{\sum_J (2J+1)V^J_{jj'jj'}}{2j+1} \cdot \frac{1+\delta_{jj'}}{2j'+1-\delta_{jj'}}.
\end{eqnarray}
Here the TBMEs of a general interaction $V$ in the coupled angular-momentum basis are defined as
\begin{equation}
V^J_{jj'jj'} = \langle (j j');J | V | (j j');J \rangle,
\end{equation}
where $|(j j') J\rangle$ are normalized coupled two-particle states.

Its contribution to the total energy is then
\begin{eqnarray}
	E_m = \sum_{j,j'} V_{jj'} \left\langle \frac{n_j (n_{j'} - \delta_{jj'})}{1+\delta_{jj'}} \right\rangle,
\end{eqnarray}
where $n_j$ denotes the number of particles in orbital $j$.

If the interaction is restricted to the pairing channel, one has
\begin{equation}
	V^{J=0}{jjjj} =V_0= -\Omega G_{jj}, \quad V_{jj} = -\frac{G_{jj}}{2j},
\end{equation}
with $G_{jj}$ representing the pairing strength for orbital $j$.

\section{Seniority conservation condition in general and number of states}
There have been extensive studies on the necessary and sufficient conditions
for an interaction to conserve seniority~\cite{talmi1993,Isacker1}. The explicit constraints are given above for certain $j$ values. In general, for a given $j$ system, they can be derived to be~\cite{qi2010alternate}
\begin{eqnarray}
\sum_J	a_{\lambda J}V_J=0
\end{eqnarray}
\begin{eqnarray}
	a_{\lambda J} = \Tilde{C} \left[ 
	\frac{4(2J + 1)}{(2j + 1)(2j - 1)} 
	- \delta_{J \lambda} 
	- 2(2J_{\alpha} + 1) 
	\begin{Bmatrix} 
		j & j & J \\ 
		j & j & \lambda 
	\end{Bmatrix} 	\right]\nonumber\\
\times \sqrt{2{\lambda}+1}~~~~
\end{eqnarray}
where $\Tilde{C}$
 is an overall scaling factor chosen for convenience so that all $a_{\lambda J}$ are integers, $J$ and $\lambda$ are angular momentum quantum numbers which can take values between 2 and $(2j-1)$, $\delta_{J \lambda}$ is the Kronecker delta function, and the $6$-$j$ symbol $\begin{Bmatrix} j & j & J_{\alpha} \\ j & j & \lambda \end{Bmatrix}$ arises from the coupling of angular momenta. This equation provides a practical way to determine whether a given two-body interaction preserves seniority in a system of three identical fermions in a single-$j$ shell, and the integer factor $C$ simplifies further algebraic analysis.
The constraint condition for different $\lambda$ may be the same which effectively reduce the total number of constraints. 

The total number of constraints equals the number of $I=j$ states for an $n=3$ system and the number of $I=0$ states for an $n=4$ system (these two numbers coincide). 
These constraints are obtained by demanding that all non-diagonal two-body matrix elements between the $v=1$ and $v=3$ ($I=j$) states for $n=3$, and between the $v=0$ and $v=4$ ($I=0$) states for $n=4$, vanish~\cite{qi2010alternate,qi2010energy}. 
As shown in Table~\ref{tab:number}, there are two such states in each case.

 If seniority is conserved, the overlap of the unique $v=1,\;I=j$ three-particle state with the coupled pair state $j^2(J)\otimes j$ reads
\begin{equation}
	\langle \Psi_{v=1} \,|\, j^2(J) j; j \rangle 
	= \sqrt{\frac{2j + 1}{2j - 1}} \left( \delta_{J, 0} - \frac{2\sqrt{2J+1}}{2j+1}\right),
\end{equation}
from which the squared overlaps entering the two-body expectation value follow:
\[
\begin{aligned}
\big|\langle \Psi_{v=1}\,|\,j^2(0)j;j\rangle\big|^2 &= \frac{2j-1}{2j+1},\\[4pt]
\big|\langle \Psi_{v=1}\,|\,j^2(J)j;j\rangle\big|^2 &= \frac{4(2J+1)}{(2j+1)(2j-1)} \quad (J>0).
\end{aligned}
\]
Those could be compared with the CFPs defined above. Consequently, the total expectation value of the two-body Hamiltonian for the three-particle state (summing over the three distinct pairs) is (see more explanation in Sec. \ref{number-p} above)
\begin{equation}
	E = j^3\Psi|V|j^3\Psi\rangle=\frac{2j - 1}{2j + 1} V_0 + \sum_{J>0} \frac{4 (2J + 1)}{(2j + 1)(2j - 1)} V_J.
\end{equation}

Consequently, the total expectation value of the two-body Hamiltonian for the three-particle state (summing over the three distinct pairs) is
\[
\langle j^3\Psi|V|j^3\Psi\rangle
= 3\left[\frac{2j-1}{2j+1}V_0+\sum_{J>0}\frac{4(2J+1)}{(2j+1)(2j-1)}V_J\right],
\]
where \(V_J=\langle j^2J|V|j^2J\rangle\). 

To end, we give the explicit linear constraints for  TBMEs to conserve seniority for some orbitals:
\begin{itemize}
	\item The TBMEs must satisfy one condition to conserve seniority for orbitals $j=9/2,11/2$ and 13/2:
	\begin{eqnarray}
			& j=9 / 2:& 65 V_2-315 V_4+403 V_6-153 V_8=0 \nonumber\\
			& j=11 / 2:& 1020 V_2-3519 V_4-637 V_6+4403 V_8\nonumber\\&&-2541 V_{10}=0 \nonumber\\
			& j=13 / 2: &1615 V_2-4275 V_4-1456 V_6+3196 V_8\nonumber\\&&-5145 V_{10}-4225 V_{12}=0 
	\end{eqnarray}
	\item For $j = 15/2$, two constraints are required:
	\begin{equation}
		\begin{aligned}
			& \begin{array}{l}
1330 V_2-2835 V_4-1807 V_6+612 V_8 \\
				+3150 V_{10}+3175 V_{12}-3625 V_{14}=0
			\end{array} \\
			& \text { and }\\ &\begin{array}{l}
				77805 V_2-169470 V_4-85527 V_6-4743 V_8 \\
				+222768 V_{10}+168025 V_{12}-208858 V_{14}=0 .
			\end{array}
		\end{aligned}
	\end{equation}
\end{itemize}

\section{M-scheme and symbolic angular momentum projection}
\label{Mscheme}
In the  $M$-scheme, only the conservation of parity and total magnetic quantum number $M=\sum_{i=1}^{n}m_{i}$ is considered.
The uncoupled basis for a single-$j$ system with $n$ particles can be expressed as
\begin{equation}\label{mf}
|M\rangle=|m_1m_2\cdots m_n\rangle,
\end{equation}
where one can require that $m_1>m_2>\cdots>m_n$. One can introduce an alternate representation starting with $\Xi$ defined by $M=M_{max}-\Xi$. Then one can further defined a set of $\xi$ values for each particle in the following manner:
\begin{equation}
m_1=j-\xi_1, m_2=j-1-\xi_2, \cdots, m_n=j-(n-1)-\xi_n.
\end{equation}
It can be easily realized that in the above definitions one has $\xi_n\geq\cdots\geq\xi_2\geq\xi_1$.
In addition, one has
$$M=\sum_{i=1}^{n}m_{i}=nj+n-\sum_{i=1}^{n}(i+\xi_i)=M_{max}-\sum_{i=1}^{n}i\xi_i$$
which leads to the condition $\xi_1+\xi_2+\cdots+\xi_n=\Xi$. The total number of such kind of combinations or partitions $[\xi_1,\xi_2,\cdots\xi_n]$ that satisfies the above restriction defines the total number of $M$-scheme bases and the partition number of $[\xi]$, which can be expressed via the Young diagram.

For a given total angular momentum value $I$, a coupled state $|\alpha IM\rangle$ with additional index $\alpha$ can be expanded in terms of the $M$-scheme bases as
\begin{equation}\label{proj2}
|\alpha IM\rangle=\sum_{\xi}c_{\xi}^{\alpha IM}P_{M}^{I}|M[\xi]\rangle,
\end{equation}
where $[\xi]$ runs over part or all the combinations and  $P_{M}^{I}$ is the angular momentum projection operator. The coefficients \(c_{\xi}\) can be constrained in various ways as long as they ensure the orthonormality relation among the resulting coupled states is satisfied~\cite{qi2007modern}. The projection operator can be expressed in a matrix form as (see, for example, Ref.~\cite{guidry2022symmetry})
\begin{align}\label{proj3a}
\langle M[\xi']|P_{M}^{I}|M[\xi]\rangle=\frac{2I+1}{8 \pi^2} \int_{\Omega} d \alpha \sin \beta d \beta d \gamma D^{*I}_{M M}(\Omega)\nonumber\\
\left\langle M[\xi']\right| \exp \left[i \alpha J_z\right] \cdot \exp \left[i \beta J_y\right] \cdot \exp \left[i \gamma J_z\right]|M [\xi]\rangle
\end{align}
where $\Omega=(\alpha,\beta, \gamma)$ are the Euler angles and $D$ is the usual Wigner D function that can be given in general as
\[
D^{\,j}_{m m'}(\alpha,\beta,\gamma)
= e^{-i m \alpha}\; d^{\,j}_{m m'}(\beta)\; e^{-i m' \gamma}.
\]
The small d function is defined as
\[
d^{\,j}_{m m'}(\beta) \;=\; \langle j m \,|\, e^{-i \beta J_y} \,|\, j m' \rangle,
\]
It can be given in closed form as
\begin{align}
d^{\,j}_{m m'}(\beta)
=&
\sum_{k}
(-1)^{\,k - m + m'}\,
\nonumber\\
&\times\frac{\sqrt{(j{+}m)!\,(j{-}m)!\,(j{+}m')!\,(j{-}m')!}}
     {(j{+}m-k)!\;k!\;(m'-m+k)!\;(j-m'-k)!}\;
\nonumber\\
&\times\bigl(\cos \tfrac{\beta}{2}\bigr)^{\,2j+m-m'-2k}\;
\bigl(\sin \tfrac{\beta}{2}\bigr)^{\,m'-m+2k},
\end{align}
where the sum runs over all integers $k$ such that every factorial argument is nonnegative.
The projector matrix can be simplified as
\begin{align}\label{proj3}
\langle M[\xi']|P_{M}^{I}|M[\xi]\rangle=&\frac{2I+1}{2}\int_{0}^{\pi}\sin\theta d\theta d_{MM}^{I}(\theta)\nonumber\\&\langle M[\xi']|\exp[i\theta I_y]|M[\xi]\rangle.
\end{align}
In practice, one can always choose the set of $M$-scheme bases with $M=J$. All the above operations can be done via symbolic calculations. At the end, one evaluates the Hamiltonian matrix in the coupled bases set $|\alpha IM\rangle$ which again can be diagonalized symbolically. It may be useful to mention that the coupled bases with conserved angular momentum $|\alpha IM\rangle$ are not uniquely defined if there is more than one state for a given $I$. One would get a new set of bases from any linear independent combinations of the old ones.

The seniority symmetry is not automatically conserved in the above projected bases. 
Besides directly solving the eigenvectors of the above projection matrix, there is another alternative to get the coefficient $c_{\xi}^{\alpha JM}$. One can rewrite Eq.~(\ref{proj2}) as
\begin{equation}\label{wave}
|\alpha IM v\rangle=\sum_{\xi''}c_{\xi''}^{\alpha IM}P_{M}^{I}|[\xi'']\rangle,
\end{equation}
where $\xi''$ denotes a set of trial wave functions  which can be determined by the requirement that $|\alpha IMv\rangle$ is an eigenstate of a non-zero pairing Hamiltonian. 
Again the Hamiltonian to be studied can be constructed and diagonalized symbolically in the set of seniority-conserving bases as obtained.
The symbolic shell-model program described above is freely available, although only the Mathematica version is currently fully implemented.

\section{Angular momentum operator and coupling of four identical fermions in a single shell}

In this section, we would like to take the system with four identical fermions in a single spherical shell with single-particle angular momentum \(j=\tfrac{9}{2}\) as an example for illustrate the angular momentum operation. Let \(a^\dagger_m\) and \(a_m\) be the usual fermionic creation and annihilation operators for the single-particle state \(|jm\rangle\) with magnetic quantum number \(m\in\{-j,-j+1,\dots,j\}\). The vacuum (no particles) is denoted \(|0\rangle\).
Again, we take total angular momentum (vector) operators for the four-particle system as
\[
\mathbf{I} = (I_x,I_y,I_z),
\]
we have
\[
\,\mathbf{I}^2 \;=\; I_0^2 \;+\; \tfrac{1}{2}\big(I_+I_- + I_-I_+\big)\,
\]
which can be obtained by writing the Cartesian Casimir $\mathbf I^2=I_x^2+I_y^2+I_z^2$ and substituting
\(
I_x=\tfrac{1}{2}(I_++I_-),\; I_y=\tfrac{1}{2i}(I_+-I_-),\; I_z=I_0
\),
then simplifying. Using $[I_+,I_-]=2I_0$ one gets the equivalent operator identities
\[
\begin{aligned}
I_+I_- &= \mathbf{I}^2 - I_0(I_0-1), \\
I_-I_+ &= \mathbf{I}^2 - I_0(I_0+1).
\end{aligned}
\]
These relations are obtained by solving the symmetric form for $I_+I_-$ and $I_-I_+$ using the commutator.
Acting on an eigenstate $\mathbf{I}^2|I,m\rangle=I(I+1)|I,m\rangle$ and $I_0|I,m\rangle=m|I,m\rangle$, one recovers the familiar ladder relations
\[
I_+I_-|I,m\rangle=(I+m)(I-m+1)\,|I,m\rangle,
\]
\[
I_-I_+|I,m\rangle=(I-m)(I+m+1)\,|I,m\rangle,
\]
which follow from the asymmetric formulas above.

Based on the single-particle commutation relations given in Ref.~\ref{eq:Ialpha}, we can determine the action of \(I_+\) on an anti-symmetrized product (Slater) state
\[
|m_1 m_2 m_3 m_4\rangle \equiv \alpha^\dagger_{m_1}\alpha^\dagger_{m_2}\alpha^\dagger_{m_3}\alpha^\dagger_{m_4}\,|0\rangle,
\]
with the convention \(j\ge m_1>m_2>m_3>m_4\ge -j\). Concretely,
\[
\begin{aligned}
I_+ |m_1 m_2 m_3 m_4\rangle
= \sum_{k=1}^4 &\bigl(\text{state obtained from }|m_1m_2m_3m_4\rangle \\
&\quad \text{by replacing } m_k\mapsto m_k+1\bigr),
\end{aligned}
\]
where any term that produces a repeated magnetic index or exceeds \(j\) vanishes, and appropriate permutation signs must be inserted to reorder the indices into decreasing order.

A state \(|\psi\rangle\) has total angular momentum \(I=0\) if it is annihilated by the raising operator:
\[
I_+|\psi\rangle = 0,\qquad I_0|\psi\rangle=0.
\]

This provides a direct linear-algebraic proof: showing two independent solutions and verifying the kernel dimension confirms that they are \(I=0\) states.

Below we consider two basis terms as examples to illustrate the operation. For the basis
\[
\big| 9/2, 7/2, -7/2, -9/2 \big\rangle
=\alpha^\dagger_{9/2}\alpha^\dagger_{7/2}\alpha^\dagger_{-7/2}\alpha^\dagger_{-9/2}|0\rangle,
\]
One gets after applying \(I_+\):
\[
\begin{aligned}
I_+&\big| 9/2, 7/2, -7/2, -9/2 \big\rangle
= \alpha^\dagger_{9/2}\alpha^\dagger_{7/2}\alpha^\dagger_{-7/2}\underbrace{[I_+,\alpha^\dagger_{-9/2}]}_{=\alpha^\dagger_{-7/2}}|0\rangle \\
&\qquad + \alpha^\dagger_{9/2}\alpha^\dagger_{7/2}\underbrace{[I_+,\alpha^\dagger_{-7/2}]}_{=\alpha^\dagger_{-5/2}}\alpha^\dagger_{-9/2}|0\rangle\\
&\qquad + \alpha^\dagger_{9/2}\underbrace{[I_+,\alpha^\dagger_{7/2}]}_{=\alpha^\dagger_{9/2}}\alpha^\dagger_{-7/2}\alpha^\dagger_{-9/2}|0\rangle \\
&\qquad + \underbrace{[I_+,\alpha^\dagger_{9/2}]}_{=0}\;\alpha^\dagger_{7/2}\alpha^\dagger_{-7/2}\alpha^\dagger_{-9/2}|0\rangle.
\end{aligned}
\]
A term-by-term evaluation, with the Pauli exclusion principle enforced, shows that most contributions vanish. The \(m=9/2\) term gives zero, as \([I_+, \alpha^\dagger_{9/2}] = 0\). The term with \([I_+, \alpha^\dagger_{7/2}] = \alpha^\dagger_{9/2}\) produces a duplicate \(\alpha^\dagger_{9/2} \alpha^\dagger_{9/2}\) and therefore vanishes, and similarly, \([I_+, \alpha^\dagger_{-9/2}] = \alpha^\dagger_{-7/2}\) leads to a duplicate \(\alpha^\dagger_{-7/2} \alpha^\dagger_{-7/2}\) and vanishes. The only nonzero contribution arises from raising \(-7/2 \to -5/2\). Consequently, one obtains
\[
I_+ \,\big| 9/2, 7/2, -7/2, -9/2 \big\rangle
= \big| 9/2, 7/2, -5/2, -9/2 \big\rangle
\]
In the same way, one can get for another basis
\[
\begin{aligned}
I_+ &\big| 9/2, 5/2, -5/2, -9/2 \big\rangle
= \underbrace{[I_+,\alpha^\dagger_{9/2}]}_{=0}\,\alpha^\dagger_{5/2}\alpha^\dagger_{-5/2}\alpha^\dagger_{-9/2}|0\rangle \\
&\quad + \alpha^\dagger_{9/2}\underbrace{[I_+,\alpha^\dagger_{5/2}]}_{=\alpha^\dagger_{7/2}}\alpha^\dagger_{-5/2}\alpha^\dagger_{-9/2}|0\rangle \\
&\quad + \alpha^\dagger_{9/2}\alpha^\dagger_{5/2}\underbrace{[I_+,\alpha^\dagger_{-5/2}]}_{=\alpha^\dagger_{-3/2}}\alpha^\dagger_{-9/2}|0\rangle \\
&\quad + \alpha^\dagger_{9/2}\alpha^\dagger_{5/2}\alpha^\dagger_{-5/2}\underbrace{[I_+,\alpha^\dagger_{-9/2}]}_{=\alpha^\dagger_{-7/2}}|0\rangle.
\end{aligned}
\]

Evaluating each nonzero term, noting that no Pauli duplicates occur, we find the following contributions. Raising \(m=5/2\) gives
\[
\alpha^\dagger_{9/2}\alpha^\dagger_{7/2}\alpha^\dagger_{-5/2}\alpha^\dagger_{-9/2}|0\rangle
= \big| 9/2, 7/2, -5/2, -9/2 \big\rangle,
\]
raising \(m=-5/2\) gives
\[
\alpha^\dagger_{9/2}\alpha^\dagger_{5/2}\alpha^\dagger_{-3/2}\alpha^\dagger_{-9/2}|0\rangle
= \big| 9/2, 5/2, -3/2, -9/2 \big\rangle,
\]
and raising \(m=-9/2\) gives
\[
\alpha^\dagger_{9/2}\alpha^\dagger_{5/2}\alpha^\dagger_{-5/2}\alpha^\dagger_{-7/2}|0\rangle
= \big| 9/2, 5/2, -5/2, -7/2 \big\rangle.
\]

Hence, the total result is the sum of these three basis states:
\[
\begin{aligned}
I_+\,\big| 9/2, 5/2, -5/2, -9/2 \big\rangle
&= \big| 9/2, 7/2, -5/2, -9/2 \big\rangle\\
&+ \big| 9/2, 5/2, -3/2, -9/2 \big\rangle \\
& + \big| 9/2, 5/2, -5/2, -7/2 \big\rangle.
\end{aligned}
\]

\begin{table*}[htbp]
\centering
\caption{$I^2$ matrix elements expressed in the 18 $M$-scheme bases with $M=0$ for four particles in $j=9/2$ system. The order of the bases are the same as in Table \ref{phi1}.}
\begin{tabular}{cccccccccccccccccc}
\hline
16 & 4 & 0 & 0 & 0 & 0 & 0 & 0 & 0 & 0 & 0 & 0 & 0 & 0 & 0 & 0 & 0 & 0 \\
64 & 46 & 9 & 3 & 0 & 0 & 0 & 3 & 1 & 0 & 0 & 0 & 0 & 0 & 0 & 0 & 0 & 0 \\
0 & 49 & 54 & 7 & 16 & 4 & 0 & 7 & 0 & 4 & 1 & 0 & 0 & 0 & 0 & 0 & 0 & 0 \\
0 & 63 & 27 & 42 & 0 & 12 & 0 & 0 & 7 & 0 & 3 & 0 & 0 & 0 & 0 & 0 & 0 & 0 \\
0 & 0 & 36 & 0 & 33 & 6 & 0 & 0 & 0 & 6 & 0 & 1 & 0 & 0 & 0 & 0 & 0 & 0 \\
0 & 0 & 54 & 42 & 36 & 64 & 10 & 0 & 0 & 0 & 6 & 4 & 2 & 0 & 0 & 0 & 0 & 0 \\
0 & 0 & 0 & 0 & 0 & 40 & 25 & 0 & 0 & 0 & 0 & 0 & 5 & 0 & 0 & 0 & 0 & 0 \\
0 & 63 & 27 & 0 & 0 & 0 & 0 & 42 & 7 & 12 & 3 & 0 & 0 & 0 & 0 & 0 & 0 & 0 \\
0 & 81 & 0 & 27 & 0 & 0 & 0 & 27 & 30 & 0 & 9 & 0 & 0 & 0 & 0 & 0 & 0 & 0 \\
0 & 0 & 54 & 0 & 36 & 0 & 0 & 42 & 0 & 64 & 6 & 4 & 0 & 10 & 2 & 0 & 0 & 0 \\
0 & 0 & 81 & 63 & 0 & 36 & 0 & 63 & 49 & 36 & 70 & 16 & 8 & 0 & 8 & 4 & 0 & 0 \\
0 & 0 & 0 & 0 & 81 & 54 & 0 & 0 & 0 & 54 & 36 & 49 & 12 & 0 & 12 & 0 & 4 & 0 \\
0 & 0 & 0 & 0 & 0 & 72 & 45 & 0 & 0 & 0 & 48 & 32 & 57 & 0 & 0 & 12 & 8 & 0 \\
0 & 0 & 0 & 0 & 0 & 0 & 0 & 0 & 0 & 40 & 0 & 0 & 0 & 25 & 5 & 0 & 0 & 0 \\
0 & 0 & 0 & 0 & 0 & 0 & 0 & 0 & 0 & 72 & 48 & 32 & 0 & 45 & 57 & 12 & 8 & 0 \\
0 & 0 & 0 & 0 & 0 & 0 & 0 & 0 & 0 & 0 & 64 & 0 & 32 & 0 & 32 & 40 & 16 & 0 \\
0 & 0 & 0 & 0 & 0 & 0 & 0 & 0 & 0 & 0 & 0 & 64 & 48 & 0 & 48 & 36 & 61 & 9 \\
0 & 0 & 0 & 0 & 0 & 0 & 0 & 0 & 0 & 0 & 0 & 0 & 0 & 0 & 0 & 0 & 49 & 21 \\
\hline
\end{tabular}
\end{table*}

\section{\(B(E2)\) value and half-life of the  $8^+$ isomeric state $^{128}$Pd}
We are interested in re-extracting the \(B(E2)\) value from the measured half-life~\cite{PhysRevLett.111.152501} here, as the number was not given in that paper and we had difficulty reproducing the value in the figure using the known simple \(T_{1/2}-B(E2)\) transformation formula.
Here we evaluated the $B(E2)$ value from experimental half-life with the formula
\begin{eqnarray}
B(E2; J_i \rightarrow J_f) = 
\frac{L\,[\,(2L+1)!!\,]^2}{8\pi(L+1)}\;
\hbar\,(\hbar c)^{2L+1}\;\frac{b_r}{T_{1/2} \cdot E_{\gamma}^5(1+\alpha)}\nonumber\\
=
\frac{\ln(2) \cdot 75}{4\pi} \cdot \hbar \, (\hbar c)^5 \frac{b_r}{T_{1/2} \cdot E_{\gamma}^5(1+\alpha)}~~~~~~~~~~~
\end{eqnarray}
 where \(\alpha\) is the internal conversion coefficient and \(b_{r}\) is the branching ratio. Inserting all the constants from the latest CODATA 2022 database, the transformation can be simplified to
\begin{equation}
B(E2)  (e^2\text{fm}^4) = \frac{564.677 \cdot b_r}{T_{1/2} (\text{ps}) \cdot [E_{\gamma} (\text{MeV})]^5 (1+\alpha)}.
\end{equation}
We further evaluate the uncertainty through common error propagation as follows:
\[
\left(\frac{\Delta B}{B}\right)^2 
= \left(\frac{\partial \ln B}{\partial E_\gamma} \, \Delta E_\gamma\right)^2 
+ \left(\frac{\partial \ln B}{\partial T_{1/2}} \, \Delta T_{1/2}\right)^2
+ \left(\frac{\partial \ln B}{\partial \alpha} \, \Delta \alpha\right)^2 .
\]
The results we got from the reported half-life and the $\alpha$ value as suggested in the paper ($\alpha=2.6(17)$) are 
\[
B(E2;8^+\rightarrow 6^+) = 11.321 \pm 5.570 e^2 \text{fm}^4,
\]
for $^{128}$Pd, which, irrespective of the large error, is much smaller in comparison with that for $^{130}$Cd:
\[
B(E2;8^+\rightarrow 6^+) = 51.284 \pm 6.996 e^2 \text{fm}^4.
\]
Here, as there is no experimental information, we assumed $b_r=1$ and $\alpha=0$, corresponding to the case where the transition proceeds entirely through the observed 
$\gamma$-ray and no internal conversion occurs. If 
$b_r<1$ or $\alpha>0$ the deduced 
$B(E2)$ value would correspondingly be a lower limit.
\section{The np coupling scheme}

The np pairing was mostly discussed within BCS-like pairing models. Within the shell-model framework, the np pair interaction often leads to complex structures. On the other hand, the spin-aligned np pair coupling bear quite some similarities  with the seniority coupling, though it deals with a significantly larger number of configurations even for single-$j$ systems. A common way to describe the coupling of a system with both neutrons and protons is to decompose it into proton and neutron blocks. We consider a simple example of two np pairs in a single-$j$ shell, the wave function of a given state with total angular momentum $I$ can
be written as (see, for example, Ref.~\cite{qi2010energy}),
\begin{equation}\label{4pn}
	|\Psi_I\rangle=\sum_{J_{p}, J_n} X_I(J_pJ_n) | j_{\pi}^2(J_p)j_{\nu}^2(J_n);I \rangle,
\end{equation}
where $X_I(J_pJ_n)$ is the amplitude of the four-body wave function and $J_p$ and $J_n$ are even numbers denoting the angular momenta of the proton and neutron pairs, respectively.

The four nucleons can couple to spin $I=0$ to $2(2j-1)$ and isospin $T=0$, 1 and $2$. 
The single-$j$ Hamiltonian can be written as,
\begin{eqnarray}\label{4h}
	\nonumber \langle  j_{\pi}^2(J_p)j_{\nu}^2(J_n);I |\hat{V}| j_{\pi}^2(J_p')j_{\nu}^2(J_n');I \rangle\\
	= (V_{J_p}+V_{J_n})\delta_{J_pJ_p'}\delta_{J_nJ_n'}
	+\sum_JM^I_J(J_pJ_n;J_p'J_n')V_J,
\end{eqnarray}
where the spin $J$ can take both even and odd values ($J=0$ to $2j$).
The symmetric matrix $M$ is given
as
\begin{eqnarray}
	\nonumber M^I_J(J_pJ_n;J_p'J_n')=\sum_{\lambda}4\hat{J}_p\hat{J}_n\hat{J}_p'\hat{J}_n'\hat{J}^2\hat{\lambda}^2\left\{
	\begin{array}{ccc}
		J_p & J_n & I \\
		\lambda & j & j \\
	\end{array}\right\}\\
	\times\left\{\begin{array}{ccc}
		J_p' & J_n' & I \\
		\lambda & j & j \\
	\end{array}
	\right\}\left\{\begin{array}{ccc}
		j & j & J \\
		\lambda & j & J_n \\
	\end{array}
	\right\}\left\{\begin{array}{ccc}
		j & j & J \\
		\lambda & j & J_n' \\
	\end{array}
	\right\},
\end{eqnarray}
where $\hat{J}=\sqrt{2J+1}$ and $\lambda$ and $j$ are half integers. 
The number of nucleon pairs in the $n=4$ system can be calculated as,
\begin{eqnarray}
	\nonumber C^I_J=|X_I(J_pJ_n)|^2(\delta_{J_pJ}+\delta_{J_nJ})\\
	+\sum_{J_p,J_n;J_p',J_n'}X_I(J_pJ_n)M^I_J(J_pJ_n;J_p'J_n')X_I(J_p'J_n'),
\end{eqnarray}
where the first and second terms in the right-hand side give the numbers of identical nucleon pairs and proton-neutron pairs, respectively.

For example, in the hole-hole channel, the ground state wave function of $^{96}$Cd is calculated to be~\cite{qi2015n},
\begin{eqnarray}
	\nonumber|\Psi_0({\rm gs})\rangle&=& 0.76|[\pi^2(0)\nu^2(0)]_I\rangle + 0.57|[\pi^2(2)\nu^2(2)]_I\rangle \\
	\nonumber &+&0.24 |[\pi^2(4)\nu^2(4)]_I\rangle+0.13|[\pi^2(6)\nu^2(6)]_I\rangle\\
	&+&0.14|[\pi^2(8)\nu^2(8)]_I\rangle.
\end{eqnarray}
In indicates that the wave function is highly mixed and the seniority coupling is largely broken as
the normal like-particle pairing coupling scheme $(\nu^2)_0 \otimes (\pi^2)_0$
accounts for only about half of the ground state wave functions.

A remarkable feature of the spin-aligned np coupling scheme is that it allows the wave function  in Eq. (\ref{4pn}) to be re-expressed in an \textit{equivalent} representation in terms of $np$ pairs. This can be done analytically with the help of the overlap matrix as
\begin{eqnarray}\label{over}
	\nonumber\langle [\nu\pi(J_1)\nu\pi(J_2)]_I |[\pi^2(J_p)\nu^2(J_n)]_I\rangle \\
	= \frac{-2}{\sqrt{N_{J_1J_2}}}\hat{J_1}\hat{J_2}
	\hat{J_p}\hat{J_n}\left\{
	\begin{array}{ccc}
		j&j&J_p\\
		j&j&J_n\\
		J_1&J_2&I
	\end{array}
	\right\},
\end{eqnarray}
where $N$ denotes the normalization factor. The wave function in Eq. (\ref{4pn}) can be rewritten as
\begin{equation}\label{4pn-aligned}
	|\Psi_I\rangle=\sum_{J_{1},J_{2}} X_I(J_{1}J_2) | (j_{\pi}j_{\nu})({J_{1}})(j_{\pi}j_{\nu})({J_{2}});I \rangle,
\end{equation}
After the transformation, one found that for the wave function for a typical $2n2p$ system like $^{96}$Cd can be represented by the spin-aligned $np$ coupling scheme, with
$X^2_{J_1=J_2=2j}=0.92-0.95$ for most systems.
An even more striking feature is that the low-lying yrast states
are calculated to be approximately equally spaced and their spin-aligned $np$ 
structure is the same for all of them. 

This can be compared with systems with identical particles which can be described by the coupling of any pairs, not necessarily the seniority coupling scheme. However, they are the seniority and spin-aligned coupling schemes starting with the $J=0$ and $J=2j+1$ pairs that reveal that richest physics and exhibit the wave function with simplest structure.

One can consider all nn, pp and np pairs on the same footing which however results in an over-complete basis~\cite{WOS:000301612400004}.
We will use the Greek letter $\gamma_n$ to label the $n$-particle 
$np$ states. The $np$ states will be
$|\gamma_2\rangle=P^+(\gamma_2)|0\rangle$ where the $np$ creation operator is
$P^+(\gamma_2)= \sum_{i,p}X(ip;\gamma_2)c^+_ic^+_p$ and $c^+_i$ ($c^+_p$)
is the neutron (proton) single-particle creation operator. In the same fashion the
two-proton (two-neutron) creation operator will be denoted as $P^+(\alpha_2)$
($P^+(\beta_2)$). The four-particle
state, $|\gamma_4\rangle=P^+(\gamma_4)\vert 0\rangle$, is
\begin{eqnarray}
	\label{eq:wf4p}
	P^+(\gamma_4)=\sum_{\alpha_2,\beta_2}X(\alpha_2\beta_2;\gamma_4)
	P^+(\alpha_2) P^+(\beta_2)+\nonumber\\
	\sum_{\gamma_2\leq \gamma_2'}X(\gamma_2\gamma_2';\gamma_4)
	P^+(\gamma_2)P^+(\gamma_2'),~~~
\end{eqnarray}
where all possible like-particle and $np$ pairs are taken into account. In the two-pair case 
the basis elements $(\nu\nu)\otimes(\pi\pi)$ and $(\nu\pi)\otimes(\nu\pi)$ may 
be proportional to each other. 

A six-particle system can be written as the coupling of one pair times
two pairs as
\begin{equation}
\label{eq:wf6p}
P^+(\gamma_6)=
\sum_{\gamma_2,\gamma_4}X(\gamma_2\gamma_4;\gamma_6)
P^+(\gamma_2)P^+(\gamma_4).
\end{equation}

The eight-particle states can be written as
\begin{equation}
P^+(\gamma_8) =
\sum_{\gamma_4\leq \gamma_4'}X(\gamma_4 \gamma_4';\gamma_8)
P^+(\gamma_4)P^+(\gamma_4')
\end{equation}
One famous example that has been extensively studied in past decade is the $N=Z$ nucleus  $^{92}$Pd with four pairs in $j=9/2$ sub-shell.
The four $J=9$ $np$ pairs in $^{92}$Pd can couple in various ways. One should be able to re-project the wave function on the different coupling of np pairs. It was found out that the dominant component is the spin-aligned coupled system. Actually it turns out that the complex wave function for the ground state
 can be well represented by a single configuration $((((\nu\pi)_9\otimes(\nu\pi)_9)_{I'=0}\otimes(\nu\pi)_9)_{I''=9}\otimes(\nu\pi)_9)_{I=0}$. 
Various quartet-like models have been developed to describe $N=Z$ systems following a similar line~\cite{PhysRevC.91.064318,Sambataro_2022,Sandulescu_2024}.

\bibliographystyle{spphys_jn}       % APS-like style for physics
\bibliography{bibliography}   % name your BibTeX data base

\end{document}